\documentclass[a4paper,11pt]{article}
\usepackage{jcappub}
\usepackage[T1]{fontenc}
\usepackage{amscd, mathrsfs}
\usepackage{amsthm}
\usepackage{cases}
\usepackage{here}
\usepackage{bm}
\usepackage{color}
\usepackage{graphics}
\usepackage{ulem}
\usepackage{tabularx}
\usepackage{multirow}
\usepackage{subcaption}
\allowdisplaybreaks

\newcommand{\ctext}[1]{\raise0.2ex\hbox{\textcircled{\scriptsize{#1}}}}

\title{\boldmath New comprehensive description of the scaling evolution of the cosmological magneto-hydrodynamic system}

\author[a,b,c,d]{Fumio Uchida,}
\author[e,f]{Motoko Fujiwara,}
\author[g,h,b]{Kohei Kamada,}
\author[d,b,a,i]{and Jun'ichi Yokoyama}

\affiliation[a]{Department of Physics, Graduate School of Science,\\The University of Tokyo, Tokyo 113-0033, Japan}
\affiliation[b]{Research Center for the Early Universe (RESCEU), Graduate School of Science,\\The University of Tokyo, Tokyo 113-0033, Japan}
\affiliation[c]{Theory Center, Institute of Particle and Nuclear Studies (IPNS), High Energy Accelerator Research Organization (KEK), Tsukuba, Ibaraki 305-0801, Japan}
\affiliation[d]{Kavli Institute for the Physics and Mathematics of the Universe (Kavli IPMU),\\WPI, UTIAS, The University of Tokyo, Kashiwa, Chiba 277-8568, Japan}
\affiliation[e]{
Physics Department, TUM School of Natural Sciences, 
Technical University of Munich, James-Franck-Str. 1, 85748 Garching, Germany}
\affiliation[f]{
Department of Physics, University of Toyama, 3190 Gofuku, Toyama 930-8555, Japan
}
\affiliation[g]{School of Fundamental Physics and Mathematical Sciences, Hangzhou Institute for Advanced Study, University of Chinese Academy of Sciences (HIAS-UCAS), Hangzhou 310024, China}
\affiliation[h]{International Centre for Theoretical Physics Asia-Pacific (ICTP-AP), Beijing/Hangzhou, China}
\affiliation[i]{Trans-scale Quantum Science Institute,\\The University of Tokyo, Tokyo 113-0033, Japan}

\emailAdd{fumio.uchida@ipmu.jp}
\emailAdd{motoko@sci.u-toyama.ac.jp}
\emailAdd{kohei.kamada@ucas.ac.cn}
\emailAdd{junichi.yokoyama@ipmu.jp}

\abstract{
    We study the evolution of primordial magnetic fields until the recombination epoch,
    which is constrained by the conservation of magnetic helicity density if they are maximally helical and by the Hosking integral if they are non-helical.
    We combine these constraints with conditions obtained by estimating time scales of energy dissipation processes to describe the evolution of magnetic field strength and magnetic coherence length analytically.
    The dissipation processes depend on whether magnetic or kinetic energy is dominant, whether the decay dynamics is linear or not, and whether the dominant dissipation term is shear viscosity or drag force.
    We apply the description to compare constraints on primordial magnetic fields at different epochs in the early universe and argue that magnetogenesis before the electroweak symmetry breaking is not feasible.
}

\begin{document}
\maketitle
\flushbottom
\section{Introduction}
\noindent
The cosmological magnetic field is a promising messenger from the early universe.
Recently, gamma-ray observations suggested the existence of the intergalactic magnetic fields in void regions and put lower bounds on their strength \cite{NeronovVovk10, 2010MNRAS.406L..70T, 2010ApJ...722L..39A, 2011MNRAS.414.3566T, 2011ApJ...727L...4D, 2011APh....35..135E, 2011A&A...529A.144T, 2011ApJ...733L..21D, 2012ApJ...747L..14V, Takahashi:2013lba, 2015ApJ...814...20F, Ackermann+18, MAGIC:2022piy}.
In addition, multi-messenger observations have the potential to determine the coherence length of the intergalactic magnetic fields \cite{2020ApJ...902L..11A}.
If they have a cosmological origin, these observations can be discussed in connection with proposed generation mechanisms of primordial magnetic fields, which involve, {\it e.g.},  inflation and/or phase transitions \cite{1988PhRvD..37.2743T, 1992ApJ...391L...1R, 1992PhRvD..46.5346G, PhysRevLett.51.1488, 1989ApJ...344L..49Q, PhysRevD.50.2421, 1991PhLB..265..258V}.
Then, we can gain insights into the early universe by analyzing the properties of, in particular, the intergalactic magnetic fields.

For this purpose, it is necessary to understand the evolution of primordial magnetic fields.
While the scenarios of magnetic field generation in the early universe determine initial conditions for the subsequent co-evolution with the plasma fluid in the early universe, 
various cosmological and observational consequences of the resultant magnetic fields at {\it e.g.}, the electroweak epoch, at the Big-Bang nucleosynthesis, pre- and post- recombination epochs, have been discussed.\footnote{We summarize cosmological and observational constraints on these magnetic fields in Appendix \ref{appx:statusofconstraints}.}
To compare these discussions on primordial magnetic fields at different epochs consistently, we have to understand their decay dynamics.

Banerjee and Jedamzik \cite{Banerjee+04} addressed this issue and provided an analytic understanding of the evolution of primordial magnetic fields.
Their methodology makes the analysis of the complicated magneto-hydrodynamic system simple.
However, numerical studies \cite{2014ApJ...794L..26Z, 2015PhRvL.114g5001B,2001PhRvE..64e6405C, Brandenburg:2016odr, Brandenburg+17, 2017PhRvE..96e3105R, 2017MNRAS.472.1628P} after Banerjee and Jedamzik have found the so-called non-helical inverse transfer of the magnetic energy that is inconsistent with their analysis.
Since the inverse transfer makes the decay of the cosmological magnetic fields slower than previously expected, primordial magnetic fields may remain unexpectedly strong in the late universe.
Then, their analytic understanding fails to relate the properties of primordial magnetic fields in the early and the late universe, consistently with the numerical studies.

Theoretical understanding of the non-helical inverse transfer was provided very recently by Hosking and Schekochihin \cite{Hosking+21}, based on the conservation of magnetic helicity.
Magnetic helicity is a quantity that measures the parity violation of a magnetic field configuration within a closed volume and is conserved in the limit of large electric conductivity \cite{1958PNAS...44..489W}.
For maximally helical magnetic fields, {\it i.e.}, when the magnetic power spectrum maximally violates parity, it has long been known that the conservation of magnetic helicity is directly rephrased as a constraint on the evolution of the magnetic field strength and the coherence length \cite{1975JFM....68..769F}.
The constraint makes the decay slower, which is nothing but the inverse transfer.
For non-helical magnetic fields, {\it i.e.}, when the magnetic power spectrum is parity even, no direct implication was thought to be obtained because the magnetic helicity is zero independently of the magnetic field strength and the coherence length.
However, according to Hosking and Schekochihin \cite{Hosking+21}, the conservation of the Hosking integral, which quantifies the fluctuations of local magnetic helicity rather than the net magnetic helicity, constrains the evolution of the magnetic field strength and the coherence length.
Their idea well explains the numerical evidence of the inverse transfer of the non-helical magnetic fields and is supported by dedicated numerical studies \cite{zhou2022scaling,2023Atmos..14..932B,2023JPlPh..89f9006B}.\footnote{Alternative explanations of the non-helical inverse transfer have been proposed in Refs.~\cite{2013PhRvD..87h3007K,2021MNRAS.501.3074B,2024OJAp....7E..75D}.
While the choice of interpretation can be a subject for future study, updating the description of the evolution of the primordial magnetic field in a manner consistent with the existing numerical results is a necessary step towards the complete understanding.
In this work, we adopt the Hosking integral conservation as a guiding principle because its physical meaning is clear and also because its conservation is supported by numerical studies.}
Therefore, we should reformulate the analytic understanding of the evolution of the cosmological magnetic fields based on the conservation of the Hosking integral.

Hosking and Schekochihin \cite{Hosking+21} also pointed out the role of magnetic reconnection in decaying magnetic turbulence.
It is reasonable to assume that different mechanisms under different circumstances process the energy of primordial magnetic fields and transfer it into heat.
They proposed that, if the magnetic energy dominates over the kinetic energy of the velocity field of the plasma fluid, and if the dynamics is non-linear, then magnetic reconnection drains the magnetic energy.
Although the idea is not fully established \cite{zhou2022scaling,2024arXiv240108569B}, it would be the best to incorporate the idea in the reformulation at this moment.
Assuming that shear viscosity dissipates the energy at small scales, magnetic reconnection has multiple regimes, which include the Sweet--Parker magnetic reconnection for a modest non-linearity and the fast magnetic reconnection for highly non-linear cases \cite{ji2011phase}.
They considered the Sweet--Parker magnetic reconnection \cite{Hosking+21}, and their subsequent work \cite{Hosking+22} considered the fast magnetic reconnection of highly non-linear magnetic fields at the recombination epoch.
In Ref.~\cite{Uchida:2022vue}, we have examined existing numerical results in the literature and confirmed that this approach explains the existing numerical results well.

In this paper, we extend their approach and analyze all the regimes with a systematic classification according to several dichotomous criteria.
The criteria of the classification are whether the magnetic or the kinetic energy dominates, whether the magnetic field is maximally helical or non-helical, whether the non-linearity is large or small, and whether a shear viscosity or a drag force is the dominant dissipation term at small scales.
While we focused on magnetically dominant and non-helical regimes in our previous work \cite{Uchida:2022vue}, we cover all the possible regimes with these criteria.
We obtain scaling formulae to describe the evolution of primordial magnetic fields until the recombination epoch in all the regimes, which enable us to track the evolution of the primordial magnetic fields with arbitrary initial conditions.

By applying the new description of the evolution of the cosmological magnetic fields, we obtain implications on the generation mechanisms of the primordial magnetic fields.
It turns out that most of the scenarios are inconsistent with the standard cosmology and the observations, according to the baryon isocurvature problem \cite{Kamada:2020bmb}.
Note that we have several options to evade the inconsistency, {\it e.g.}, an inflationary magnetogenesis at very low energy scales even below the electroweak scale $\sim 100\,{\rm GeV}$ \cite{Yanagihara:2023qvx}.
We describe the evolution of the typical strength and the coherence length of primordial magnetic fields in detail by choosing representative initial conditions and explicitly show that magnetogenesis before the electroweak symmetry breaking is not feasible.

\section{Outline of the analysis \label{sec:Outline}}
In this section, we explain the general outline of the analysis, and prepare all the ingredients and physical constraints to derive the time evolution of the system.
The problem is to solve the set of equations of magneto-hydrodynamics (MHD) explained in Sec.~\ref{sec:MHD_Eqs}.
However, we do not have to tackle it directly as long as we are interested in the evolution of typical strength and coherence length of primordial magnetic fields.
Following the methodology initiated by Banerjee and Jedamzik~\cite{Banerjee+04}, we reduce the problem to find a solution that is consistent with relevant conserved quantities and decay time scales introduced in Sec.~\ref{sec:conserved_quantities} and Sec.~\ref{sec:decay_time_scale}, respectively.

\subsection{MHD equations \label{sec:MHD_Eqs}}
Primordial magnetic fields before the recombination epoch are coupled with the background expansion of the universe and the plasma fluid that consists of the Standard-Model particles.

First, the expansion of the universe is accounted for by the following rescaling of the physical quantities into the comoving quantities \cite{PhysRevD.54.1291}, which are defined as
\begin{align}
    {\bm B}
        =a^2 {\bm B}_\text{p},\quad 
    {\bm v}
        ={\bm v}_\text{p},\quad
    \rho
        =a^4\rho_\text{p},\quad
    p
        =a^4 p_\text{p},\notag\\
    \sigma
        =a \sigma_\text{p},\quad
    \eta
        =a^{-1}\eta_\text{p},\quad
    \alpha
        =a \alpha_\text{p},
    \label{eq:Relations of comoving and physical quantities}
\end{align}
where $a$ is the scale factor normalized to be unity in the present epoch, and a  subscript $_{\text p}$ denotes physical quantities.
Hereafter, $\bm{B}$ denotes the comoving magnetic field, $\bm{v}$ the comoving velocity of the fluid motion, $\rho$ and $p$ the comoving energy density and the pressure of the fluid, respectively, and $\sigma$, $\eta$, and $\alpha$ the comoving transport coefficients defined shortly.

In terms of the comoving quantities, the coupled system of the magnetic field ${\bm B}$ and the velocity field ${\bm v}$ are governed by the MHD equations in the Minkowski spacetime \cite{PhysRevD.54.1291,Banerjee+04}.
We neglect the impact on the MHD dynamics through the metric perturbations.
The MHD equations consist of the induction equation,
\begin{align}
    \partial_\tau {\bm B}-{\bm \nabla}\times({\bm v}\times{\bm B})
        =\frac{1}{\sigma}{\nabla}^2 {\bm B},
    \label{eq:Faraday's induction equation}
\end{align}
and the Navier--Stokes equation,
\begin{align}\hspace{-1.5mm}
    \partial_\tau {\bm v}+({\bm v}\cdot{\bm\nabla}){\bm v}-\frac{1}{\rho+p} ({\bm\nabla}\times{\bm B})\times{\bm B}+\frac{{\bm\nabla}\delta p}{\rho+p}
        =\eta\left[{\nabla}^2 {\bm v}+\frac{1}{3} {\bm\nabla}({\bm\nabla}\cdot{\bm v})\right]-\alpha\bm{v}.
    \label{eq:Navier--Stokes equation}
\end{align}
The energy density $\rho$ and the pressure $p$ have their homogeneous contribution as the background, and we denote the perturbations from their background components by $\delta\rho$ and $\delta p$.
While the terms except the ones with derivatives with respect to the conformal time, $\partial_\tau$, in the left-hand sides of these equations transfer comoving energy density between the magnetic field and the fluid, the terms in the right-hand sides dissipate the energy away from the system.
The comoving electric conductivity, $\sigma$, and the comoving shear viscosity, $\eta$, dissipate the energy of colliding particles of the fluid into heat.
The comoving coefficient of the drag force, $\alpha$, also dissipates the energy of the system and has contributions from the friction of the fluid with free-streaming particles as a background of the fluid motion and from the Hubble expansion in the matter-dominated era.
Their typical values range 
$\sigma\sim 10^{-15}\text{--}10^{-11}\,{\rm GeV}$, 
$\eta\sim 10^{15}\text{--}10^{25}\,{\rm GeV}^{-1}$, 
and 
$\alpha\sim 10^{-40}\text{--}10^{-20}\,{\rm GeV}$.
For precise values of these dissipation coefficients, see Appendix \ref{appx:DissipationCoefficients}.
The continuity equation
\begin{align}
    \partial_\tau \delta\rho+{\bm\nabla}\cdot\left[(\rho+p){\bm v}\right]
        ={\bm E}\cdot({\bm\nabla}\times{\bm B}),
    \label{eq:Continuity}
\end{align}
where
\begin{align}
    {\bm E}
        =\left(\frac{1}{\sigma}{\bm \nabla}-{\bm v}\right)\times{\bm B},
    \label{eq:Electric field in terms of magnetic field}
\end{align}
and the equation of state, $\delta p=w\delta\rho$, close the equations of motions.
We assume $w=1/3$ in the radiation-dominated universe and $w=0$ in the matter-dominated regime.

\subsection{Physical requirements on time evolution\label{sec:Methodology}}
Instead of solving the set of MHD equations directly, we parametrize the system by a few quantities and judiciously carry out dimensional analysis.
We first define five variables $B$, $\xi_{\rm M}$, $v$, $\xi_{\rm K}$, and $\epsilon$ in terms of magnetic and velocity energy spectra to parametrize the system.
By assuming the homogeneity and isotropy of the ensemble averages, power spectra of the magnetic and velocity fields, $P_B(\tau,k)$ and $P_v(\tau,k)$, are defined as
\begin{align}
    \langle\boldsymbol{B}(\boldsymbol{k})\cdot\boldsymbol{B}(\boldsymbol{k}')\rangle
        &=(2\pi)^3\delta^3(\boldsymbol{k}+\boldsymbol{k}') P_B(\tau,k),
    \label{eq:MagneticEnergySpectrum_Def}\\
    \langle\boldsymbol{v}(\boldsymbol{k})\cdot\boldsymbol{v}(\boldsymbol{k}')\rangle
        &=(2\pi)^3\delta^3(\boldsymbol{k}+\boldsymbol{k}') P_v(\tau,k),
    \label{eq:KineticEnergySpectrum_Def}
\end{align}
where $\boldsymbol{B}(\boldsymbol{k})$ and $\boldsymbol{v}(\boldsymbol{k})$ denote Fourier modes of the fields with the convention
\begin{align}
    f({\bm k})
        :=\int d^3xe^{-i{\bm k}\cdot{\bm x}}f({\bm x}),\quad
    f({\bm x})
        =\int\dfrac{d^3k}{(2\pi)^3}e^{i{\bm k}\cdot{\bm x}}f({\bm k}),
    \label{eq:FourierTrsf_Def}
\end{align}
for general $f$.

We define the typical magnetic field strength, $B$, the magnetic coherence length, $\xi_{\text M}$, the typical velocity, $v$, and the kinetic coherence length, $\xi_{\text K}$, as
\begin{align}
    B(\tau)^2
        &:=\langle\bm{B}({\bm x})^2\rangle=\int\frac{d^3k}{(2\pi)^3} P_B(\tau,k),\quad
    \xi_{\text{M}}(\tau)
        :=\frac{1}{B^2} \int\frac{d^3k}{(2\pi)^3} \frac{2\pi}{k} P_B(\tau,k) ,
    \label{eq:Brms_XiM_Def}\\
    v(\tau)^2
        &:=\langle\bm{v}({\bm x})^2\rangle=\int\frac{d^3k}{(2\pi)^3} P_v(\tau,k),\quad\;\;
    \xi_{\text{K}}(\tau)
        :=\frac{1}{v^2} \int\frac{d^3k}{(2\pi)^3} \frac{2\pi}{k} P_v(\tau,k),
    \label{eq:vrms_XiK_Def}
\end{align}
where the infrared cutoff of the integrals is implicit, assuming peaky spectra.\footnote{When the spectra are peaky, they are blue in the small-$k$ range, and we do not care the infrared cutoff because the infrared contributions to the integrals are subdominant.
On the other hand, a class of inflationary magnetogenesis scenarios may generate a nearly scale-invariant or even red magnetic energy spectrum, namely $P_B\sim k^{n_B}$, $n_B\leq-2$ ($n_B=-3$ for the exact scale-invariance) \cite{2008JCAP...01..025M}.
In this case, the infrared cutoff determines the magnetic coherence length $\xi_{\rm M}$, which must be much shorter than the current Hubble scale for the sake of applicability of our analysis.}

We also define the magnetic helicity fraction $\epsilon$ by
\begin{align}
    \epsilon(\tau)
        :=\dfrac{\langle\bm{A}({\bm x})\cdot\bm{B}({\bm x})\rangle}{B^2\xi_{\rm M}/(2\pi)},
\end{align}
where ${\bm A}$ is the vector potential.
The numerator is the net magnetic helicity density, which is independent of the magnetic energy spectrum $P_B$ but bound by it, according to the Schwartz inequality, which leads to the realizability condition \cite{1978mfge.book.....M}
\begin{align}
    \left\vert\epsilon\right\vert
        \leq1.
    \label{eq:Realizability}
\end{align}

The magnetic field is non-helical if $\epsilon=0$ and maximally helical if $\vert\epsilon\vert=1$.
When the system is not frozen but decaying under a physical mechanism, the kinetic variables $v$ and $\xi_{\rm K}$ are determined by the magnetic variables $B$ and $\xi_{\rm M}$.
See Appendix \ref{appx:RegimeDependentAnalyses} for a detailed explanation of this point.
Then, our task is to find $B(\tau)$, $\xi_{\rm M}(\tau)$, and $\epsilon(\tau)$ as functions of time before the recombination epoch, which is accomplished by incorporating various conservation laws and decay time scales of the magnetic field as we see below.

\subsubsection{Conserved quantities}
\label{sec:conserved_quantities}
The MHD evolution can be characterized by some conserved quantities at each epoch. We summarize the conserved quantities and the constraints they put on the above-introduced parameters.
With the large electric conductivity in the early universe, the net magnetic helicity density is approximately conserved \cite{1958PNAS...44..489W,Banerjee+04}, implying
\begin{align}
    \epsilon(\tau) B(\tau)^2\xi_{\rm M}(\tau)
        ={\rm const.}
    \label{eq:HelicityConservation}
\end{align}

Once we know the evolution of the strength and the coherence length of the magnetic field, Eq.~\eqref{eq:HelicityConservation} determines the evolution of the helicity fraction.
As turns out later, $\vert\epsilon(\tau)\vert$ increases unless it saturates the realizability condition \eqref{eq:Realizability}.
When the realizability condition is saturated and the magnetic field is maximally helical, the net magnetic helicity conservation constrains the evolution of $B$ and $\xi_{\rm M}$ as
\begin{align}
    B(\tau)^2\xi_{\rm M}(\tau)
        ={\rm const.}
    \label{eq:HelicityConservation_MH}
\end{align}

If the helicity fraction is tiny, not the net magnetic helicity but rather the local fluctuations of magnetic helicity density constrains the evolution of $B$ and $\xi_{\rm M}$.
The local fluctuations of magnetic helicity density are quantified by the Hosking integral \cite{Hosking+21}, $I_{\rm H}\sim B^4\xi_{\rm M}^5$.\footnote{This expression omits a time-independent numerical coefficient, where we assume Gaussianity of the probability distribution and a self-similar scaling of the power spectra's decay. For more explanation about these assumptions, see Appendix \ref{sec:Hosking_integral}.}
For the definition of and discussion on the Hosking integral, see Appendix \ref{sec:Hosking_integral}.
Conservation of the Hosking integral is numerically established for magnetically dominated and non-helical turbulence \cite{zhou2022scaling}, implying
\begin{align}
    B(\tau)^4\xi_{\rm M}(\tau)^5
        ={\rm const.},
    \label{eq:HelicityConservation_NH}
\end{align}
which constrains the evolution of $B$ and $\xi_{\rm M}$.
We assume that this constraint applies whenever the magnetic energy dominates over the kinetic energy and the helicity fraction is less than unity.
With this constraint, Eq.~\eqref{eq:HelicityConservation} implies $\epsilon\propto B^{-6/5}$, which increases when magnetic strength decays.

For completeness, we also consider the cases where the kinetic energy is comparable to or dominates over the magnetic energy.
Following the discussion in Ref.~\cite{Hosking+21}, we consider conservation of the so-called Saffman cross-helicity integral $\sim B^2v^2\xi_{\rm M}^3$ together with the constraints \eqref{eq:HelicityConservation_MH} and \eqref{eq:HelicityConservation_NH} for the equipartition case, where the magnetic and kinetic energy densities are comparable.

Then, we obtain $v\propto B^2$ for maximally helical magnetic fields and $v\propto B^{1/5}$ for non-helical magnetic fields and suspect that equipartition is not maintained as the magnetic energy decays.
Because of the differences of decay rates of $B$ and $v$, maximally helical equipartition soon turns to the magnetically dominated regime, and non-helical equipartition to the kinetically dominated regime, respectively \cite{Hosking+21}.

When the kinetic energy dominates over the magnetic energy, we approximate that the decay dynamics is governed by pure hydrodynamics.
In pure hydrodynamics, the kinetic energy at large scales is conserved \cite{Llor11}, which implies $v^2\xi_{\rm K}^{n+3}={\rm const.}$, where $n$ is the power-law index of the kinetic energy spectrum.
Numerical simulations suggest that the sub-dominant magnetic field is enhanced by small-scale dynamo, it quickly becomes quasi-equipartition, and the decay law of the magnetic field follows the purely hydrodynamic one of the velocity field \cite{Hosking+21,Brandenburg+17}.
We then use
\begin{align}
    B(\tau)^2\xi_{\rm M}(\tau)^{n+3}
        ={\rm const.}\quad
    \label{eq:IRKineticConservation}
\end{align}
as the condition that constrains the evolution of $B$ and $\xi_{\rm M}$ for non-helical (quasi-)equipartition.
As already explained, scaling regimes are not maximally helical equipartition nor kinetically dominated far away from equipartition because quick transitions to either magnetically dominated regimes or non-helical (quasi-)equipartition regimes are expected.

Let us collectively denote the relevant conserved quantities by $C(B,\xi_{\rm M})$, which are expressed in terms of $B$ and $\xi_{\rm M}$.
Equations \eqref{eq:HelicityConservation_MH}, \eqref{eq:HelicityConservation_NH}, or \eqref{eq:IRKineticConservation} imply a constraint
\begin{align}
    &C\left(B,\xi_{\rm M}\right)
        =C(B_{\rm ini},\xi_{\rm M,ini}),
    \label{eq:Constraint_general}\\
    &C(B,\xi_{\rm M})
        :=\left\{\begin{matrix}
        B^2\xi_{\rm M}\hspace{3mm}&\hspace{0mm}\text{maximally helical and magnetically dominated}\\[3pt]
        B^4\xi_{\rm M}^5\hspace{3mm}&\hspace{11.5mm}\text{non-helical and magnetically dominated}\\[3pt]
        B^2\xi_{\rm M}^{n+3}\hspace{0mm}&\hspace{19.7mm}\text{non-helical and quasi-equipartition}
        \end{matrix}\right.
        \label{eq:ConservedQuantities}
\end{align}
where the subscript $_{\rm ini}$ implies variables evaluated at the beginning.
In Table \ref{tab:conserved_quantities}, we summarize the above approximate conserved quantities in the system.
Equation \eqref{eq:Constraint_general} constrains the point $(\xi_{\rm M}(\tau),B(\tau))$ on a fixed line (orange lines in Fig.~\ref{fig:EvolutionPlot1}) in the $\xi_{\rm M}$-$B$ plane.

\begin{table}[t]
    \centering
    \begin{tabular}{c|c|c}
        Conserved quantities & $C$ & Condition 
        \\\hline\hline\vspace{-4mm}&&\\
        Net magnetic helicity density \cite{1958PNAS...44..489W} 
        & $\epsilon B^2\xi_{\rm M}$ 
        & $\begin{matrix}{\text{always}}
        \\
        {\text{$(\epsilon = 1$ for maximally helical)}}\end{matrix}$
        \\\hline\vspace{-4mm}&&\\
        $\begin{matrix}\text{Hosking integral \cite{Hosking+21}}
        \\
        \text{(local fluctuations of magnetic helicity density)}\end{matrix}$ 
        & $B^4\xi_{\rm M}^5$ & non-helical
        \\\hline\vspace{-4mm}&&\\
        Saffman cross-helicity integral \cite{Hosking+21} & $B^2v^2\xi_{\rm M}^{3}$ & equipartition
        \\\vspace{-4.5mm}&&\\\hline\vspace{-4mm}&&\\
        Kinetic energy at large scales \cite{1967JFM....27..581S,birkhoff1954fourier,1956RSPTA.248..369B,1987flme.book.....L} & $v^2\xi_{\rm K}^{n+3}$ & kinetically dominated
    \end{tabular}
    \caption{Approximate conserved quantities of MHD. When the magnetic energy dominates over the kinetic energy, either of magnetic helicity density \cite{1958PNAS...44..489W} or the Hosking integral \cite{Hosking+21} conserves and constrains the evolution of $B$ and $\xi_{\rm M}$, with the condition $C={\rm const.}$ When the magnetic energy does not dominate, we take into account the conservation of the Saffman cross-helicity integral \cite{Hosking+21} and the kinetic energy at large scales \cite{1967JFM....27..581S,birkhoff1954fourier,1956RSPTA.248..369B,1987flme.book.....L}, where $n$ is the power-law index of the kinetic energy spectrum, as well. For given $B(\tau)$ and $\xi_{\rm M}(\tau)$, the evolution of the helicity fraction $\epsilon$ is determined by the magnetic helicity density conservation.}
    \label{tab:conserved_quantities}
\end{table}

\subsubsection{Decay time scale}
\label{sec:decay_time_scale}
Decay time scale of a magnetic field is another physical input required to clarify its time evolution.
Strong magnetic fields corresponding to short decay time scales, which should have decayed earlier in the history of the universe, are excluded.
We first estimate decay time scales of the magnetic field which is controlled by different mechanisms in different regimes.

One of the relevant mechanism is the magnetic reconnection.
It is reasonable to assume that magnetic reconnection transfers the magnetic energy into kinetic energy and heat if the magnetic energy dominates over the kinetic energy and if the decay process of the magnetic energy is non-linear \cite{Hosking+21,Hosking+22}.
Magnetic reconnection has sub-regimes \cite{ji2011phase}, and we assume the Sweet--Parker magnetic reconnection \cite{1958IAUS....6..123S,Parker57,park+84,Uchida:2022vue} for moderate non-linearity and the fast magnetic reconnection \cite{2022JPlPh..88e1501S,Hosking+22} for substantial non-linearity.

On the other hand, if the magnetic energy dominates over the kinetic energy and if the decay process is linear in the sense that dissipation into heat at the magnetic coherence length is significant, then the decay rate of the magnetic energy is determined by the balance between the excitation by the Lorentz force and the dissipation of the kinetic energy \cite{Banerjee+04}.

In the non-helical quasi-equipartition, where pure hydrodynamics governs the energy decay, we assume that the decay time scale is given by the eddy turnover time \cite{1993noma.book.....B} if non-linear and the time scales of the dissipation terms in the Navier--Stokes equation if linear.
The expression of each time scale changes according to whether the viscosity $\eta$ or the friction $\alpha$ is more significant.
For the detailed derivation of each time scale, see Appendix \ref{appx:RegimeDependentAnalyses}.

Let us unifiedly denote the relevant decay time scales by $\tau_{\rm decay}(\tau, B, \xi_{\rm M})$.
By comparing it with the conformal time $\tau$, which is nothing but the time scale of the evolution of the early universe, conditions
\begin{align}
    \tau_{\rm decay}(\tau,B(\tau),\xi_{\rm M}(\tau))
        \gg\tau,\quad
    {\text{frozen}}
    \label{eq:FrozenCondition_general}
\end{align}
for the cases where the relevant magneto-hydrodynamic mechanism is too slow to process the magnetic field, and
\begin{align}
    \tau_{\rm decay}(\tau,B(\tau),\xi_{\rm M}(\tau))
    \sim\tau,\quad
    {\text{under the process}}
    \label{eq:TimeScaleCondition_general}
\end{align}
for the cases where the magnetic field is not frozen but under the process of decay hold \cite{Banerjee+04}.
These conditions \eqref{eq:FrozenCondition_general} and \eqref{eq:TimeScaleCondition_general} exclude the region with stronger magnetic fields in the $\xi_{\rm M}$-$B$ plane (the region above each gray line in Fig.~\ref{fig:EvolutionPlot1}).
Sometimes $\tau_{\rm decay}(\tau, B, \xi_{\rm M})/\tau$ can be an increasing function of $\tau$ for fixed $B$ and $\xi_{\rm M}$.
In such cases, re-entrance into frozen regimes after experiencing the decay process is expected.
Then, we propose a condition to determine the line in the $B$-$\xi_{\rm M}$ plane for a given conformal time $\tau$ (see Fig.~\ref{fig:EvolutionPlot1}),
\begin{align}
    &\min_{\tau_{\rm ini}\leq\tau'\leq\tau} \frac{\tau_{\rm decay}(\tau',B,\xi_{\rm M})}{\tau'}
        \sim1,
    \label{eq:TimeScaleCondition_general2}
\end{align}
where
\begin{align}    
    &\tau_{\rm decay}(\tau',B,\xi_{\rm M})
        :=\left\{\begin{matrix}
        (\rho+p)^{\frac{1}{4}}\sigma^{\frac{3}{4}}\eta^{\frac{1}{4}}B^{-\frac{1}{2}}\xi_{\rm M}^{\frac{3}{2}}\hspace{4mm}\text{}&\hspace{3mm}\text{viscous Sweet--Parker reconnection}\\[1pt]
        (\rho+p)^{\frac{1}{2}}\sigma^{\frac{1}{2}}\alpha^{\frac{1}{2}}B^{-1}\xi_{\rm M}^{2}\hspace{4mm}\text{}&\hspace{1.5mm}\text{dragged Sweet--Parker reconnection}\\[0pt]
        S_{\rm c}^{\frac{1}{2}}(\rho+p)^{\frac{1}{2}}\sigma^{\frac{1}{2}}\eta^{\frac{1}{2}}B^{-1}\xi_{\rm M}\text{}&\hspace{19.5mm}\text{viscous fast reconnection}\\[0pt]
        S_{\rm c}^{\frac{1}{2}}(\rho+p)\alpha B^{-2}\xi_{\rm M}^2\hspace{9mm}\text{}&\hspace{17.6mm}\text{dragged fast reconnection}\\[3pt]
        (\rho+p)\eta B^{-2}\hspace{18mm}\text{}&\hspace{31mm}\text{linear and viscous}\\[4pt]
        (\rho+p)\alpha B^{-2}\xi_{\rm M}^2\hspace{13mm}\text{}&\hspace{29.7mm}\text{linear and dragged}\\[2pt]
        (\rho+p)^{\frac{1}{2}}B^{-1}\xi_{\rm M}\hspace{13mm}\text{}&\hspace{31mm}\text{kinetic turbulence}\\[3pt]
        \eta^{-1}\xi_{\rm M}^2\hspace{27mm}\text{}&\hspace{-9mm}\text{kinetically dominated, linear, and viscous}
        \end{matrix}\right.
        \label{eq:DecayTimeScales}
\end{align}
where $S_{\rm c}$ is the critical Lundquist number, which 
is numerically evaluated as $S_{\rm c} \sim 10^4$ \cite{1986PhFl...29.1520B,2005PhRvL..95w5003L,2007PhPl...14j0703L,ji2011phase}.
If the solution of conditions \eqref{eq:Constraint_general} and \eqref{eq:TimeScaleCondition_general2} gives a weaker magnetic field than the initial value, $B< B_{\rm ini}$, then we interpret the solution as the wanted $B(\tau)$ and $\xi_{\rm M}(\tau)$.
Otherwise, the system is frozen all the time since $\tau_{\rm ini}$ until $\tau$, and therefore $B(\tau)=B_{\rm ini}$ and $\xi_{\rm M}(\tau)=\xi_{\rm M,ini}$.

\section{Regime-dependent description \label{sec:RegimeDependent}}
\begin{figure}[t]
    \begin{minipage}[h]{1.0\hsize}
    \includegraphics[keepaspectratio, width=1.0\textwidth]{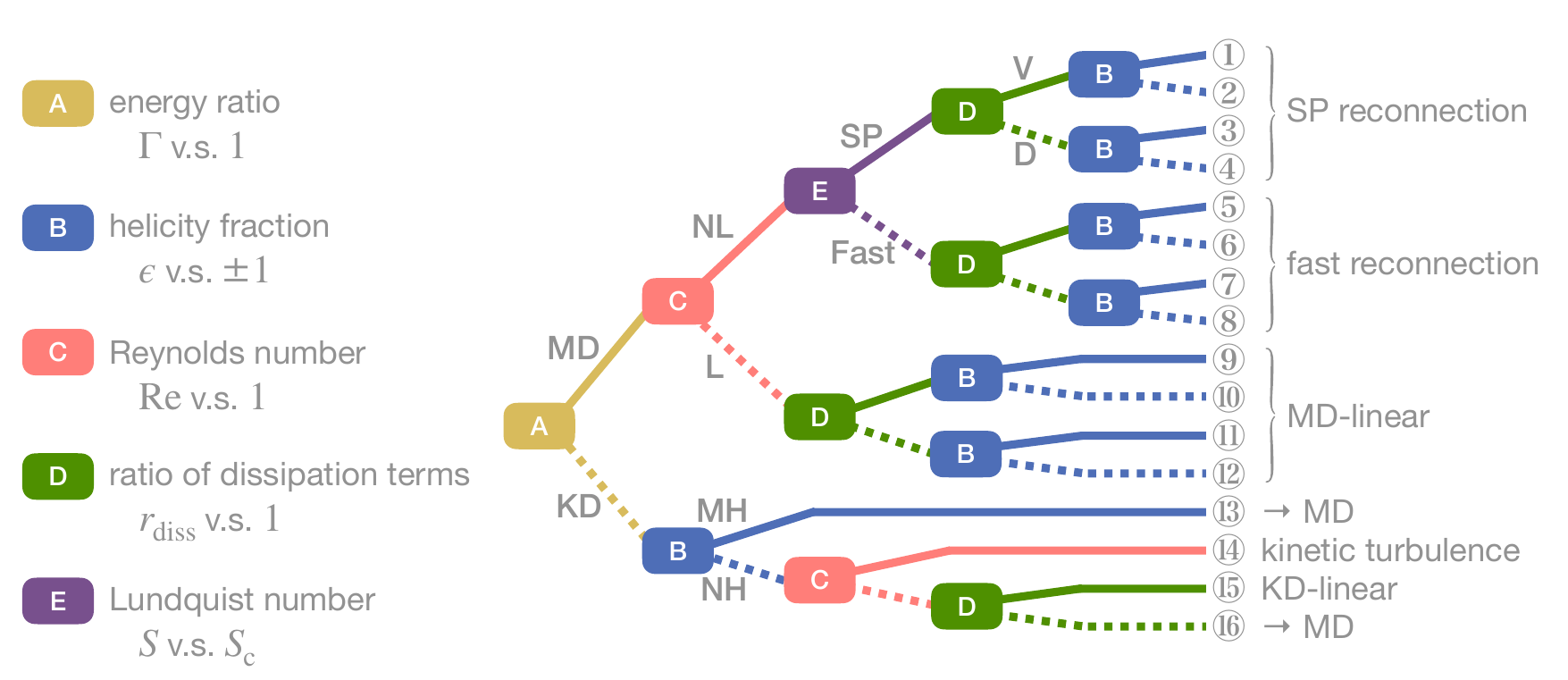}
    \end{minipage}
    \caption{\label{fig:Classification}Classification of regimes according to the criteria A--E. We have four branches of regimes, magnetically dominated (MD) and non-linear (NL) branch, magnetically dominated and linear (L) branch, kinetically dominated (KD) and maximally helical (MH) branch, and kinetically dominated and non-helical (NH) branch. Each branch has one or several regimes, according to whether viscous (V) or dragged (D) and whether the Sweet--Parker (SP) or the fast (Fast) magnetic reconnection is operating, which adds up to sixteen in total.
    }
\end{figure}

In this section, we classify regimes according to which conserved quantity and which decay time scale are relevant.
We propose a systematic way of the regime classification, based on several dichotomous criteria.
Scaling solutions are obtained by solving Eqs.~\eqref{eq:Constraint_general} and \eqref{eq:TimeScaleCondition_general} in each regime, which are summarized in Table \ref{tb:ResultsSummary}.
We classify each regime by the following five criteria that are characterized by several parameters introduced below (see also Fig.~\ref{fig:Classification}).

\begin{itemize}
\item  \textbf{Criterion A: Energy ratio between magnetic field and velocity field}

Whether the magnetic energy or the kinetic energy dominates is quantified by defining the ratio of magnetic and kinetic energy densities~\cite{Banerjee+04}
\begin{align}
    \Gamma
        :=\dfrac{(\rho+p)v^2}{B^2},
    \label{eq:EnergyRatio_Def}
\end{align}
which is smaller (larger) than unity when the magnetic (kinetic) energy dominates over the other. 

\item  \textbf{Criterion B: Helicity fraction (parity violation)}

Whether the magnetic field is maximally helical or not is, by definition, rephrased as whether the helicity fraction $\epsilon$ is $\pm1$ or not.

\item  \textbf{Criterion C: Reynolds numbers (non-linearity)}

The non-linearity of the decay dynamics is quantified by the Reynolds numbers, which can be defined as \cite{Uchida:2022vue}
\begin{align}
    {\rm Re}
        :=\left\{\begin{matrix}
        {\rm Re}_{\rm M}:=\dfrac{\sigma v \xi_{\rm M}^2}{\xi_{\rm K}}\hspace{22.5mm}
        \text{when $\Gamma<1$,}\vspace{-3mm}\\\\
        {\rm Re}_{\rm K}:=\dfrac{v\xi_{\rm K}}{\max\{\eta,\alpha\xi_{\rm K}^2\}}\hspace{10mm}
        \text{when $\Gamma>1$},
        \end{matrix}\right.
\end{align}
by comparing the magnitudes of the non-linear and linear terms in the induction equation and the Navier--Stokes equation.
We here estimated $\left\vert{\bm\nabla}\times\left({\bm v}\times{\bm B}\right)\right\vert\sim vB/\min\left\{\xi_{\text M},\xi_{\rm K}\right\}\sim vB/\xi_{\rm K}$, $\left\vert{\nabla}^2{\bm B}/\sigma\right\vert\sim B/\left(\sigma\xi_{\rm M}^2\right)$, $\left\vert\left({\bm v}\cdot{\bm\nabla}\right){\bm v}\right\vert\sim v^2/\xi_{\rm K}$, and $\left\vert\eta\left[\cdots\right]-\alpha{\bm v}\right\vert\sim\max\left\{\eta v/\xi_{\rm K}^2, \alpha v\right\}$.

\item  \textbf{Criterion D: Dominant dissipation source}

Whether the viscosity or the friction is the dominant dissipation term corresponds to whether \cite{Uchida:2022vue}
\begin{align}
    r_{\rm diss}
        :=\dfrac{\alpha\xi_{\rm K}^2}{\eta}
\end{align}
is smaller or larger than unity.

\item  \textbf{Criterion E: Lundquist number (reconnection mechanism)}

Finally, as for the magnetic reconnection, how significant the diffusion by the finite electric resistivity is matters.
It is quantified by the Lundquist number \cite{2022JPlPh..88e1501S, Hosking+22}
\begin{align}
    S
        ={\rm Re}_{\rm M}^{\frac{4}{5}}.
\end{align}
For the definition of the Lundquist number, which is the ratio between the timescales of the outflow of magnetic reconnection and magnetic diffusion, and the derivation of this relationship with the magnetic Reynolds number, see Appendix \ref{appx:RegimeDependentAnalyses}.
We assume that the Sweet--Parker reconnection is relevant if $S\ll S_{\rm c}$ and that the fast reconnection is relevant if $S\gg S_{\rm c}$, following Ref.~\cite{Hosking+22}.
\end{itemize}

Based on these criteria A--E, we classify the situation into sixteen regimes.
See Fig.~\ref{fig:Classification}.
We first consider criterion A.
If $\Gamma\ll1$, then the system is magnetically dominated.
Otherwise, if $\Gamma\gtrsim 1$, then the system is kinetically dominated.
We include $\Gamma\sim1$ in the kinetically dominated regime because $\Gamma\gg1$ as the initial condition results in quasi-equipartition by the small-scale dynamo.

For magnetically dominated regimes, we next consider criterion C.
When ${\rm Re}_{\rm M}\gg1$, then the system is magnetically dominated and non-linear.
We expect that magnetic reconnection is the dominant process in this branch of regimes.
The reconnection mechanism is the Sweet--Parker one if $S\ll S_{\rm c}$ and the fast one if $S\gg S_{\rm c}$.
When ${\rm Re}_{\rm M}\ll1$, then the system is magnetically dominated and linear.
Each of these three branches of the Sweet-Parker reconnection, the fast reconnection, and the magnetically dominated and linear regimes is classified into four according to criteria D and B.

For kinetically dominated regimes, we consider criterion B.
When $\vert\epsilon\vert=1$, the magnetic field is kinetically dominated and maximally helical, and we expect a transition to magnetically dominated regimes \cite{Hosking+21} as explained in the last section.
When $\vert\epsilon\vert\ll1$, the magnetic field follows the evolution laws of kinetically dominated and non-helical fields.
If ${\rm Re}_{\rm K}\gg1$, then kinetic turbulence occurs.
If ${\rm Re}_{\rm K}\ll1$, then decay processes are linear.
Linear regimes are classified according to criterion D.
If $r_{\rm diss}\gg1$, then the system is kinetically dominated and linearly dissipated by the drag force.
However, we expect a transition from this regime to magnetically dominated regimes as explained below.

For each regime in Fig.~\ref{fig:Classification}, one can specify the relevant conserved quantity and decay time scale as shown in Table \ref{tb:ResultsSummary}.
We solve Eqs.~\eqref{eq:Constraint_general} and \eqref{eq:TimeScaleCondition_general} and obtain scaling solutions.
In (magnetically dominated, non-linear, and) the Sweet--Parker magnetic reconnection regimes, we obtain
\begin{align}
    B
        &=B_{\rm ini}^{\frac{6}{7}}\xi_{\rm M,ini}^{\frac{3}{7}}(\rho+p)^{\frac{1}{14}}\sigma^{\frac{3}{14}}\eta^{\frac{1}{14}}\tau^{-\frac{2}{7}},&
    \xi_{\rm M}
        &=B_{\rm ini}^{\frac{2}{7}}\xi_{\rm M,ini}^{\frac{1}{7}}(\rho+p)^{-\frac{1}{7}}\sigma^{-\frac{3}{7}}\eta^{-\frac{1}{7}}\tau^{\frac{4}{7}}\notag\\
    &&&\hspace{3mm}\text{viscous and maximally helical (\ctext{1})},\label{eq:MDNLSPVMH}\\
    B
        &=B_{\rm ini}^{\frac{12}{17}}\xi_{\rm M,ini}^{\frac{15}{17}}(\rho+p)^{\frac{5}{34}}\sigma^{\frac{15}{34}}\eta^{\frac{5}{34}}\tau^{-\frac{10}{17}},&
    \xi_{\rm M}
        &=B_{\rm ini}^{\frac{4}{17}}\xi_{\rm M,ini}^{\frac{5}{17}}(\rho+p)^{-\frac{2}{17}}\sigma^{-\frac{6}{17}}\eta^{-\frac{2}{17}}\tau^{\frac{8}{17}}\notag\\
    &&&\hspace{14mm}\text{viscous and non-helical (\ctext{2})},\label{eq:MDNLSPVNH}\\
    B
        &=B_{\rm ini}^{\frac{4}{5}}\xi_{\rm M,ini}^{\frac{2}{5}}(\rho+p)^{\frac{1}{10}}\sigma^{\frac{1}{10}}\alpha^{\frac{1}{10}}\tau^{-\frac{1}{5}},&
    \xi_{\rm M}
        &=B_{\rm ini}^{\frac{2}{5}}\xi_{\rm M,ini}^{\frac{1}{5}}(\rho+p)^{-\frac{1}{5}}\sigma^{-\frac{1}{5}}\alpha^{-\frac{1}{5}}\tau^{\frac{2}{5}}\notag\\
    &&&\hspace{1.5mm}\text{dragged and maximally helical (\ctext{3})},\label{eq:MDNLSPDMH}\\
    B
        &=B_{\rm ini}^{\frac{8}{13}}\xi_{\rm M,ini}^{\frac{10}{13}}(\rho+p)^{\frac{5}{26}}\sigma^{\frac{5}{26}}\alpha^{\frac{5}{26}}\tau^{-\frac{5}{13}},&
    \xi_{\rm M}
        &=B_{\rm ini}^{\frac{4}{13}}\xi_{\rm M,ini}^{\frac{5}{13}}(\rho+p)^{-\frac{2}{13}}\sigma^{-\frac{2}{13}}\alpha^{-\frac{2}{13}}\tau^{\frac{4}{13}}\notag\\
    &&&\hspace{13mm}\text{dragged and non-helical (\ctext{4})}.\label{eq:MDNLSPDNH}
\end{align}
As for the velocity field, we obtain 
\begin{align}
    v
        =\sigma^{\frac{1}{2}}\tau^{-\frac{3}{2}}\xi_{\rm M}^2,\quad
    \xi_{\rm K}
        =\sigma^{-1}\tau\xi_{\rm M}^{-1}
\end{align}
See Appendix \ref{appx:RegimeDependentAnalyses} for the derivation, where the mechanisms of the Sweet--Parker magnetic reconnection is explained.

In (magnetically dominated, non-linear, and) the fast magnetic reconnection regimes, we obtain
\begin{align}
    B
        &=B_{\rm ini}^{\frac{2}{3}}\xi_{\rm ini}^{\frac{1}{3}}S_{\rm c}^{\frac{1}{6}}(\rho+p)^{\frac{1}{6}}\sigma^{\frac{1}{6}}\eta^{\frac{1}{6}}\tau^{-\frac{1}{3}},&
    \xi_{\rm M}
        &=B_{\rm ini}^{\frac{2}{3}}\xi_{\rm ini}^{\frac{1}{3}}S_{\rm c}^{-\frac{1}{3}}(\rho+p)^{-\frac{1}{3}}\sigma^{-\frac{1}{3}}\eta^{-\frac{1}{3}}\tau^{\frac{2}{3}}\notag\\
    &&&\hspace{0mm}\text{viscous and maximally helical (\ctext{5})},\label{eq:MDNLFVMH}\\
    B
        &=B_{\rm ini}^{\frac{4}{9}}\xi_{\rm ini}^{\frac{5}{9}}S_{\rm c}^{\frac{5}{18}}(\rho+p)^{\frac{5}{18}}\sigma^{\frac{5}{18}}\eta^{\frac{5}{18}}\tau^{-\frac{5}{9}},&
    \xi_{\rm M}
        &=B_{\rm ini}^{\frac{4}{9}}\xi_{\rm ini}^{\frac{5}{9}}S_{\rm c}^{-\frac{2}{9}}(\rho+p)^{-\frac{2}{9}}\sigma^{-\frac{2}{9}}\eta^{-\frac{2}{9}}\tau^{\frac{4}{9}}\notag\\
    &&&\hspace{11.5mm}\text{viscous and non-helical (\ctext{6})},\label{eq:MDNLFVNH}\\
    B
        &=B_{\rm ini}^{\frac{2}{3}}\xi_{\rm ini}^{\frac{1}{3}}S_{\rm c}^{\frac{1}{12}}(\rho+p)^{\frac{1}{6}}\alpha^{\frac{1}{6}}\tau^{-\frac{1}{6}},&
    \xi_{\rm M}
        &=B_{\rm ini}^{\frac{2}{3}}\xi_{\rm ini}^{\frac{1}{3}}S_{\rm c}^{-\frac{1}{6}}(\rho+p)^{-\frac{1}{3}}\alpha^{-\frac{1}{3}}\tau^{\frac{1}{3}}\notag\\
    &&&\hspace{-1mm}\text{dragged and maximally helical (\ctext{7})},\label{eq:MDNLFDMH}\\
    B
        &=B_{\rm ini}^{\frac{4}{9}}\xi_{\rm ini}^{\frac{5}{9}}S_{\rm c}^{\frac{5}{36}}(\rho+p)^{\frac{5}{18}}\alpha^{\frac{5}{18}}\tau^{-\frac{5}{18}},&
    \xi_{\rm M}
        &=B_{\rm ini}^{\frac{4}{9}}\xi_{\rm ini}^{\frac{5}{9}}S_{\rm c}^{-\frac{1}{9}}(\rho+p)^{-\frac{2}{9}}\alpha^{-\frac{2}{9}}\tau^{\frac{2}{9}}\notag\\
    &&&\hspace{10.5mm}\text{dragged and non-helical (\ctext{8})}.\label{eq:MDNLFDNH}
\end{align}
As for the velocity field, as explained in Appendix \ref{appx:RegimeDependentAnalyses},
\begin{align}
    v=S_{\rm c}^{\frac{3}{8}}\sigma^{-\frac{1}{4}}\tau^{-\frac{3}{4}}\xi_{\rm M}^{\frac{1}{2}},\quad
    \xi_{\rm K}=S_{\rm c}^{-\frac{1}{4}}\sigma^{-\frac{1}{2}}\tau^{\frac{1}{2}}
\end{align}
hold.
Note that the conventional inverse cascade scaling law for 
the maximally helical magnetic field, $B \propto \tau^{-1/3}, \xi_\mathrm{M} \propto \tau^{2/3}$, 
which is numerically supported~\cite{Banerjee+04,Brandenburg:2016odr}, 
is obtained for the regime of viscous and maximally helical case (\ctext{5}). 
In the conventional approach, the time scale is identified with the eddy-turnover time, which is proportional to the time scale 
for the fast reconnection, and hence the same time dependence is obtained. We suppose that the previous numerical simulations
look at this regime. While the conventional treatment supposes the equipartition between the magnetic 
and kinetic energy to reach at the time scale, strictly speaking, numerical simulations do not support the equipartition.
Therefore, we believe that our formula for the regime of viscous and maximally helical case (\ctext{5})
is a better explanation of the numerical simulations.

In magnetically dominated and linear regimes, we obtain
\begin{align}
    B
        &=(\rho+p)^{\frac{1}{2}}\eta^{\frac{1}{2}}\tau^{-\frac{1}{2}},&
    \xi_{\rm M}
        &=B_{\rm ini}^2\xi_{\rm M,ini}(\rho+p)^{-1}\eta^{-1}\tau\notag\\
    &&&\hspace{0mm}\text{viscous and maximally helical (\ctext{9})},\label{eq:MDLVMH}\\
    B
        &=(\rho+p)^{\frac{1}{2}}\eta^{\frac{1}{2}}\tau^{-\frac{1}{2}},&
    \xi_{\rm M}
        &=B_{\rm ini}^{\frac{4}{5}}\xi_{\rm M,ini}(\rho+p)^{-\frac{2}{5}}\eta^{-\frac{2}{5}}\tau^{\frac{2}{5}}\notag\\
    &&&\hspace{11.5mm}\text{viscous and non-helical (\ctext{10})},\label{eq:MDLVNH}\\
    B
        &=B_{\rm ini}^{\frac{2}{3}}\xi_{\rm M,ini}^{\frac{1}{3}}(\rho+p)^{\frac{1}{6}}\alpha^{\frac{1}{6}}\tau^{-\frac{1}{6}},&
    \xi_{\rm M}
        &=B_{\rm ini}^{\frac{2}{3}}\xi_{\rm M,ini}^{\frac{1}{3}}(\rho+p)^{-\frac{1}{3}}\alpha^{-\frac{1}{3}}\tau^{\frac{1}{3}}\notag\\
    &&&\hspace{-1mm}\text{dragged and maximally helical (\ctext{11})},\label{eq:MDLDMH}\\
    B
        &=B_{\rm ini}^{\frac{4}{9}}\xi_{\rm M,ini}^{\frac{5}{9}}(\rho+p)^{\frac{5}{18}}\alpha^{\frac{5}{18}}\tau^{-\frac{5}{18}},&
    \xi_{\rm M}
        &=B_{\rm ini}^{\frac{4}{9}}\xi_{\rm M,ini}^{\frac{5}{9}}(\rho+p)^{-\frac{2}{9}}\alpha^{-\frac{2}{9}}\tau^{\frac{2}{9}}\notag\\
    &&&\hspace{10.5mm}\text{dragged and non-helical (\ctext{12})}.\label{eq:MDLDNH}
\end{align}
As for the velocity field, as explained in Appendix \ref{appx:RegimeDependentAnalyses}, we assume
\begin{align}
    v=\tau^{-1}\xi_{\rm M},\quad
    \xi_{\rm K}=\xi_{\rm M}.
\end{align}

In the kinetically dominated and maximally helical regime (\ctext{13}), we expect a transition into a magnetically dominated regime due to the conservation of the net magnetic helicity density and the Saffman cross-helicity integral, as explained in the last section.

In the kinetically dominated, non-helical, and non-linear regime (\ctext{14}), we obtain
\begin{align}
    B
        &=B_{\rm ini}^{\frac{2}{n+5}}\xi_{\rm M, ini}^{\frac{2(n+3)}{n+5}}(\rho+p)^{\frac{n+3}{2n+10}}\tau^{-\frac{n+3}{n+5}},&
    \xi_{\rm M}
        &= B_{\rm ini}^{\frac{2}{n+5}}\xi_{\rm M,ini}^{\frac{2(n+3)}{n+5}}(\rho+p)^{-\frac{1}{n+5}}\tau^{\frac{2}{n+5}}.&\label{eq:KDNHNL}
\end{align}
In the kinetically dominated, non-helical, linear, and viscous regime (\ctext{15}), we obtain
\begin{align}
    B
        &= B_{\rm ini}\xi_{\rm M,ini}^{\frac{n+3}{2}}\eta^{-\frac{n+3}{4}}\tau^{-\frac{n+3}{4}},&
    \xi_{\rm M}
        &=\tau^{\frac{1}{2}}\eta^{\frac{1}{2}}.&\label{eq:KDNHLV}
\end{align}
In these kinetically dominated and non-helical regimes, we expect quasi-equipartition
\begin{align}
    v\sim (\rho+p)^{-\frac{1}{2}}B,\quad
    \xi_{\rm K}\sim\xi_{\rm M}.
\end{align}
Finally in the kinetically dominated, non-helical, linear, and dragged regime (\ctext{16}), we doubt that any scaling solution exists because the characteristic decay time scale, $\tau_{\rm diss}=\alpha^{-1}$, is independent of $B$ and $\xi_{\rm M}$.
When $\alpha\tau\ll1$, the system is frozen.
On the other hand, when $\alpha\tau\gg1$, the velocity field is exponentially damped, and we expect a transition into a magnetically dominated regime.

\begin{table}\vspace{-10mm}
    \hspace{-4mm}
      \begin{tabularx}{0.975\textwidth}{c|c|c|c|c|c}
        \multicolumn{3}{c|}{Regime}
        &$\begin{matrix}
            C(B,\xi_{\rm M})\end{matrix}$
        &$\begin{matrix}
            \tau_{\rm decay}(\tau,B,\xi_{\rm M})\end{matrix}$
        &$\begin{matrix}
            \text{Scaling}\\
            \text{solution}\end{matrix}$\\
        \cline{1-6}\vspace{-4mm}\\\cline{1-6}\vspace{-4mm}&&&&\\
        \ctext{1}&\multirow{5}{*}{$\begin{matrix}
                \Gamma\ll1,\\
                {\rm Re}_{\rm M}\gg1,\\
                S\ll S_{\rm c}\\
                \text{SP reconnection}\end{matrix}$}
        &$\begin{matrix}
            r_{\rm diss}\ll1,\;
            \vert\epsilon\vert=1\end{matrix}$
        &$B^2\xi_{\rm M}$
        &\multirow{2}{*}{$(\rho+p)^{\frac{1}{4}}\sigma^{\frac{3}{4}}\eta^{\frac{1}{4}}B^{-\frac{1}{2}}\xi_{\rm M}^{\frac{3}{2}}$}
        &$\begin{matrix}\text{\eqref{eq:MDNLSPVMH}}\end{matrix}$\\
        \cline{1-1}\cline{3-4}\cline{6-6}\vspace{-4mm}&&&&\\
        \ctext{2}&&$\begin{matrix}
            r_{\rm diss}\ll1,\;
            \vert\epsilon\vert\ll1\end{matrix}$
        &$B^4\xi_{\rm M}^5$
        &&$\begin{matrix}\text{\eqref{eq:MDNLSPVNH}}\end{matrix}$\\
        \cline{1-1}\cline{3-6}\vspace{-4mm}&&&&\\
        \ctext{3}&&$\begin{matrix}
            r_{\rm diss}\gg1,\;
            \vert\epsilon\vert=1\end{matrix}$
        &$B^2\xi_{\rm M}$
        &\multirow{2}{*}{$(\rho+p)^{\frac{1}{2}}\sigma^{\frac{1}{2}}\alpha^{\frac{1}{2}}B^{-1}\xi_{\rm M}^{2}$}
        &$\begin{matrix}\text{\eqref{eq:MDNLSPDMH}}\end{matrix}$\\
        \cline{1-1}\cline{3-4}\cline{6-6}\vspace{-4mm}&&&&\\
        \ctext{4}&&$\begin{matrix}
            r_{\rm diss}\gg1,\;
            \vert\epsilon\vert\ll1\end{matrix}$ 
        &$B^4\xi_{\rm M}^5$
        &&$\begin{matrix}\text{\eqref{eq:MDNLSPDNH}}\end{matrix}$\\
        \cline{1-6}\vspace{-4mm}&&&&\\
        \ctext{5}&\multirow{5}{*}{$\begin{matrix}
            \Gamma\ll1,\\
            {\rm Re}_{\rm M}\gg1,\\
            S\gg S_{\rm c}\\
            \text{fast reconnection}\end{matrix}$}
        &$\begin{matrix}
            r_{\rm diss}\ll1,\;
            \vert\epsilon\vert=1\end{matrix}$
        &$B^2\xi_{\rm M}$
        &\multirow{2}{*}{$S_{\rm c}^{\frac{1}{2}}(\rho+p)^{\frac{1}{2}}\sigma^{\frac{1}{2}}\eta^{\frac{1}{2}}B^{-1}\xi_{\rm M}$}
        &$\begin{matrix}\text{\eqref{eq:MDNLFVMH}}\end{matrix}$\\
        \cline{1-1}\cline{3-4}\cline{6-6}\vspace{-4mm}&&&&\\
        \ctext{6}&&$\begin{matrix}
            r_{\rm diss}\ll1,\;
            \vert\epsilon\vert\ll1\end{matrix}$
        &$B^4\xi_{\rm M}^5$
        &&$\begin{matrix}\text{\eqref{eq:MDNLFVNH}}\end{matrix}$\\
        \cline{1-1}\cline{3-6}\vspace{-4mm}&&&&\\
        \ctext{7}&&$\begin{matrix}
            r_{\rm diss}\gg1,\;
            \vert\epsilon\vert=1\end{matrix}$
        &$B^2\xi_{\rm M}$
        &\multirow{2}{*}{$S_{\rm c}^{\frac{1}{2}}(\rho+p)\alpha B^{-2}\xi_{\rm M}^2$}
        &$\begin{matrix}\text{\eqref{eq:MDNLFDMH}}\end{matrix}$\\
        \cline{1-1}\cline{3-4}\cline{6-6}\vspace{-4mm}&&&&\\
        \ctext{8}&&$\begin{matrix}
            r_{\rm diss}\gg1,\;
            \vert\epsilon\vert\ll1\end{matrix}$
        &$B^4\xi_{\rm M}^5$
        &&$\begin{matrix}\text{\eqref{eq:MDNLFDNH}}\end{matrix}$\\
        \cline{1-6}\vspace{-4mm}&&&&\\
        \ctext{9}&\multirow{5}{*}{$\begin{matrix}
            \vspace{-3mm}\\
            \Gamma\ll1,\\
            {\rm Re}_{\rm M}\ll1\\\vspace{-3mm}\\
            \text{MD-linear}\end{matrix}$}
        &$\begin{matrix}
            r_{\rm diss}\ll1,\;
            \vert\epsilon\vert=1\end{matrix}$
        &$B^2\xi_{\rm M}$
        &\multirow{2}{*}{$(\rho+p)\eta B^{-2}$}
        &$\begin{matrix}\text{\eqref{eq:MDLVMH}}\end{matrix}$\\
        \cline{1-1}\cline{3-4}\cline{6-6}\vspace{-4mm}&&&&\\
        \ctext{10}&&$\begin{matrix}
            r_{\rm diss}\ll1,\;
            \vert\epsilon\vert\ll1\end{matrix}$
        &$B^4\xi_{\rm M}^5$
        &&$\begin{matrix}\text{\eqref{eq:MDLVNH}}\end{matrix}$\\
        \cline{1-1}\cline{3-6}\vspace{-4mm}&&&&\\
        \ctext{11}&&$\begin{matrix}
            r_{\rm diss}\gg1,\;
            \vert\epsilon\vert=1\end{matrix}$
        &$B^2\xi_{\rm M}$
        &\multirow{2}{*}{$(\rho+p)\alpha B^{-2}\xi_{\rm M}^2$}
        &$\begin{matrix}\text{\eqref{eq:MDLDMH}}\end{matrix}$\\
        \cline{1-1}\cline{3-4}\cline{6-6}\vspace{-4mm}&&&&\\
        \ctext{12}&&$\begin{matrix}
            r_{\rm diss}\gg1,\;
            \vert\epsilon\vert\ll1\end{matrix}$
        &$B^4\xi_{\rm M}^5$
        &&$\begin{matrix}\text{\eqref{eq:MDLDNH}}\end{matrix}$\\
        \cline{1-6}\vspace{-4mm}&&&&\\
        \ctext{13}&$\begin{matrix}
            \Gamma\gtrsim1,\vert\epsilon\vert=1\\
            \text{KD-MH}\end{matrix}$
        &&&&$\to$ MD\\\cline{1-6}\vspace{-4mm}&&&&\\
        \ctext{14}&\multirow{4}{*}{$\begin{matrix}
            \Gamma\gtrsim1,\\
            \vert\epsilon\vert\ll1\\
            \text{KD-NH}\end{matrix}$}
        &${\rm Re}_{\rm K}\gg1$
        &\multirow{3}{*}{$B^2\xi_{\rm M}^{n+3}$}
        &$(\rho+p)^{\frac{1}{2}}B^{-1}\xi_{\rm M}$
        &$\begin{matrix}\text{\eqref{eq:KDNHNL}}\end{matrix}$\\
        \cline{1-1}\cline{3-3}\cline{5-6}\vspace{-4mm}&&&&\\
        \ctext{15}&&$\begin{matrix}
            {\rm Re}_{\rm K}\ll1,\;
            r_{\rm diss}\ll1\end{matrix}$
        &&$\eta^{-1}\xi_{\rm M}^2$
        &$\begin{matrix}\text{\eqref{eq:KDNHLV}}\end{matrix}$\\
        \cline{1-1}\cline{3-3}\cline{5-6}\vspace{-4mm}&&&&\\
        \ctext{16}&&$\begin{matrix}
            {\rm Re}_{\rm K}\ll1,\;
            r_{\rm diss}\gg1\end{matrix}$
        &&$\alpha^{-1}$
        &$\to$ MD
      \end{tabularx}
      \caption{Classification (See also Fig.~\ref{fig:Classification}) and the scaling solutions in the sixteen regimes. Scaling solutions are obtained by taking into account the relevant conserved quantities and decay time scales in terms of $B$ and $\xi_{\rm M}$.}
      \label{tb:ResultsSummary}
\end{table}

\section{Application to primordial magnetic fields\label{sec:application}}
In this section, we clarify which regime is relevant in which epoch in the early universe.
We also make use of our condition \eqref{eq:TimeScaleCondition_general2} with $\tau_{\rm ini}=\tau_{\rm EW}$, where $\tau_{\rm EW}$ refers to the time right after the electroweak phase transition.
We mostly focus only on magnetically dominated regimes, since the transitions from kinetically dominated to magnetically dominated regimes are one-way.
We express magnetic field strength and magnetic coherence length in the comoving units, {\it i.e.}, $1\,{\rm G}$ is one comoving $\rm{G}$ and $1\,{\rm Mpc}$ is one comoving $\rm{Mpc}$ below.

If we start from a kinetically dominated regime, transition into a magnetically dominated regime is expected when either of
\begin{align}
    \vert\epsilon\vert=1\quad
    \text{or}\quad
    \alpha\tau\gg1
\end{align}
is satisfied.
The former condition can be satisfied if we start from a partially helical magnetic field with $\epsilon_{\rm ini}$ because the helicity fraction increases as the magnetic field decays, according to Eq.~\eqref{eq:HelicityConservation}.
When the helicity fraction reaches $\pm1$, the system is in the kinetically dominated and maximally helical regime, which enters magnetically dominated regimes.
As for the latter condition, the kinetically dominated, non-helical, linear, and viscous regime becomes the dragged regime because $r_{\rm diss}=\alpha\tau$.
Also, the kinetically dominated, non-helical, and non-linear regime becomes the linear and dragged regime because ${\rm Re}_{\rm K}=(\alpha\tau)^{-1}$.
Therefore, we expect a transition into a magnetically dominated regime.

\begin{figure}[t]
    \begin{minipage}[h]{0.95\hsize}
    \includegraphics[keepaspectratio,width=1.0\textwidth]{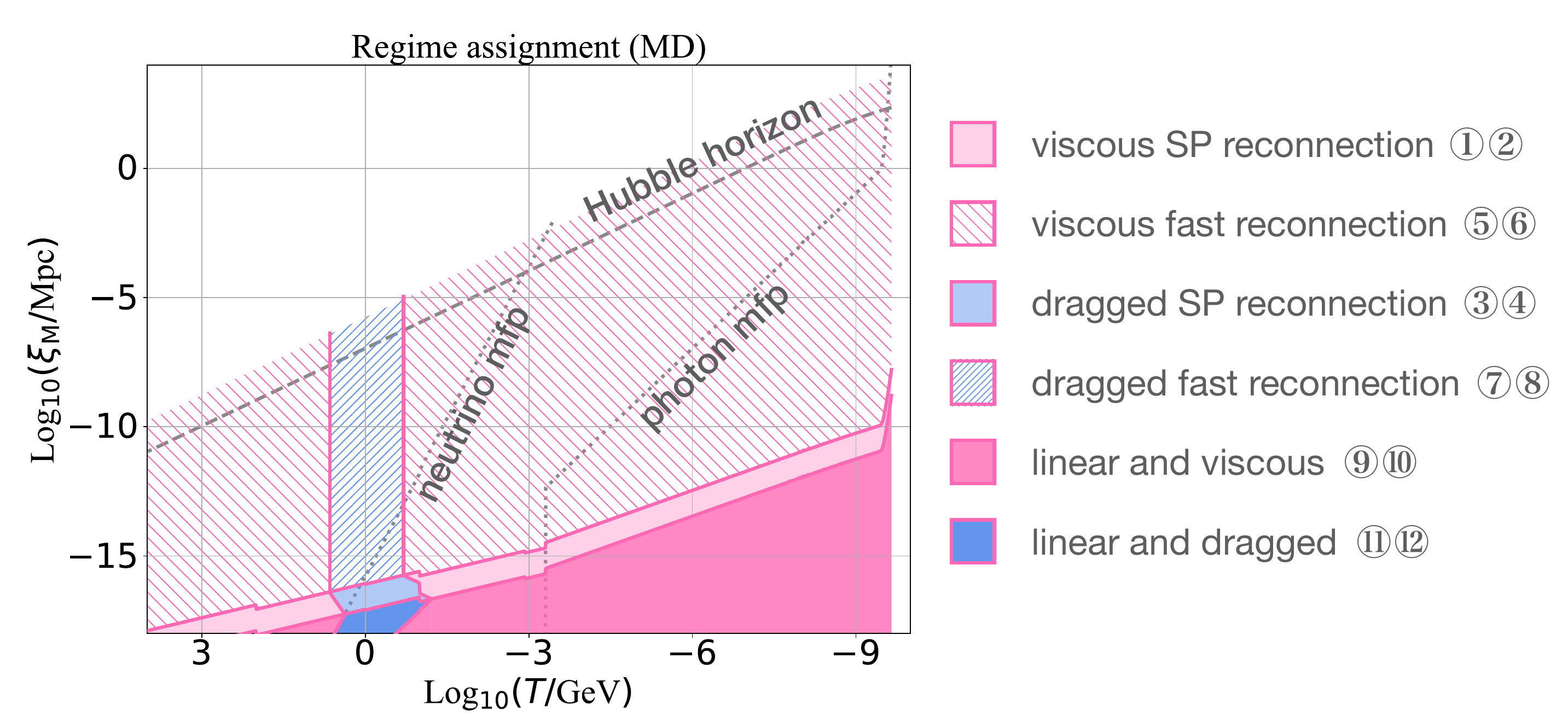} 
    \end{minipage}
    \caption{\label{fig:RegimeAssignment} Regime assignment of magnetic fields that satisfy the condition Eq.~\eqref{eq:TimeScaleCondition_general}. 
    Note that for given $T$ and $\xi_\mathrm{M}$, the magnetic field 
    strength $B$ and the corresponding regime that satisfy Eq.~\eqref{eq:TimeScaleCondition_general} are uniquely determined.
    Pink solid lines are the boundaries of different regimes.
    At every temperature, the decay process is the magnetically dominated, linear, and viscous one in the pink-shaded region, the viscous Sweet--Parker reconnection in the pale pink-shaded region, the viscous fast reconnection in the pink-hatched region, the magnetically dominated, linear, and dragged one in the blue-shaded region, the dragged Sweet--Parker reconnection in the pale blue-shaded region, and the viscous fast reconnection in the blue-hatched region.
    The gray dashed line is the Hubble horizon, and the gray-dotted lines are mean free paths of neutrinos and photons.
    }\vspace{3mm}
\end{figure}

On the other hand, we find $\Gamma\ll1$ in any magnetically dominated scaling regimes in the early universe (see Appendix \ref{appx:RegimeDependentAnalyses} for the explanation), and we do not expect transitions from magnetically dominated into kinetically dominated regimes.
Which magnetically dominated regimes are relevant in which epoch in the early universe is summarized in Fig.~\ref{fig:RegimeAssignment}.
If we assume that the system is in a scaling regime, 
a larger magnetic coherence length corresponds to stronger magnetic field.
For given $T$ and $\xi_\mathrm{M}$, we identify the magnetic field strength $B$ as well as the corresponding regime that satisfies Eq.~\eqref{eq:TimeScaleCondition_general}.
The system is linear for weak magnetic fields, and the Sweet--Parker and the fast magnetic reconnection are relevant for stronger magnetic fields.
Around ${\rm GeV}$ scale, the mean free path of neutrinos is significantly large, and they act on the plasma as drag force, which becomes dominant over shear viscosity.
Note that Eq.~\eqref{eq:TimeScaleCondition_general} does not depend 
on the helicity that the magnetic field carries, and hence the 
corresponding regimes are also independent of them.

Let us explain the explicit form of Eq.~\eqref{eq:TimeScaleCondition_general} at the electroweak scale $T\sim 100\,{\rm GeV}$.
If the coherence length is smaller than $\sigma^{-1/2}\tau^{1/2}\sim9\times 10^{-19}\,{\rm Mpc}$, then the system is in the linear and viscous regime, and the viscosity dissipation of the subdominant velocity field is the rate-determining process.
Longer coherence lengths are subject to the magnetic reconnection processes.
The boundary, $\xi_{\rm M}=S_{\rm c}^{1/4}\sigma^{-1/2}\tau^{1/2}\sim9\times 10^{-18}\,{\rm Mpc}$, separates the Sweet--Parker and the fast reconnection regimes.
Drag force is absent in this epoch because any particles have small mean free paths, and therefore the shear viscosity is the dominant dissipation term.
The relationship between $\xi_{\rm M}$ and corresponding $B$ that satisfy Eq.~\eqref{eq:TimeScaleCondition_general} is given by
\begin{align}
    B
        =\left\{\vspace{2mm}\begin{matrix}
            1\times10^{-10}{\,\rm G}\left(\dfrac{\xi_{\rm M,ini}}{10^{-17}\,{\rm Mpc}}\right),
            &\xi_{\rm M,ini}>9\times 10^{-18}\,{\rm Mpc}\vspace{3mm}\\
            2\times10^{-13}{\,\rm G}\left(\dfrac{\xi_{\rm M,ini}}{10^{-18}\,{\rm Mpc}}\right)^3,
            &9\times 10^{-19}\,{\rm Mpc}<\xi_{\rm M,ini}<9\times 10^{-18}\,{\rm Mpc}\vspace{5mm}\\
            1\times10^{-13}{\,\rm G},
            &\xi_{\rm M,ini}<9\times10^{-19}\,{\rm Mpc},\vspace{1mm}
        \end{matrix}\right.
    \label{eq:CondiitonEW}
\end{align}
where the first line is for the viscous fast reconnection regime (\ctext{5}, \ctext{6}), the second line is for the viscous Sweet--Parker reconnection regime (\ctext{1}, \ctext{2}), and the third line is for the linear and viscous regime (\ctext{9}, \ctext{10}).
At ${\rm GeV}$ scale, the mean free path of neutrinos is larger than $\xi_{\rm K}$, and they act on the plasma as a drag force.
If the coherence length is smaller than $\eta^{1/2}\alpha^{-1/2}\sim7\times 10^{-20}\,{\rm Mpc}$, then the shear viscosity without neutrinos is dominant over the drag force with neutrinos.
If the coherence length is larger than $\sigma^{-1/2}\tau^{1/2}\sim8\times 10^{-18}\,{\rm Mpc}$, the system is in the magnetic reconnection regimes.
The boundary between the dragged Sweet--Parker reconnection and the dragged fast reconnection regimes is at $\xi_{\rm M}=S_{\rm c}^{1/4}\sigma^{-1/2}\tau^{1/2}\sim8\times 10^{-17}\,{\rm Mpc}$.
With the condition Eq.~\eqref{eq:TimeScaleCondition_general}, $\xi_{\rm M}$ and the corresponding $B$ are related as
\begin{align}
    B
        =\left\{\vspace{2mm}\begin{matrix}
            2\times10^{-10}{\,\rm G}\left(\dfrac{\xi_{\rm M}}{10^{-16}\,{\rm Mpc}}\right),
            &\xi_{\rm M}>8\times 10^{-17}\,{\rm Mpc}\vspace{3mm}\\
            2\times10^{-12}{\,\rm G}\left(\dfrac{\xi_{\rm M}}{10^{-17}\,{\rm Mpc}}\right)^2,
            &8\times10^{-18}\,{\rm Mpc}<\xi_{\rm M}<8\times10^{-17}\,{\rm Mpc}\vspace{5mm}\\
            2\times10^{-13}{\,\rm G}\left(\dfrac{\xi_{\rm M}}{10^{-18}\,{\rm Mpc}}\right),
            &7\times10^{-20}\,{\rm Mpc}<\xi_{\rm M}<8\times10^{-18}\,{\rm Mpc},\vspace{5mm}\\
            1\times10^{-14}{\,\rm G},
            &\xi_{\rm M}<7\times10^{-20}\,{\rm Mpc}\vspace{1mm}
        \end{matrix}\right.
    \label{eq:CondiitonGeV}
\end{align}
where the first line is for the dragged fast reconnection regime (\ctext{7}, \ctext{8}), the second line is for the dragged Sweet--Parker reconnection regime (\ctext{3}, \ctext{4}), the third line is for the linear and dragged regime (\ctext{11}, \ctext{12}), and the final line is for the linear and viscous regime (\ctext{9}, \ctext{10}).
At the QCD scale $\sim 100\,{\rm MeV}$, we have five regimes.
If the coherence length is smaller than $\eta^{1/2}\alpha^{-1/2}\sim8\times10^{-18}\,{\rm Mpc}$, then the shear viscosity without neutrinos is dominant over the drag force with neutrinos.
If the coherence length is larger than $\sigma^{-1/2}\tau^{1/2}\sim2\times10^{-17}\,{\rm Mpc}$, the system is in the magnetic reconnection regimes.
While neutrinos contribute to the drag term as long as the coherence length is smaller than $\sigma^{-1}\eta^{-1/2}\alpha^{1/2}\tau\sim4\times10^{-17}\,{\rm Mpc}$, they contribute to the shear viscosity otherwise.
The boundary between the viscous Sweet--Parker reconnection and the viscous fast reconnection regimes is at $\xi_{\rm M}=S_{\rm c}^{1/4}\sigma^{-1/2}\tau^{1/2}\sim2\times10^{-16}\,{\rm Mpc}$.
$B$ and $\xi_{\rm M}$ that satisfy Eq.~\eqref{eq:TimeScaleCondition_general} are related as
\begin{align}
    B
        =\left\{\vspace{2mm}\begin{matrix}
            7\times10^{-12}{\,\rm G}\left(\dfrac{\xi_{\rm M}}{10^{-16}\,{\rm Mpc}}\right),
            &\xi_{\rm M}>2\times10^{-16}\,{\rm Mpc}\vspace{3mm}\\
            2\times10^{-12}{\,\rm G}\left(\dfrac{\xi_{\rm M}}{10^{-16}\,{\rm Mpc}}\right)^3,
            &4\times10^{-17}\,{\rm Mpc}<\xi_{\rm M}<2\times10^{-16}\,{\rm Mpc}\vspace{5mm}\\
            8\times10^{-15}{\,\rm G}\left(\dfrac{\xi_{\rm M}}{10^{-17}\,{\rm Mpc}}\right)^2,
            &2\times10^{-17}\,{\rm Mpc}<\xi_{\rm M}<4\times10^{-17}\,{\rm Mpc},\vspace{5mm}\\
            1\times10^{-14}{\,\rm G}\left(\dfrac{\xi_{\rm M}}{10^{-17}\,{\rm Mpc}}\right),
            &8\times10^{-18}\,{\rm Mpc}<\xi_{\rm M}<2\times10^{-17}\,{\rm Mpc},\vspace{5mm}\\
            1\times10^{-14}{\,\rm G},
            &\xi_{\rm M}<8\times10^{-18}\,{\rm Mpc},\vspace{1mm}
        \end{matrix}\right.
    \label{eq:CondiitonQCD}
\end{align}
where the first line is for the viscous fast reconnection regime (\ctext{5}, \ctext{6}), the second line is for the viscous Sweet--Parker reconnection regime (\ctext{1}, \ctext{2}), the third line is for the dragged Sweet--Parker reconnection regime (\ctext{3}, \ctext{4}), the fourth line is for the linear and dragged regime (\ctext{11}, \ctext{12}), and the final line is for the linear and viscous regime (\ctext{9}, \ctext{10}).
Just before the BBN at $\sim 1\,{\rm MeV}$, neutrinos contribute to the drag term, but the drag term is subdominant.
If the coherence length is smaller than $\sigma^{-1/2}\tau^{1/2}\sim1\times10^{-16}\,{\rm Mpc}$, the system is in the linear regime.
Otherwise, the viscous magnetic reconnection regimes are relevant.
The boundary between the viscous Sweet--Parker reconnection and the viscous fast reconnection regimes is at $\xi_{\rm M}=S_{\rm c}^{1/4}\sigma^{-1/2}\tau^{1/2}\sim1\times10^{-15}\,{\rm Mpc}$.
$B$ and $\xi_{\rm M}$ that satisfy Eq.~\eqref{eq:TimeScaleCondition_general} are related as
\begin{align}
    B
        =\left\{\begin{matrix}
            1\times10^{-12}{\,\rm G}\left(\dfrac{\xi_{\rm M}}{10^{-15}\,{\rm Mpc}}\right),
            &\xi_{\rm M}>1\times10^{-15}\,{\rm Mpc}\vspace{3mm}\\
            7\times10^{-16}{\,\rm G}\left(\dfrac{\xi_{\rm M}}{10^{-16}\,{\rm Mpc}}\right)^3,
            &1\times10^{-16}\,{\rm Mpc}<\xi_{\rm M}<1\times10^{-15}\,{\rm Mpc}\vspace{5mm}\\
            2\times10^{-15}{\,\rm G},
            &\xi_{\rm M}<1\times10^{-16}\,{\rm Mpc},\vspace{1mm}
        \end{matrix}\right.
    \label{eq:CondiitonBBN}
\end{align}
where the first line is for the viscous fast reconnection regime (\ctext{5}, \ctext{6}), the second line is for the viscous Sweet--Parker reconnection regime (\ctext{1}, \ctext{2}), and the third line is for the linear and viscous regime (\ctext{9}, \ctext{10}).
At $\sim 0.5\,{\rm MeV}$, electrons become massive, the mean free path of photons grows, and photons start to contribute as the drag force.
However, $r_{\rm diss}$ is always smaller than unity below this scale.
At the recombination epoch $\sim 0.1\,{\rm eV}$,
if the coherence length is smaller than $\sigma^{-1/2}\tau^{1/2}\sim2\times10^{-11}\,{\rm Mpc}$, the system is in the linear regime.
Otherwise, the viscous magnetic reconnection regimes are relevant.
The boundary between the viscous Sweet--Parker reconnection and the viscous fast reconnection regimes is at $\xi_{\rm M}=S_{\rm c}^{1/4}\sigma^{-1/2}\tau^{1/2}\sim2\times10^{-10}\,{\rm Mpc}$.
$B$ and $\xi_{\rm M}$ that satisfy Eq.~\eqref{eq:TimeScaleCondition_general} are related as
\begin{align}
    B
        =\left\{\begin{matrix}
            5\times10^{-14}{\,\rm G}\left(\dfrac{\xi_{\rm M}}{10^{-10}\,{\rm Mpc}}\right),
            &\xi_{\rm M}>2\times 10^{-10}\,{\rm Mpc}\vspace{3mm}\\
            1\times10^{-14}{\,\rm G}\left(\dfrac{\xi_{\rm M}}{10^{-10}\,{\rm Mpc}}\right)^3,
            &2\times10^{-11}\,{\rm Mpc}<\xi_{\rm M}<2\times10^{-10}\,{\rm Mpc}\vspace{5mm}\\
            1\times10^{-16}{\,\rm G},
            &\xi_{\rm M}<2\times10^{-11}\,{\rm Mpc},\vspace{1mm}
        \end{matrix}\right.
    \label{eq:Condiitonrec}
\end{align}
where the first line is for the viscous fast reconnection regime (\ctext{5}, \ctext{6}), the second line is for the viscous Sweet--Parker reconnection regime (\ctext{1}, \ctext{2}), and the third line is for the linear and viscous regime (\ctext{9}, \ctext{10}).
Since the charged plasma disappears at recombination, we assume that the magneto-hydrodynamic evolution of the magnetic field is negligible in the late universe \cite{Durrer+13}.

\begin{figure}[t]
    \begin{minipage}[h]{0.95\hsize}
    \includegraphics[keepaspectratio, width=1.0\textwidth]{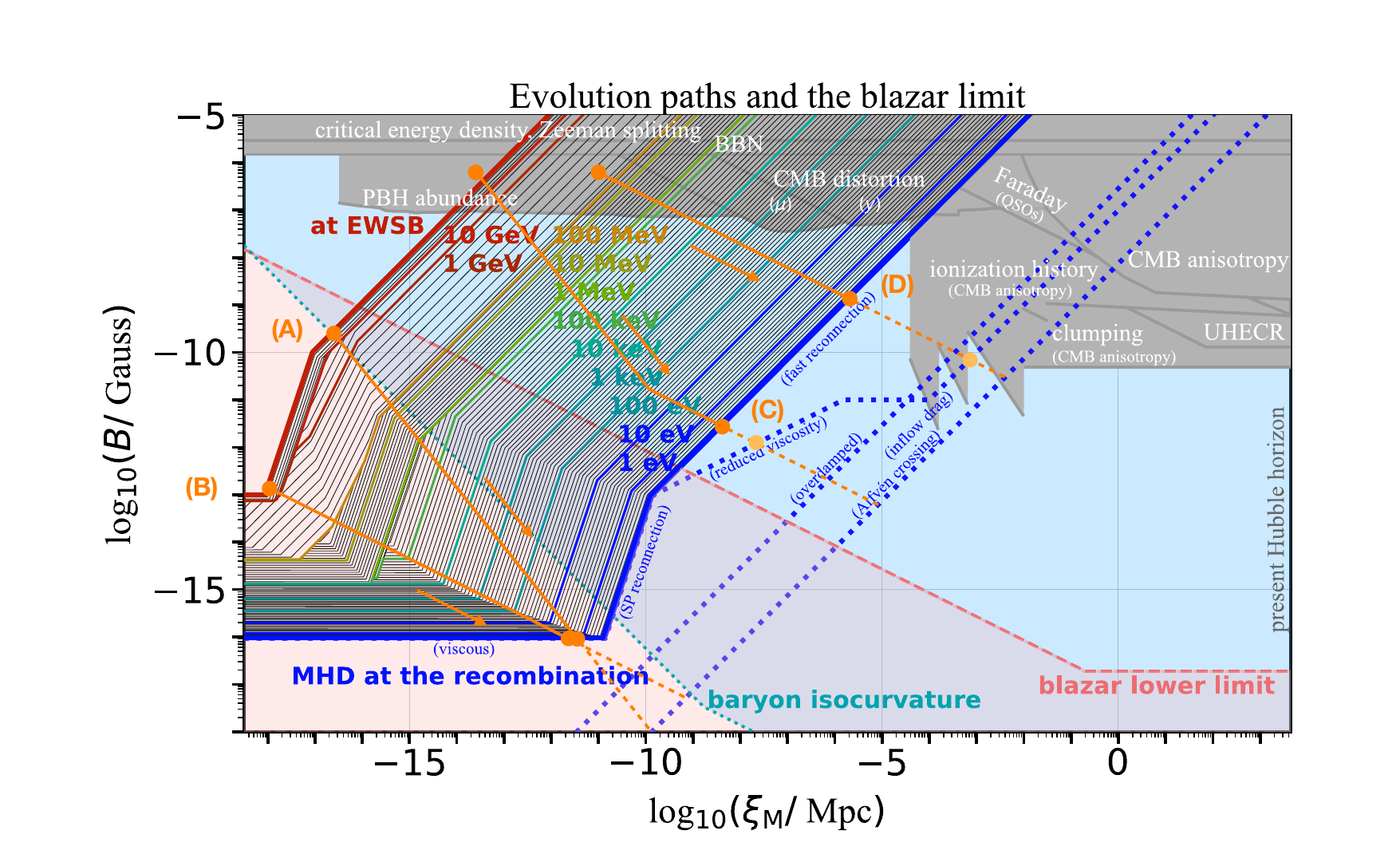}
    \end{minipage}
    \caption{\label{fig:EvolutionPlot1} The summary plot in the $\xi_{\rm M}$-$B$ plane.
    Shaded regions imply the constraints on primordial magnetic fields.
    See Appendix \ref{appx:statusofconstraints} for the constraints represented by the gray-shaded regions. The constraint from CMB spectral distortions for helical magnetic fields are shown for the comparison with the evolution path (D). Here we focus on the}
    lower bound on the strength of the observed intergalactic magnetic fields in void regions (pink-shaded region) \cite{MAGIC:2022piy} and the upper bound at the electroweak scale that arises from the baryon isocurvature problem (blue-shaded region) \cite{Kamada:2020bmb}.
    Orange solid lines are the evolution paths for the initial conditions (A--D). They evolve as the time-scale lines (thin-gray and thick-colored solid lines) moves downward, where the colored (gray) lines represents the time-scale lines at temperatures of every one ($0.1$) order of magnitude.
    Other possible time-scale lines at the recombination epoch have been discussed in the literature \cite{2000PhRvL..85..700J,Banerjee+04,Hosking+22} (blue-dotted lines). In particular, if we adopt Ref.~\cite{Hosking+22}, (C) and (D) result in the pale orange dots.
    \vspace{3mm}
\end{figure}
We plot the $B$-$\xi_{\rm M}$ relation determined by Eq.~\eqref{eq:TimeScaleCondition_general2} at temperatures of every $0.1$ order of magnitude in Fig.~\ref{fig:EvolutionPlot1}.
For the practical purpose, one can use this plot in the following way.
\begin{enumerate}
    \item Specify the initial conditions $B_{\rm ini}, \xi_{\rm ini},\epsilon_{\rm ini},$ and $T_{\rm ini}$.
    \item Put a point at ($\xi_{\rm M,ini}, B_{\rm ini}$) in the $\xi_{\rm M}$-$B$ plane in the figure.
    \item Draw a line $B^2\xi_{\rm M}=\epsilon_{\rm ini}B_{\rm ini}^2\xi_{\rm M, ini}$ for $\epsilon_{\rm ini} \not =0$.
    \item Look for the time-scale line (thin gray line) corresponding to the temperature $T<T_{\rm ini}$ of interest.
    \item If $\epsilon_{\rm ini}=1$ initially, find the intersection of the line $B^2\xi_{\rm M}=B_{\rm ini}^2\xi_{\rm M, ini}$ and the time-scale line at $T$.
    \item Otherwise, draw a line $B^4\xi^5_{\rm M}=B^4_{\rm ini}\xi^5_{\rm M, ini}$ and find the intersection of this line and the time-scale line at $T$. If the intersection is below the line $B^2\xi_{\rm M}=\epsilon_{\rm ini}B_{\rm ini}^2\xi_{\rm M, ini}$, forget about that intersection and rather find the intersection of the line $B^2\xi_{\rm M}=\epsilon_{\rm ini}B_{\rm ini}^2\xi_{\rm M, ini}$ and the time-scale line at $T$.
\end{enumerate}
Then, the point in the $\xi_{\rm M}$-$B$ plane that one finds as the intersection of two lines represents the solution, $B(T),\xi_{\rm M}(T)$, which may be discussed in connection with whatever conditions one has at temperature $T$.
In particular, if one would like to obtain the present magnetic field strength and coherence length, one can just find the intersection between the line at the recombination determined by Eq.~\eqref{eq:Condiitonrec} and the line drawn by the procedure described above.
For the analytic expression, one should consult Table \ref{tb:ResultsSummary} and find the formula for the relevant regime.

\section{Discussion}
In the previous section, we have explained how one can describe the evolution of primordial magnetic fields with arbitrary initial conditions.
In this section, applying this general description, we provide a few explicit examples of magnetic field evolution in the early universe.
Note that magnetic field strength and magnetic coherence length are expressed in comoving units.

\begin{table}\begin{center}
      \begin{tabularx}{0.99\textwidth}{c||c|c|c|c||c|c}
        &$T_{\rm ini}$&$B_{\rm ini}$&$\xi_{\rm M,ini}$&$\epsilon_{\rm ini}$& $B_{\rm rec}$&$\xi_{\rm M,rec}$\\\hline
        &&&&&&\vspace{-4mm}\\
        (A)&$100\,{\rm GeV}$&$3\times 10^{-10}\,{\rm G}$&$3\times 10^{-17}\,{\rm Mpc}$&$0$&$9\times 10^{-17}\,{\rm G}$&$2\times 10^{-12}\,{\rm Mpc}$\\&&&&&&\vspace{-4mm}\\
        (B)&$100\,{\rm GeV}$&$1\times 10^{-13}\,{\rm G}$&$1\times 10^{-18}\,{\rm Mpc}$&$1$&$9\times 10^{-17}\,{\rm G}$&$4\times 10^{-12}\,{\rm Mpc}$\\&&&&&&\vspace{-4mm}\\
        (C)&$100\,{\rm GeV}$&$7\times 10^{-7}\,{\rm G}$&$3\times 10^{-14}\,{\rm Mpc}$&$3\times 10^{-6}$&$3\times 10^{-12}\,{\rm G}$&$4\times 10^{-9}\,{\rm Mpc}$\\&&&&&&\vspace{-4mm}\\
        (D)&$1\,{\rm MeV}$&$7\times 10^{-7}\,{\rm G}$&$1\times 10^{-11}\,{\rm Mpc}$&$1$&$1\times 10^{-9}\,{\rm G}$&$2\times 10^{-6}\,{\rm Mpc}$
      \end{tabularx}
      \caption{Four choices of initial conditions and the resultant magnetic fields at the recombination epoch. (A) Non-helical magnetic fields generated before the electroweak scale. The strength is taken as strong as possible for the baryon isocurvature problem not to arise. Non-helical magnetic field is assumed as an extreme choice. (B) Maximally helical magnetic fields generated before the electroweak scale. The strength is taken as strong as possible for the baryon isocurvature problem nor the baryon overproduction \cite{Kamada:2016cnb,Kamada:2020bmb} not to arise. (C) A low-scale inflationary magnetogenesis scenario \cite{Yanagihara:2023qvx} at the electroweak scale. (D) Another low-scale inflationary magnetogenesis scenario \cite{Yanagihara:2023qvx} at just above the big-bang nucleosynthesis.}
      \label{tb:InitialConditions}
      \end{center}
    \end{table}
We take four initial conditions, (A--D), given in Table \ref{tb:InitialConditions}.
We first consider primordial magnetic fields generated above the electroweak scale.
Then, at the electroweak symmetry breaking, the magnetic fields should satisfy the condition,
\begin{align}
    \left(\dfrac{B(T_{\rm EW})}{10^{-13}\,{\rm G}}\right)\left(\dfrac{\xi_{\rm M}(T_{\rm EW})}{10^{-13}\,{\rm Mpc}}\right)\lesssim 1
    \quad
    \text{at}
    \quad
    \xi_{\rm M}\ll10^{-9}\,{\rm Mpc}
    \label{eq:BaryonIsocurvatureShort}
\end{align}
to avoid the baryon isocurvature problem (blue-shaded region bounded by a dotted line in Fig.~\ref{fig:EvolutionPlot1}) \cite{Kamada:2020bmb}.
On the other hand, the magneto-hydrodynamic processes make the magnetic fields on or below the line determined by the condition, Eq.~\eqref{eq:TimeScaleCondition_general2}, evaluated at the electroweak scale (the leftmost blue solid line in Fig.~\ref{fig:EvolutionPlot1} gives the upper limit).

For Case (A), we choose the initial strength and coherence length
\begin{align}
    B_{\rm ini}^{({\text{A}})}=3\times 10^{-10}\,{\rm G},\quad
    \xi_{\rm M,ini}^{({\text{A}})}=3\times 10^{-17}\,{\rm Mpc}
\end{align}
at $T_{\rm ini}^{({\text{A}})}=100\,{\rm GeV}$, so that the magnetic field is as strong as possible below the upper limits.
We assume a non-helical magnetic field because, if the magnetic field is helical, not only the baryon isocurvature problem but also the baryon overproduction problem \cite{Kamada:2016cnb,Kamada:2020bmb} should be avoided.
The allowed maximum of the magnetic helicity fraction for $B_{\rm ini}^{({\text{A}})}, \xi_{\rm M,ini}^{({\text{A}})}$ is $\epsilon\sim10^{-7}$, assuming that the dominant contribution to the baryon asymmetry of the universe is generated independently of the baryon number generation by helical primordial magnetic fields.
If we assume maximally helical magnetic field generated before the electroweak scale, baryon overproduction is avoided only when \cite{Kamada:2020bmb}
\begin{align}
    10^{-10}\left(\dfrac{B(T_{\rm EW})}{10^{-13}\,{\rm G}}\right)^2\left(\dfrac{\xi_{\rm M}(T_{\rm EW})}{10^{-16}\,{\rm Mpc}}\right)\ll \eta_{B,\rm obs}\sim 10^{-10}.
    \label{eq:BaryonOverproduction}
\end{align}

For Case (B),
we choose the initial strength and coherence length
\begin{align}
    B_{\rm ini}^{({\text{B}})}=1\times 10^{-13}\,{\rm G},\quad
    \xi_{\rm M,ini}^{({\text{B}})}=1\times 10^{-18}\,{\rm Mpc}
\end{align}
at $T_{\rm ini}^{({\text{B}})}=100\,{\rm GeV}$, so that the magnetic field is as strong as possible below the upper limits.

The other initial conditions are based on the fact that magnetogenesis scenarios in the broken phase of the electroweak symmetry are free from the baryon isocurvature problem \cite{Yanagihara:2023qvx}.
They introduced the Chern-Simons coupling between the inflaton and the gauge field, through which the inflaton velocity amplifies helical magnetic fields.
For a choice of parameters, their models have a potential to generate a magnetic field with
\begin{align}
    B_{\rm ini}^{({\text{C}})}=7\times 10^{-7}\,{\rm G},\quad
    \xi_{\rm M,ini}^{({\text{C}})}=3\times 10^{-14}\,{\rm Mpc},\quad
    \epsilon_{\rm ini}^{({\text{C}})}=3\times 10^{-6}
\end{align}
at $T_{\rm ini}^{({\text{C}})}=100\,{\rm GeV}$, by taking the magnetogenesis temperature as high as possible, avoiding the baryon isocurvature problem.
Their models also have a potential to generate a magnetic field with
\begin{align}
    B_{\rm ini}^{({\text{D}})}=7\times 10^{-7}\,{\rm G},\quad
    \xi_{\rm M,ini}^{({\text{D}})}=1\times 10^{-11}\,{\rm Mpc},\quad
    \epsilon_{\rm ini}^{({\text{D}})}=1
\end{align}
at $T_{\rm ini}^{({\text{D}})}=1\,{\rm MeV}$, by taking the magnetogenesis temperature as low as possible \cite{Yanagihara:2023qvx}.
Note that this initial condition does not satisfy Eq.~\eqref{eq:TimeScaleCondition_general2}. We assume that the magnetic field decays and the coherence length grows immediately, conserving its magnetic helicity density, until Eq.~\eqref{eq:TimeScaleCondition_general2} is satisfied.\footnote{If the coherence length 
is too short, the conservation of the magnetic helicity or the Hosking integral can be violated.
See the discussion in Sec.~\ref{sec:MagneticHelicity}.}

\begin{figure}[t]
    \begin{tabular}{cc}
        \begin{minipage}[h]{0.494\hsize}
            \includegraphics[keepaspectratio, width=1.0\textwidth]{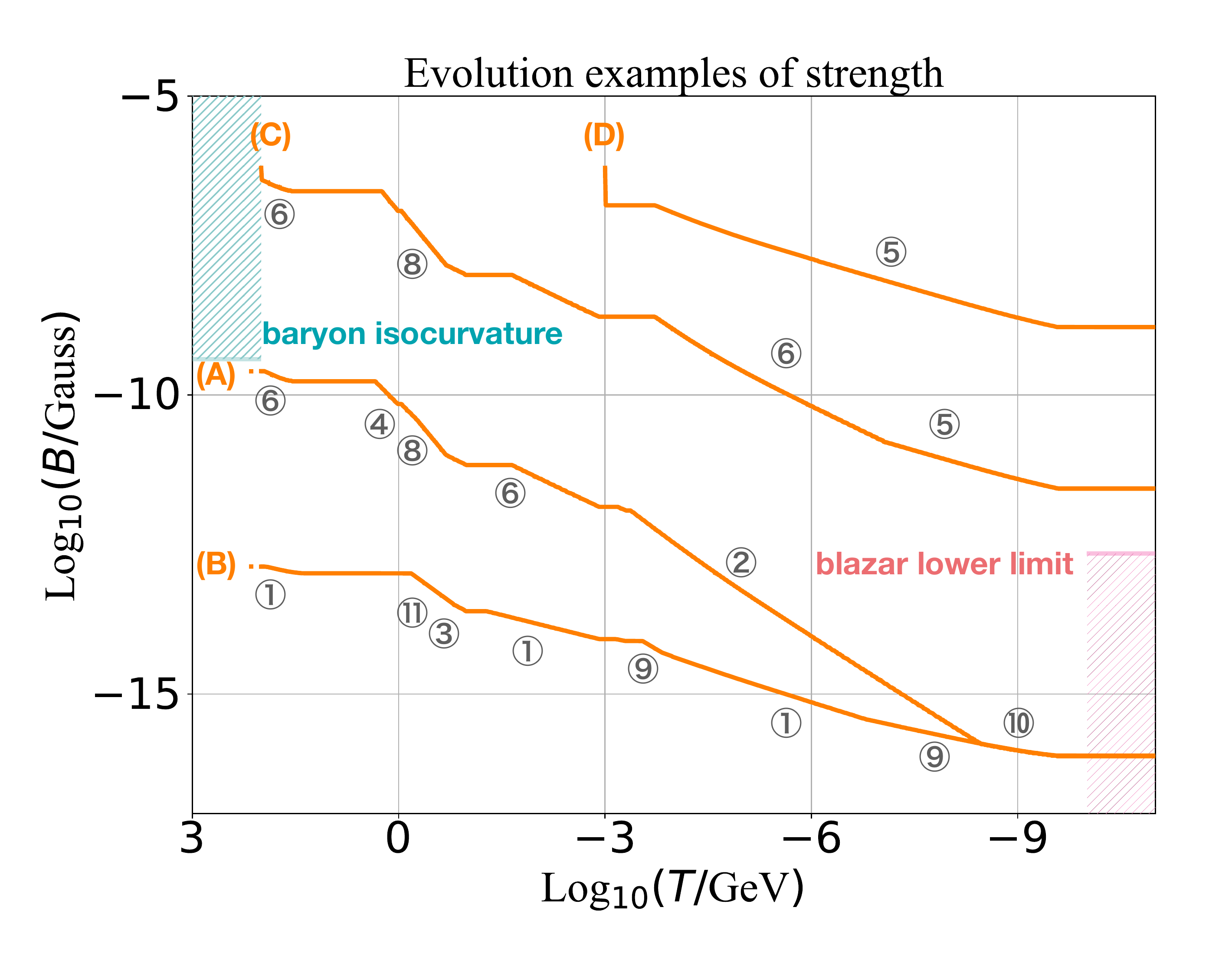}
        \end{minipage}&
        \begin{minipage}[h]{0.494\hsize}
            \includegraphics[keepaspectratio, width=1.0\textwidth]{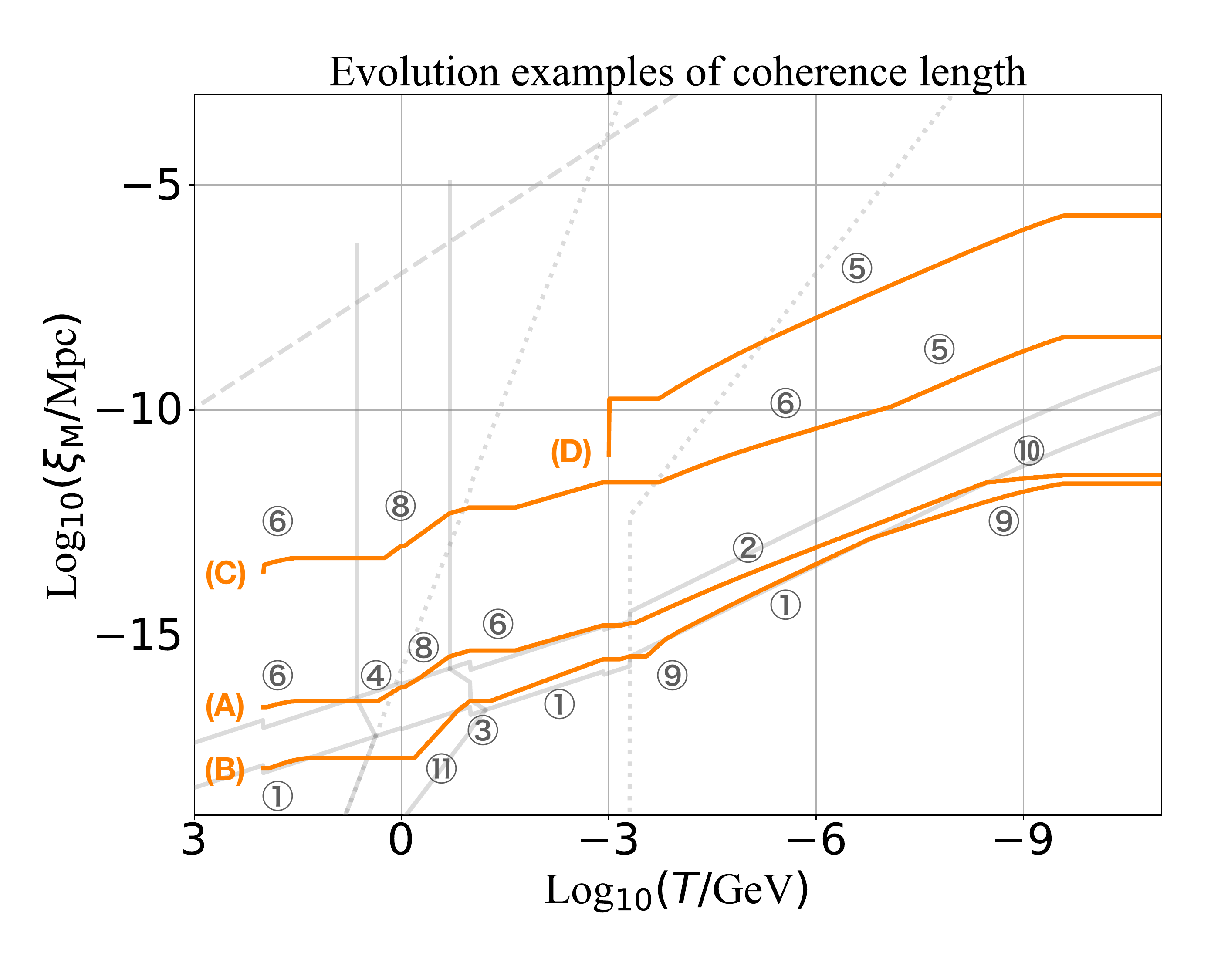}
        \end{minipage}
    \end{tabular}
    \caption{\label{fig:EvolutionPlot4} Evolution of the strength (orange solid lines in the left panel) and the coherence length (orange solid lines in the right panel) of magnetic fields for the initial conditions (A--D). The pale blue-hatched region is the exclusion at the electroweak temperature by the baryon isocurvature problem. Pale pink-hatched regions are below the observational lower limit of the intergalactic magnetic fields in void regions. The pale gray lines in the right panel are for the comparison with Fig.~\ref{fig:RegimeAssignment}, from which we can read off the regimes the system is in (denoted by numbers with circles).}
    \vspace{3mm}
\end{figure}

Figure~\ref{fig:EvolutionPlot4} depicts evolution of the strength and the coherence length of magnetic fields for various initial conditions which are obtained by solving the 
conservation laws, Eqs.~\eqref{eq:HelicityConservation_MH} and \eqref{eq:HelicityConservation_NH}, and the $B$-$\xi_{\rm M}$ relation determined by the lines of time scales in Fig.~\ref{fig:EvolutionPlot1}.
The left panel describes the evolution of magnetic field strength, and the right panel is for the evolution of magnetic coherence length.

By comparing the right panel with Fig.~\ref{fig:RegimeAssignment}, we can read off what mechanism is operating at each temperature.
Magnetic fields for the initial conditions (A--C) are almost frozen before the neutrino-drag fast reconnection (\ctext{8}) is involved at around ${\rm GeV}$ scale.
The decay of magnetic field strength is milder for the maximally helical field (B) compared with the non-helical fields (A, C).

For (A), the magnetic field is mostly in the reconnection regimes (\ctext{6} $\to$ \ctext{4} $\to$ \ctext{8} $\to$ \ctext{6} $\to$ \ctext{2}) and finally enters the linear and viscous regime (\ctext{10}) at $\sim 10\,{\rm eV}$.
For (B), the magnetic field is mostly in the reconnection regimes (\ctext{1} and \ctext{3}) but experiences both of the linear regimes (\ctext{11} and \ctext{9}).
For (C) after that, and for (D), the viscous fast reconnection regime (\ctext{5} and \ctext{6}) is the only relevant decay process.
The flat regions in the plots imply that magnetic fields sometimes freeze before the reconnection rate becomes fast enough again.

The bending of the plots for (C) at around $100\,{\rm eV}$ reflects the transition from the non-helical to the maximally helical regime (\ctext{6} $\to$ \ctext{5}).
The magnetic field (C) initially has only a tiny magnetic helicity fraction, but the magnetic field decays so much, while magnetic helicity is conserved, that it becomes maximally helical eventually.
Strictly speaking, the Hosking integral is ill-defined for fractionally helical magnetic fields. Recently, however, Ref.~\cite{2025JPlPh..91E...5B} discussed that the evolution of fractionally helical magnetic fields is well-approximated as a transition from the one governed by the Hosking integral conservation to the one governed by the helicity conservation, which supports the validity of our treatment.

The resultant magnetic fields at the recombination epoch are estimated as
\begin{align}
    B_{\rm rec}^{({\text{A}})}&=9\times 10^{-17}\,{\rm G},\quad
    \xi_{\rm M,rec}^{({\text{A}})}=2\times 10^{-12}\,{\rm Mpc},\\
    B_{\rm rec}^{({\text{B}})}&=9\times 10^{-17}\,{\rm G},\quad\,
    \xi_{\rm M,rec}^{({\text{B}})}=4\times 10^{-12}\,{\rm Mpc},\\
    B_{\rm rec}^{({\text{C}})}&=3\times 10^{-12}\,{\rm G},\quad\,
    \xi_{\rm M,rec}^{({\text{C}})}=4\times 10^{-9}\,{\rm Mpc},\\
    B_{\rm rec}^{({\text{D}})}&=1\times 10^{-9}\,{\rm G},\quad\,
    \xi_{\rm M,rec}^{({\text{D}})}=2\times 10^{-6}\,{\rm Mpc}.
\end{align}
An important consequence is that the resultant magnetic fields for (A, B) do not explain the observed strength of the intergalactic magnetic fields in void regions.
This fact implies that any magnetogenesis scenario above the electroweak scale is not a viable option.

On the other hand, magnetogenesis scenarios at or below the electroweak scale, (C) and (D), are promising in the sense that they are both safe from the baryon isocurvature constraint and able to account for the origin of the intergalactic magnetic field.\footnote{Note that, to investigate the low-scale inflationary magnetogenesis models \cite{Yanagihara:2023qvx} further, we need to avoid the constraints from primordial black hole abundances \cite{2020JCAP...05..039S,2024JCAP...12..012K} (See Appendix \ref{appx:statusofconstraints}), which still have several uncertainties. 
}

We plot the evolution paths of these initial conditions, (A--D), in the $\xi_{\rm M}$-$B$ plane in Fig.~\ref{fig:EvolutionPlot1}.
If we assume the allowed maximum helicity fraction for the initial condition (A), it reaches maximally helical before the recombination and results in the same resultant magnetic fields as the case (B) (also shown in Fig.~\ref{fig:EvolutionPlot1}).
We also note that, for strong magnetic fields, the estimate of the viscosity coefficient could be modified, and also the kinetic reconnection regime could be taken into account because the width of reconnection current sheets become comparable to typical scales of particle collision.
We show a possibility of these modifications proposed in Ref.~\cite{Hosking+22} with a faint blue solid line in the figure. 
If we accept this estimate, the estimate of the resultant magnetic fields for (C, D) changes, but the conclusion that magnetogenesis scenarios at or below the electroweak scale are promising is the same.

\section{Summary}
To conclude, let us summarize the new results obtained in our study.\\
\indent
First, we would like to emphasize that
\begin{itemize}
    \item this is the first study with which one can describe the evolution of primordial magnetic fields comprehensively as a function of time before the recombination epoch, consistently with the existing numerical studies.
\end{itemize}
Banajee and Jedamzik \cite{Banerjee+04} addressed very the same problem, but they did not consider the non-helical inverse transfer and the reconnection-driven turbulence, which are suggested mostly by later numerical studies.
Hosking and Schekochihin \cite{Hosking+22} did take into account these ideas and discussed the $B-\xi_{\rm M}$ relation at the recombination epoch.
However, since constraints on primordial magnetic fields are imposed at different epochs (Appendix \ref{appx:statusofconstraints}), understanding the magnetic field only at the recombination epoch is insufficient if one wishes to discuss how those constraints constrain magnetogenesis scenarios.

The general importance of our study lies in its potential applicability. With the help of our results,
\begin{itemize}
    \item if one sets an initial condition of primordial magnetic fields, one can discuss their effect on events in the early universe at an arbitrary temperature afterward.
    Conversely, when one has an implication of the existence of primordial magnetic fields, one can trace back their evolution and quantitatively discuss possible magnetogenesis scenarios.
\end{itemize}

Second, while either magnetic energy dominance or an equipartition of magnetic and kinetic energy is often assumed in the literature, 
\begin{itemize}
    \item we have systematically exhausted magnetically dominated, kinetically dominated, and equipartition regimes,
\end{itemize}
only a part of which has been done by Ref.~\cite{Hosking+21}.

Third, as an application of our result, we have updated the constraints on magnetogenesis scenarios. It has already been discussed that causally generated and non-helical primordial magnetic fields result in tiny magnetic fields today, which are insufficient to explain the origin of the void magnetic field \cite{2016JCAP...01..002W, Brandenburg+17}. 
In addition, some of us recently showed that general primordial magnetic fields generated above the electroweak scale, which may be either causally-generated or super-horizon and either helical or non-helical, are severely constrained and cannot explain the origin of the void magnetic field \cite{Kamada:2020bmb}.
In these studies, however, the evolution of non-helical magnetic fields after their generation was either parametrized with an uncertainty or treated without considering the non-helical inverse transfer.
In this regard, our description of the magnetic field evolution is readily applicable to quantitatively reassess the conclusions in the literature.
\begin{itemize}
    \item {
    We confirmed that the conclusions in the literature hold qualitatively, namely that magnetogenesis above the electroweak scale does not reproduce the observed void magnetic field and that helical magnetogenesis at around or below the electroweak scale may be the origin of the void magnetic field.}
\end{itemize}

As a final remark, the analysis in this article is purely analytic. 
Although the existing numerical simulations support the formulae we developed here for some of the regimes~\cite{Uchida:2022vue}, there exists room to be explored further by numerical simulations.
For example, the validity of the reconnection time scales with shear viscosity is under debate \cite{zhou2022scaling, 2024arXiv240108569B}.
No numerical simulations have been performed for some of the regimes.
The late-time behavior of the scaling regimes has not been addressed numerically.
To wholly establish the validity of our formulae, comprehensive numerical studies are desired.

\acknowledgments
FU acknowledges Masahiro Takada, Yasushi Suto, Masahiro Kawasaki, Shigeki Matsumoto, and Aya Bamba for their helpful comments on his work.
KK thanks Axel Brandenburg for useful comments.
The work of FU was supported by JSPS KAKENHI Grant No.~23KJ0642 and the Forefront Physics and Mathematics Program to Drive Transformation (FoPM).
This work of MF was supported by the Collaborative Research Center SFB1258 and by the Deutsche Forschungsgemeinschaft (DFG, German Research Foundation) under Germany's Excellence Strategy - EXC-2094 - 390783311.
KK was supported by the National Natural Science Foundation of China (NSFC) under Grant No. 12347103 and the JSPS KAKENHI Grant-in-Aid for Challenging Research (Exploratory) JP23K17687.
JY was supported by JSPS KAKENHI Grant Nos.\ 20H05639 and 25K07296.

\appendix
\section{Constraints on primordial and the void magnetic fields\label{appx:statusofconstraints}}
\begin{figure}[ht]
    \begin{minipage}[h]{0.99\hsize}
        \includegraphics[keepaspectratio, width=0.99\textwidth]{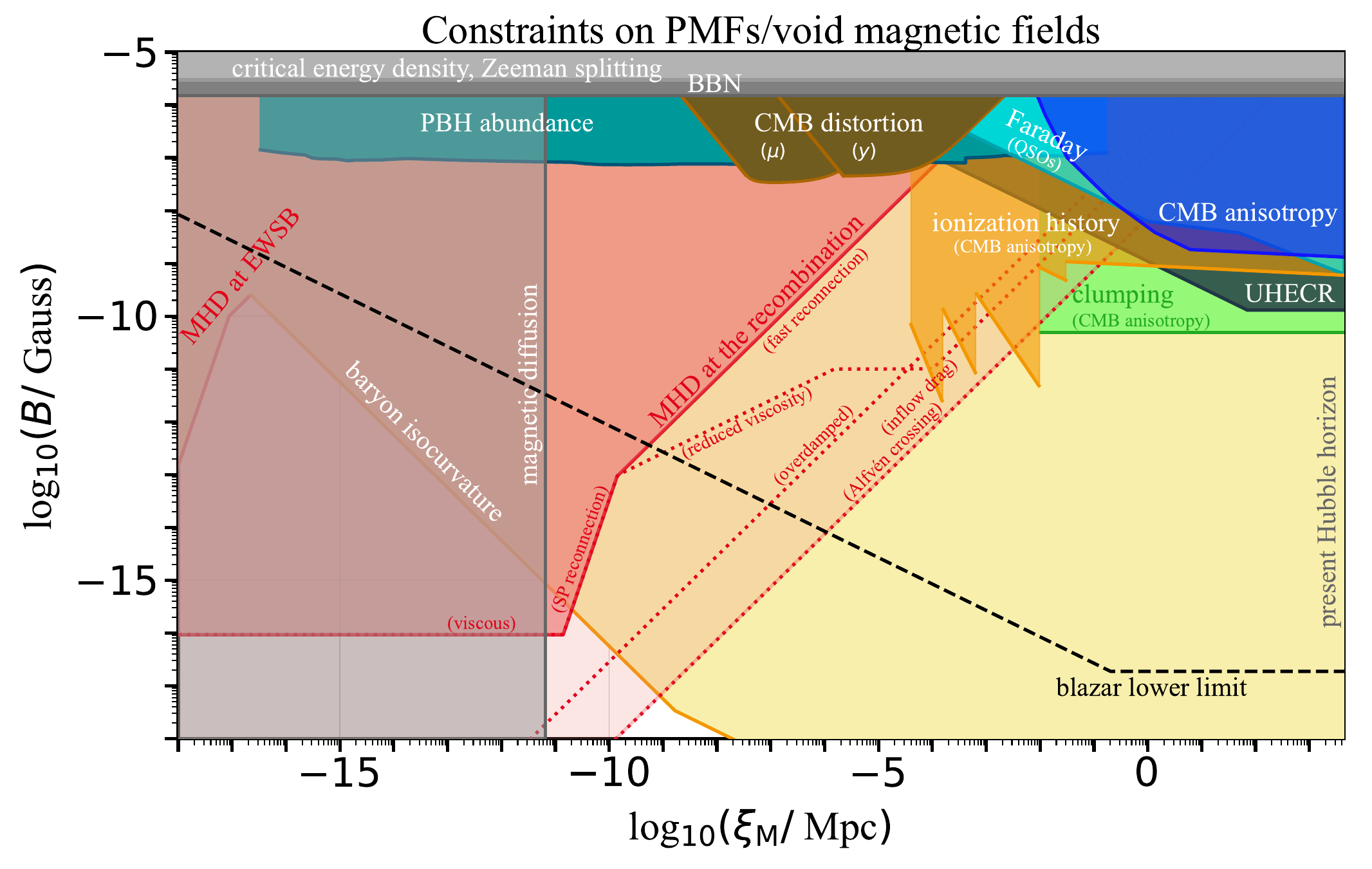}
    \end{minipage}
    \vspace{3mm}
    \caption{\label{fig:PMFLimits}Several limits on primordial or the void magnetic fields in the comoving unit. Each constraint exclude a color-shaded region at a certain epoch. 
    The BBN \cite{2012PhRvD..86f3003K} (gray-shaded), PBH abundance \cite{2024JCAP...12..012K} (teal-shaded), CMB spectral distortions \cite{2024arXiv241103183U} (brown-filled), and the baryon isocurvature perturbations \cite{Kamada:2020bmb} (yellow-filled) place upper bounds on the primordial magnetic field in the early universe.
    The constraint from CMB spectral distortions for non-helical magnetic fields are shown here.
    The MHD considerations in the early universe (our result and Refs.~\cite{2000PhRvL..85..700J,Banerjee+04,Hosking+22}) exclude the red-shaded regions.
    The CMB anisotropy constrains primordial magnetic fields at the recombination epoch (blue-shaded \cite{2016A&A...594A..19P}, orange-shaded \cite{2022MNRAS.517.3916P}, and green-shaded \cite{2019PhRvL.123b1301J}).
    Magnetic dissipation after the recombination exclude short-range magnetic fields \cite{2001PhR...348..163G} (gray-shaded, transparent).
    The Zeeman splitting \cite{2009PhRvD..80l3012N} (light gray-shaded), the ultra high-energy cosmic rays \cite{2023PhRvD.108j3008N} (navy-filled), and the Faraday rotation \cite{1999ApJ...514L..79B} (turquoise-filled) put upper bounds on large-scale magnetic fields in the local universe.
    These upper bounds may be compared with the blazar lower bound \cite{MAGIC:2022piy} (black-dashed line).}
\end{figure}

\noindent
In this appendix, we briefly review the current status of the constraints on primordial and the void magnetic fields.
In Fig.~\ref{fig:PMFLimits}, we plot some of the constraints as representatives, while it should be noted that different constraints are imposed on magnetic fields at different epochs, implying that one should not interpret the plot as a direct comparison of different constraints.

Since there exist numerous discussions on constraining primordial magnetic fields in the literature, we classify the constraints into several classes, according to how the evolution of the primordial magnetic field affects them, and introduce a few representative ones for each class.
For detailed discussion on each constraint, see reviews {\it e.g.}, \cite{2001PhR...348..163G,2002RvMP...74..775W,Durrer+13,2021Galax...9..109V,2021Univ....7..223A}.

The first class of constraints are those concerning primordial magnetic fields in the early universe.
Strong primordial magnetic fields $\gtrsim 10^{-9}\,{\rm G}$ at recombination on large scales can affect it in several ways \cite{Durrer+13} and are excluded by the observed CMB anisotropy (blue-shaded \cite{2016A&A...594A..19P}, orange-shaded \cite{2022MNRAS.517.3916P}, and green-shaded \cite{2019PhRvL.123b1301J} regions in Fig.~\ref{fig:PMFLimits} for different mechanisms. The constraints also depend on the spectral tilt of the magnetic energy spectrum. To let the plot be a conservative exclusion, we have plotted the envelope for different cases on the magnetic energy spectral tilt, $0.05\leq n_B+3\leq5$.);
a substantial magnetic energy relative to the total energy density at the Big-Bang Nucleosynthesis (BBN) (light gray-shaded) is excluded mainly because it enhances the expansion rate of the universe around that epoch and alters the primordial light element abundances \cite{1969Natur.223..938G,1970ApJ...160..451M,1996PhRvD..54.7207K, 2001PhR...348..163G}. 
A recent estimate \cite{2012PhRvD..86f3003K} constrains the magnetic field strength at the BBN epoch to $B\lesssim 1.5\times10^{-6}\,{\rm G}$ (gray-shaded);
the anisotropic stress of the primordial magnetic fields should be small not to generate large super-horizon curvature perturbations that result in too much primordial black hole abundance \cite{2020JCAP...05..039S}, leading to $B\lesssim 1\times10^{-7}\,{\rm G}$ \cite{2024JCAP...12..012K} in the radiation-dominated era (teal-shaded); spatial fluctuations of ${\rm U}(1)_Y$ magnetic helicity should be small because they generate baryon isocurvature perturbations that would affect the BBN \cite{1998PhRvL..80...22G}, excluding the yellow-shaded region even for non-helical primordial magnetic field \cite{Kamada:2020bmb} at the electroweak symmetry breaking.

The second class of the constraints are those concerning the magnetic energy decay.
This class of constraints crucially rely on the understanding of the magneto-hydrodynamic evolution of the primordial magnetic fields.
Strong magnetic fields should have decayed earlier than the recombination epoch (red-shaded regions. Our result is shown in the solid edge, and other possibilities \cite{2000PhRvL..85..700J,Banerjee+04,Hosking+22} in the dotted edges);
short-range magnetic fields should have dissipated by today because of the finite electric resistivity in the late universe \cite{2001PhR...348..163G} (gray-filled, transparent);
magnetic fields should not inject a lot of energy into and distort the black-body spectrum of the CMB photons \cite{2000PhRvL..85..700J} (brown-filled \cite{2024arXiv241103183U} based on the up-to-date understanding of the magneto-hydrodynamic evolution).

The third class is the observational ones in the present universe, which often involve cosmological hydrodynamic simulations.
The observed energy spectra of ${\rm TeV}$ blazars suggest lower bounds of the void magnetic field in the local universe \cite{1989ApJ...339..629H,1995Natur.374..430P,NeronovVovk10, 2010MNRAS.406L..70T, 2010ApJ...722L..39A, 2011MNRAS.414.3566T, 2011ApJ...727L...4D, 2011APh....35..135E, 2011A&A...529A.144T, 2011ApJ...733L..21D, 2012ApJ...747L..14V, Takahashi:2013lba, 2015ApJ...814...20F, Ackermann+18, MAGIC:2022piy, 2021Univ....7..223A}, conservatively $B\gtrsim10^{-17}\,{\rm G}$ \cite{MAGIC:2022piy} (black-dashed line); the bright and nearby gamma-ray burst, GRB 221009A, enables us to place similar lower bounds using temporal information as well \cite{2023ApJ...955L..10H,2024MNRAS.527L..95D}; the Zeeman splitting of quasar absorption lines, which is interpreted as the consequence of magnetic fields in galaxies, places an upper bound on the strength of intergalactic magnetic fields \cite{2009PhRvD..80l3012N} (light gray-shaded);
ultra-high-energy cosmic ray (UHECR) events arriving from their suggested source direction place upper limits on the strength of the void magnetic fields \cite{2023PhRvD.108j3008N} (navy-shaded); 
the Faraday rotation measure places upper limits \cite{1999ApJ...514L..79B} (turquoise-shaded);
cosmological magneto-hydrodynamical simulations enable statistical inference of the magnetogenesis scenarios, {\it i.e.}, primordial or astrophysical, based on these observational data \cite{2017CQGra..34w4001V,2020MNRAS.495.2607O,2022MNRAS.515..256P}.

Note that we have by no means exhausted all the relevant constraints here.
A number of different strategies to place upper bounds on the primordial and the void magnetic fields can be found in the literature.
For instance, constraints on the filament magnetic fields, {\it e.g.}, Refs.~\cite{2021MNRAS.505.4178V,2022MNRAS.512..945C,2023MNRAS.518.2273C}, can also be relevant to primordial magnetic fields.
 
\section{Dissipation coefficients}
\label{appx:DissipationCoefficients}
\begin{figure}[ht]
    \begin{tabular}{ll}
        \begin{minipage}[h]{0.48\hsize}
            \includegraphics[keepaspectratio, width=0.85\textwidth]{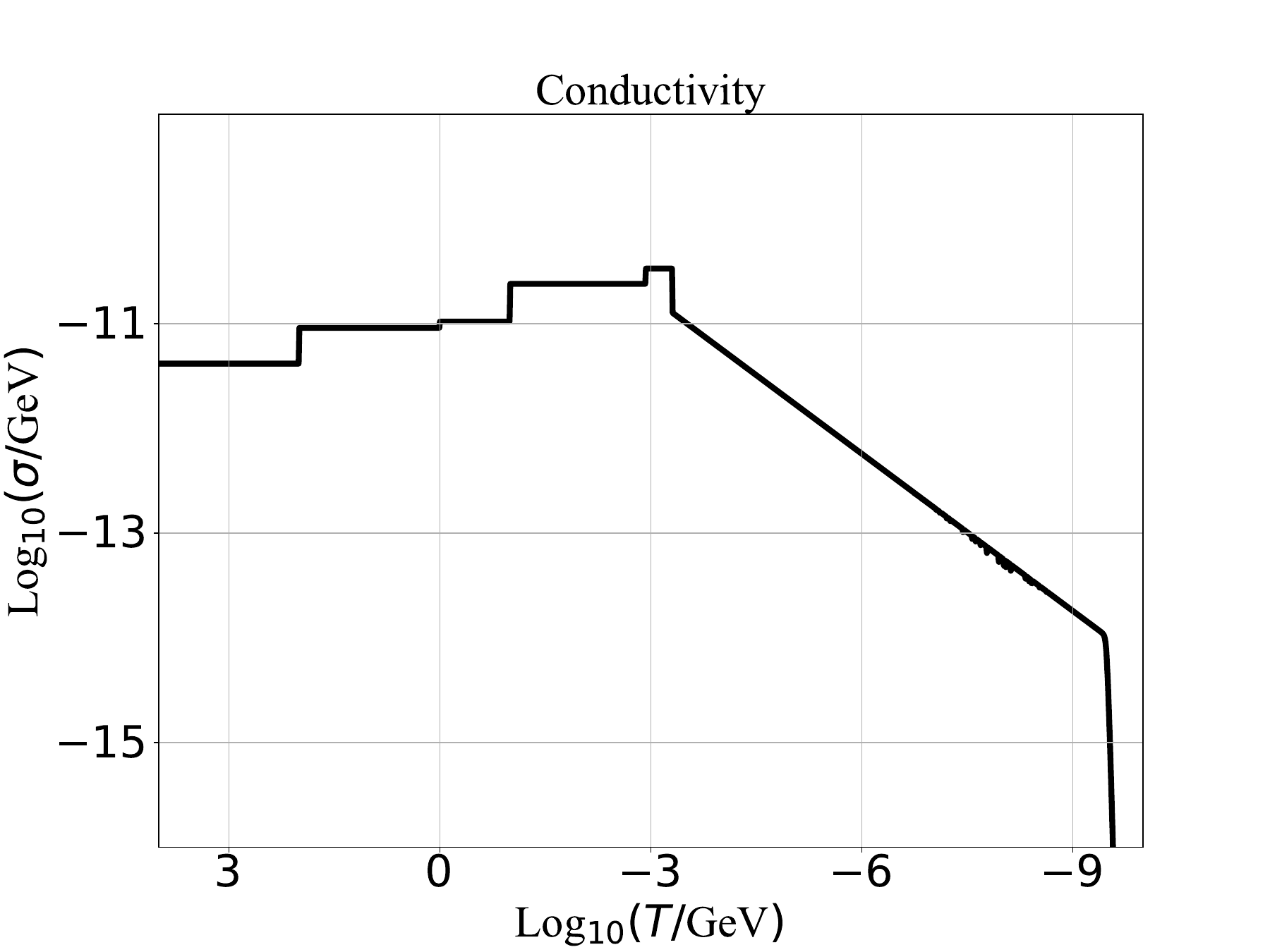}
        \end{minipage}&
        \begin{minipage}[h]{0.48\hsize}
            \includegraphics[keepaspectratio, width=0.85\textwidth]{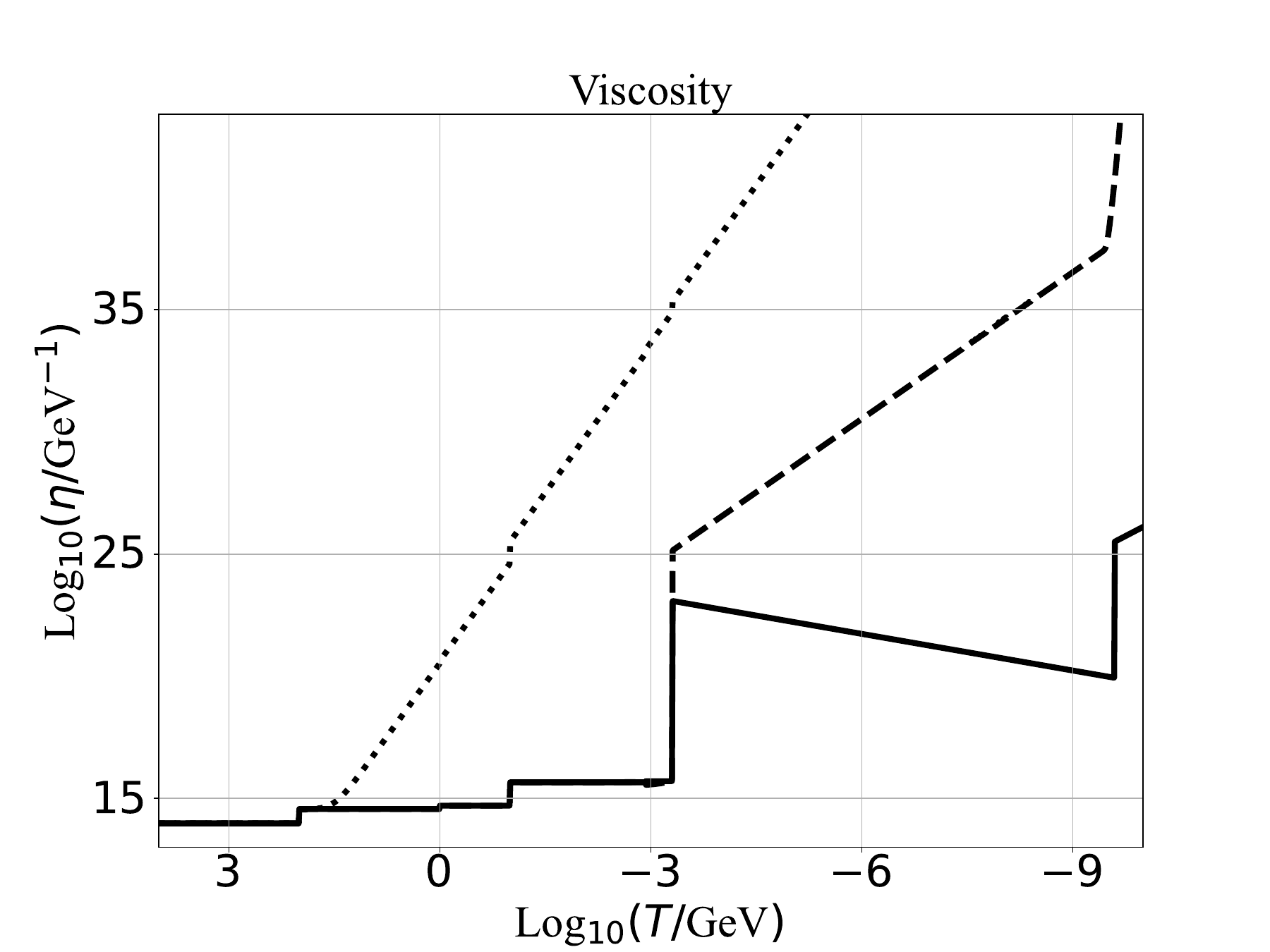}
        \end{minipage}\\
        \begin{minipage}[h]{0.48\hsize}
            \includegraphics[keepaspectratio, width=0.85\textwidth]{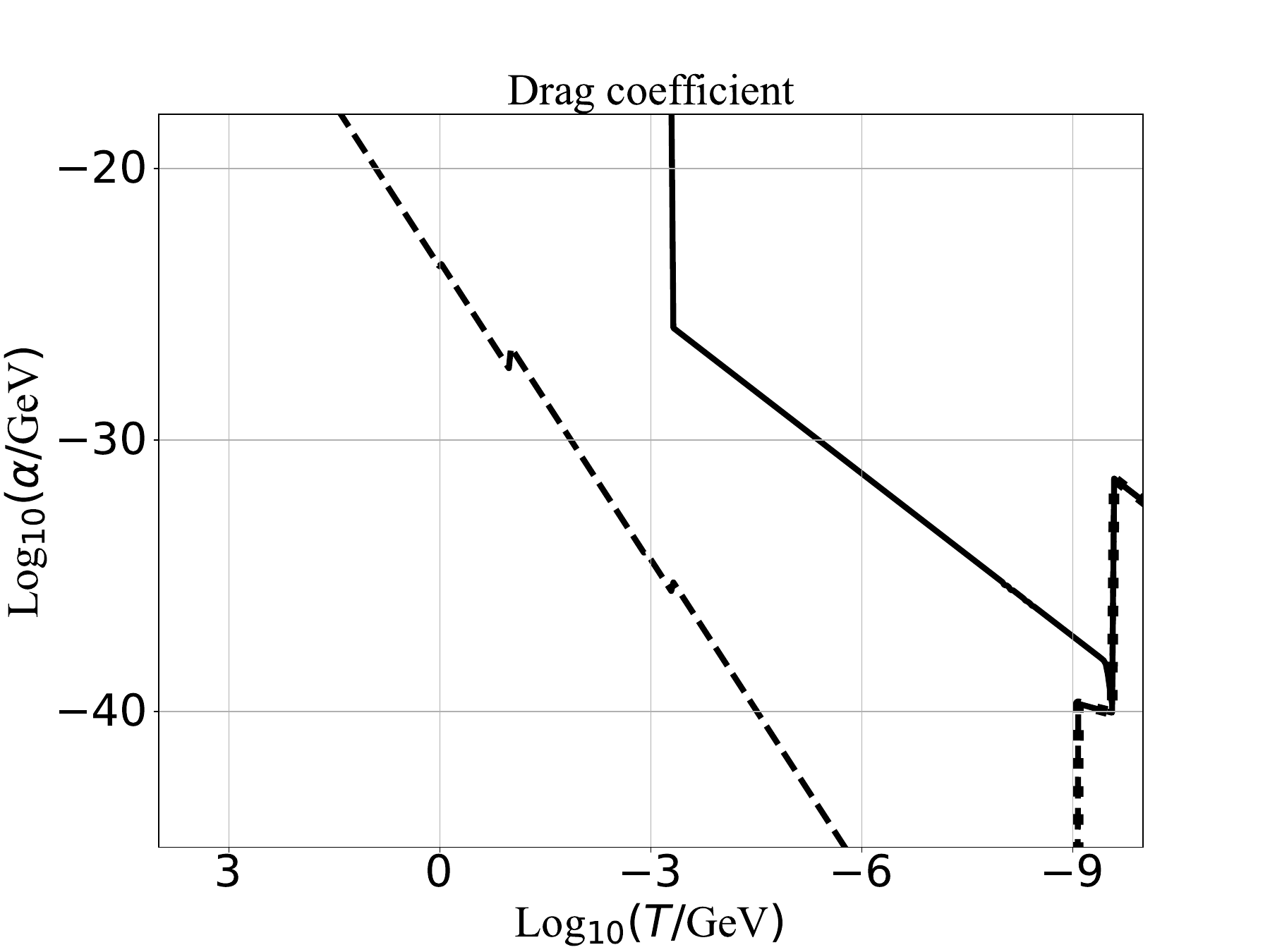}
        \end{minipage}&
        \begin{minipage}[h]{0.48\hsize}
            \includegraphics[keepaspectratio, width=0.85\textwidth]{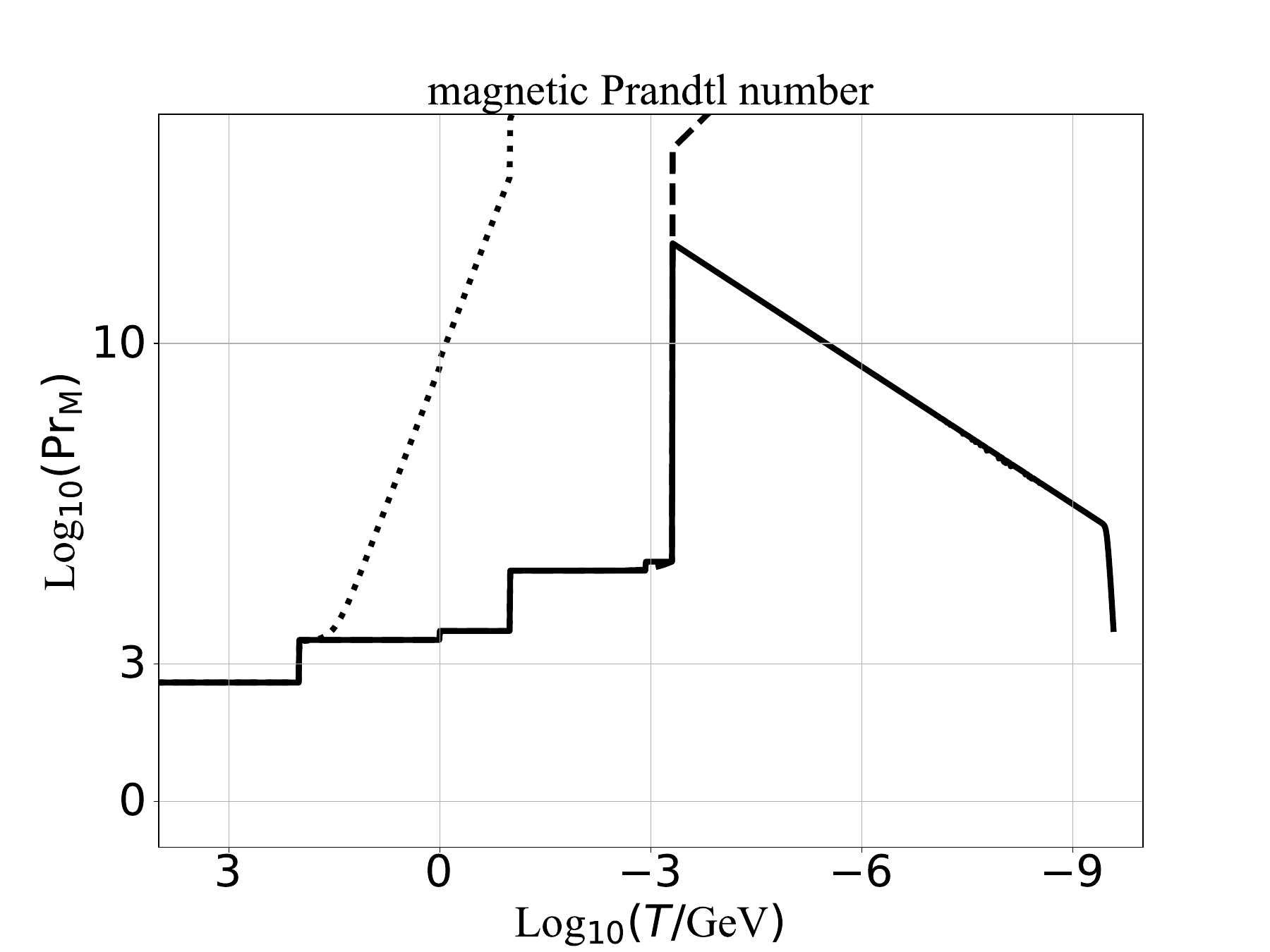}
        \end{minipage}
    \end{tabular}
    \vspace{8mm}
    \caption{\label{fig:DissipatioCoefficients}Values of the dissipation coefficients, namely the electric conductivity $\sigma$ (top left), the shear viscosity $\eta$ (top right), and the drag coefficient $\alpha$ (bottom left). In addition, the magnetic Prandtl number ${\rm Pr}_{\rm M}$ is plotted (bottom right). For each plot, the solid line is for the case when both neutrinos and photons decouple from the fluid, the dashed line is for the cases when only neutrinos decouple, and the dotted line is for the cases when both neutrinos and photons are still coupled with the fluid.}
\end{figure}

\begin{figure}[t]
    \begin{tabular}{ll}
        \begin{minipage}[h]{0.50\hsize}\vspace{0mm}
            \includegraphics[keepaspectratio, width=0.85\textwidth]{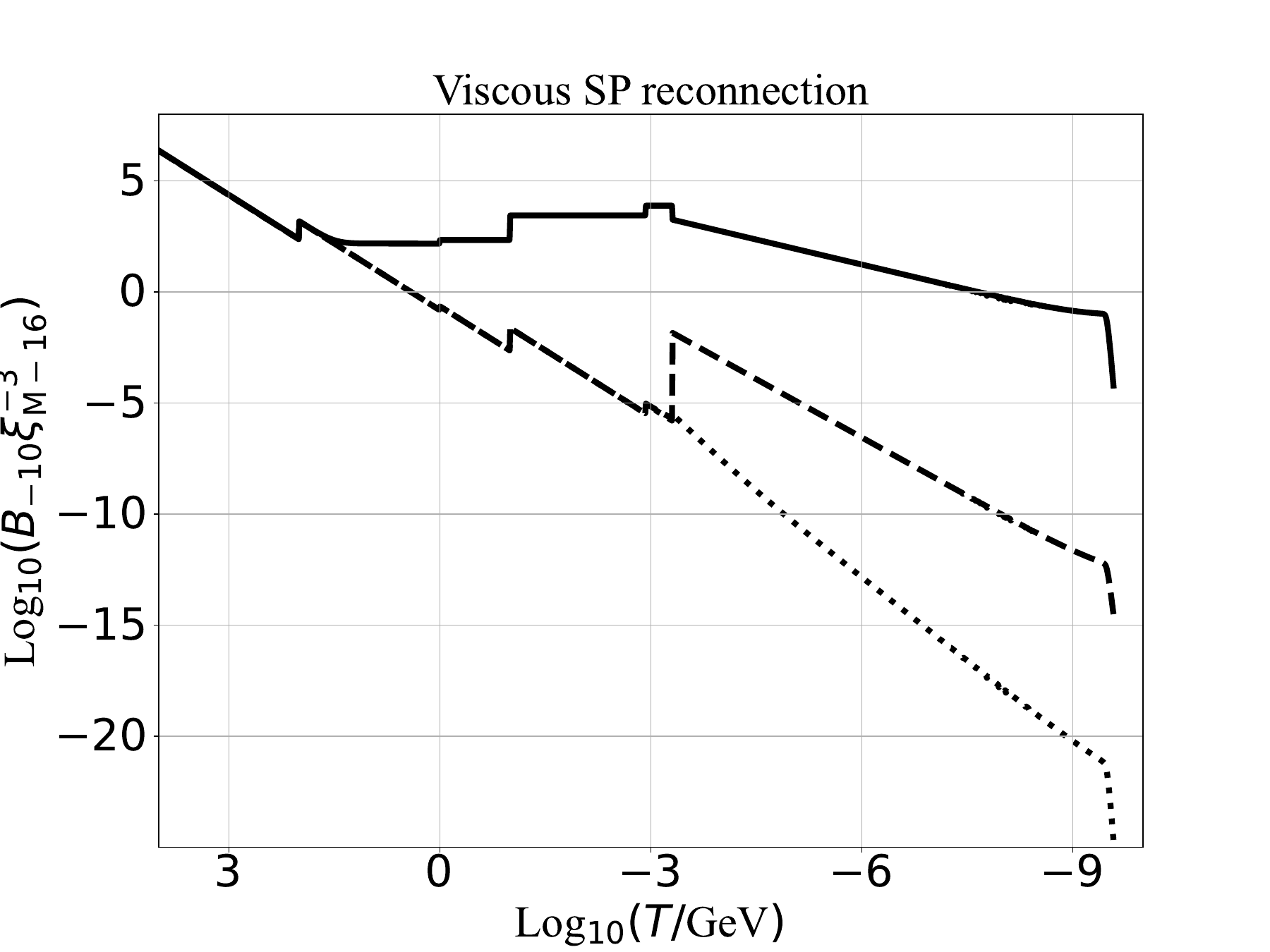}
        \end{minipage}&
        \begin{minipage}[h]{0.50\hsize}\vspace{0mm}
            \includegraphics[keepaspectratio, width=0.85\textwidth]{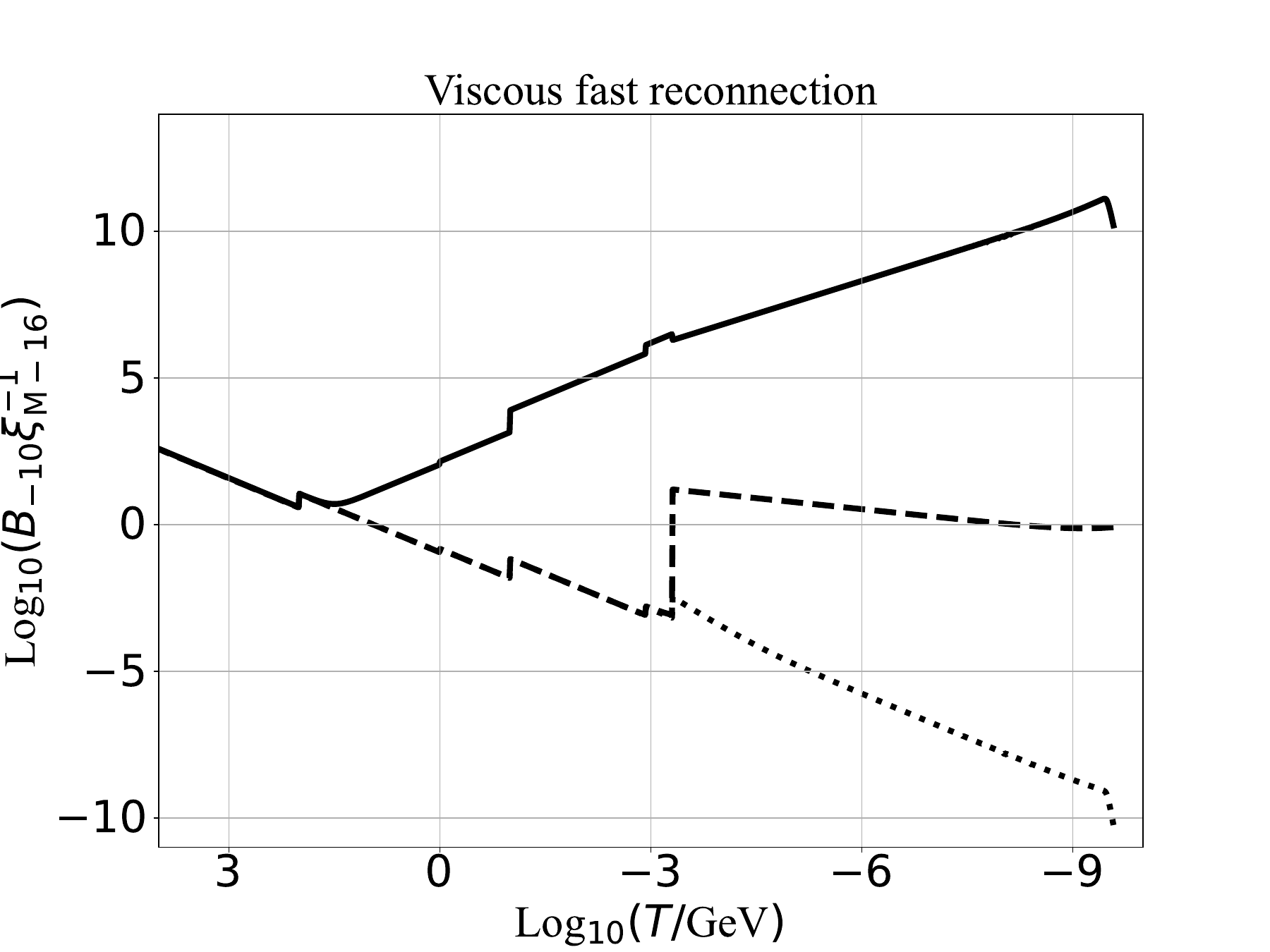}
        \end{minipage}\\
        \begin{minipage}[h]{0.50\hsize}
            \includegraphics[keepaspectratio, width=0.85\textwidth]{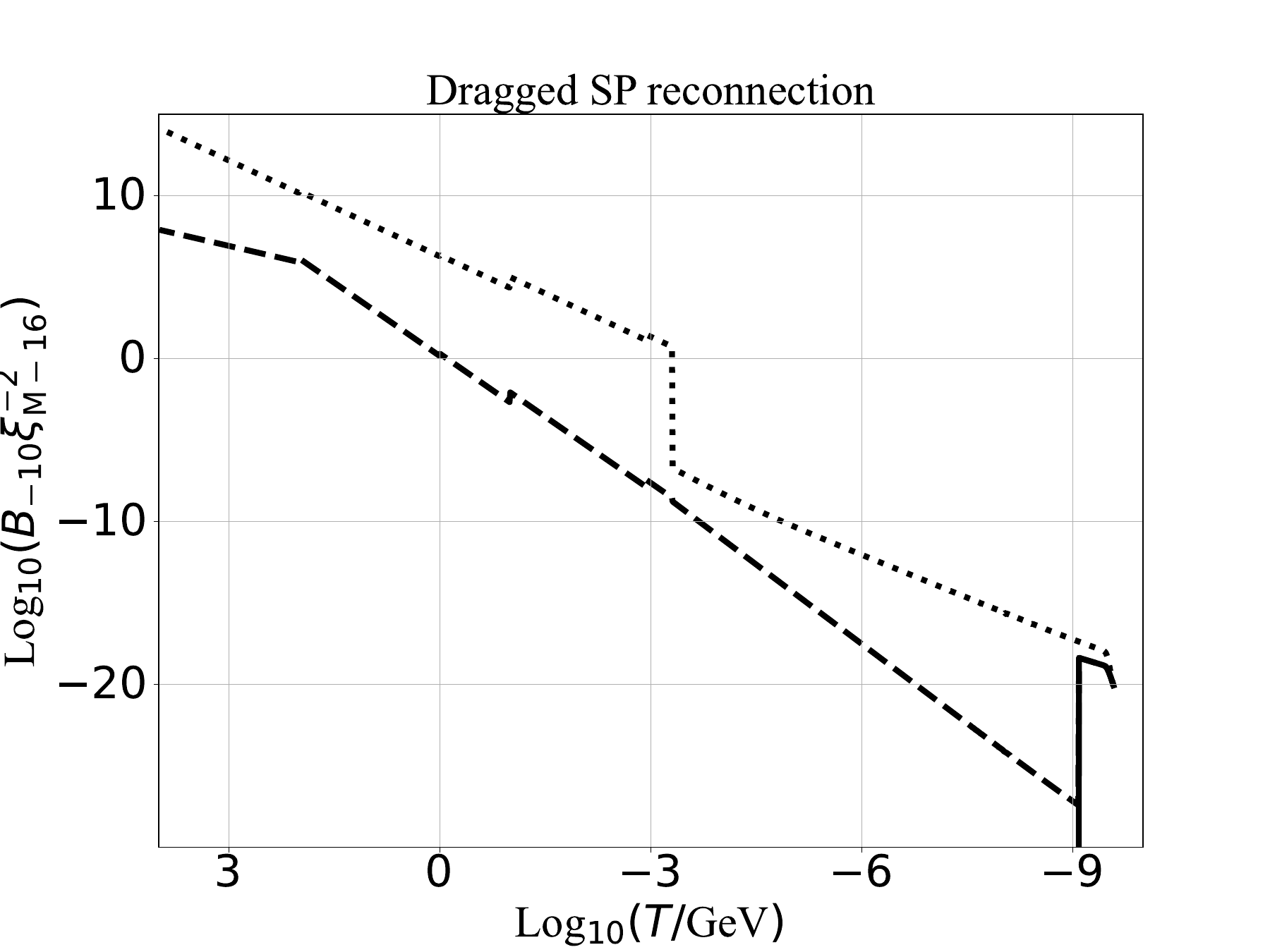}
        \end{minipage}&
        \begin{minipage}[h]{0.50\hsize}
            \includegraphics[keepaspectratio, width=0.85\textwidth]{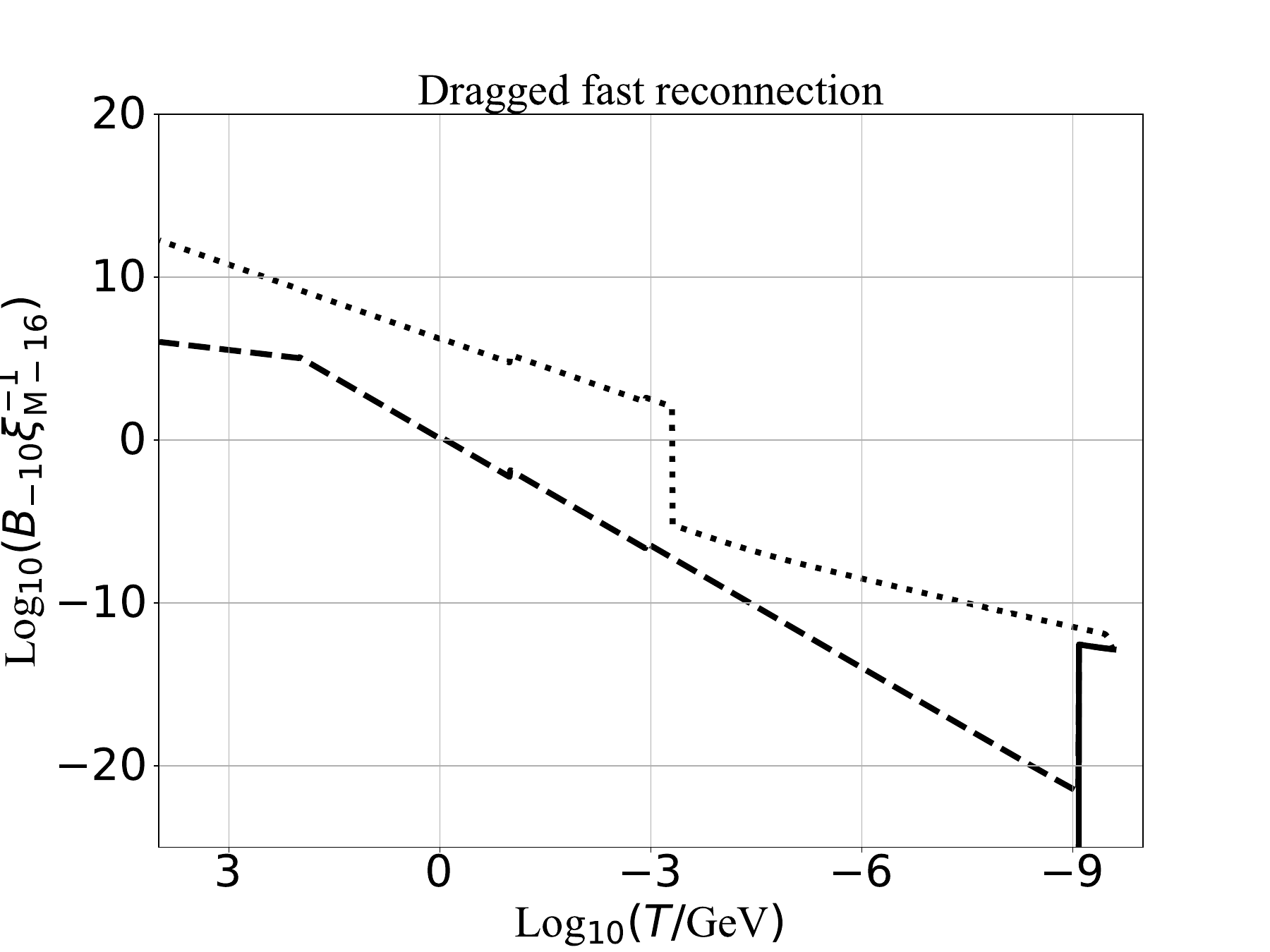}
        \end{minipage}\\
        \begin{minipage}[h]{0.50\hsize}
            \includegraphics[keepaspectratio, width=0.85\textwidth]{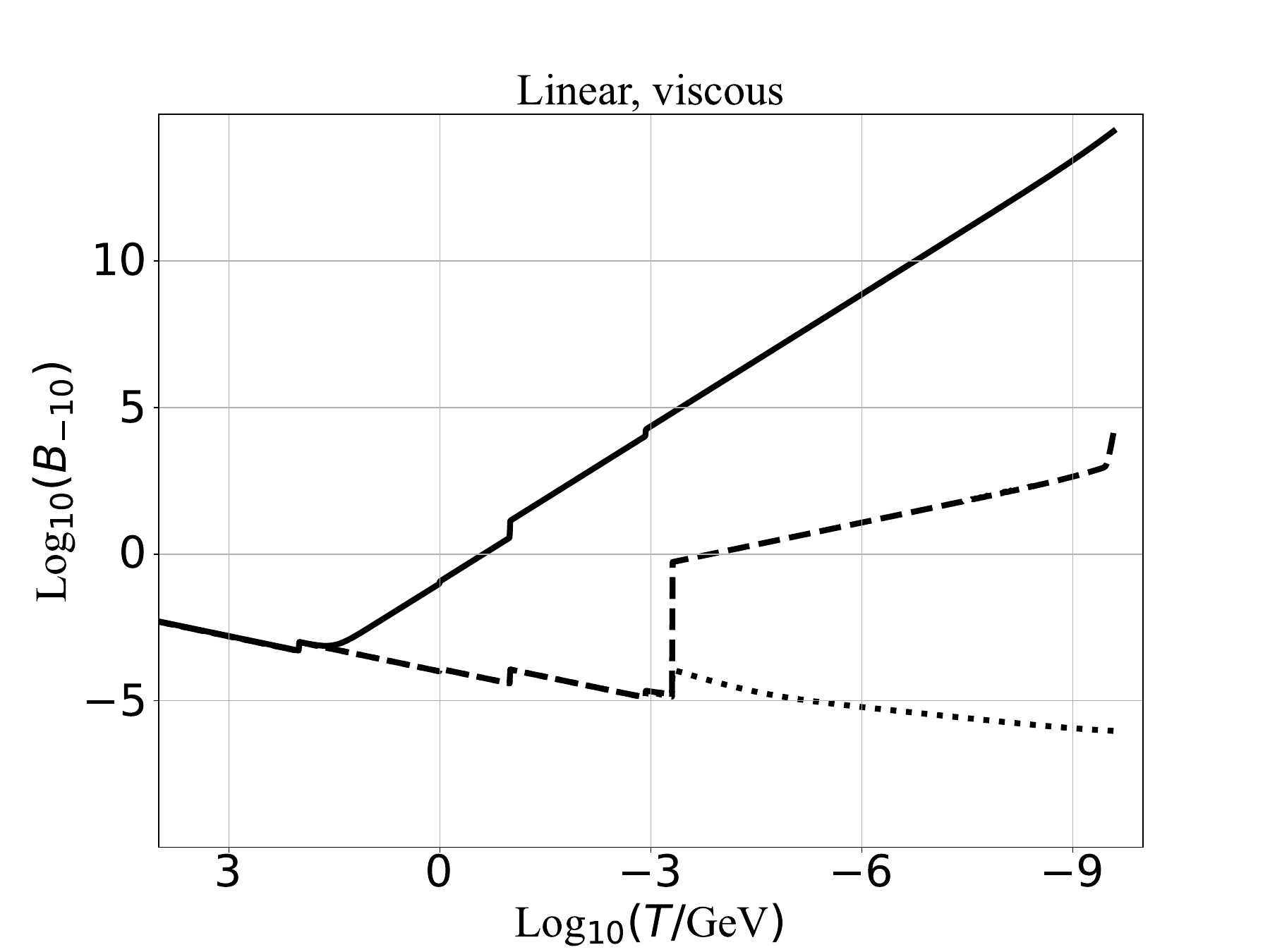}
        \end{minipage}&
        \begin{minipage}[h]{0.50\hsize}
            \includegraphics[keepaspectratio, width=0.85\textwidth]{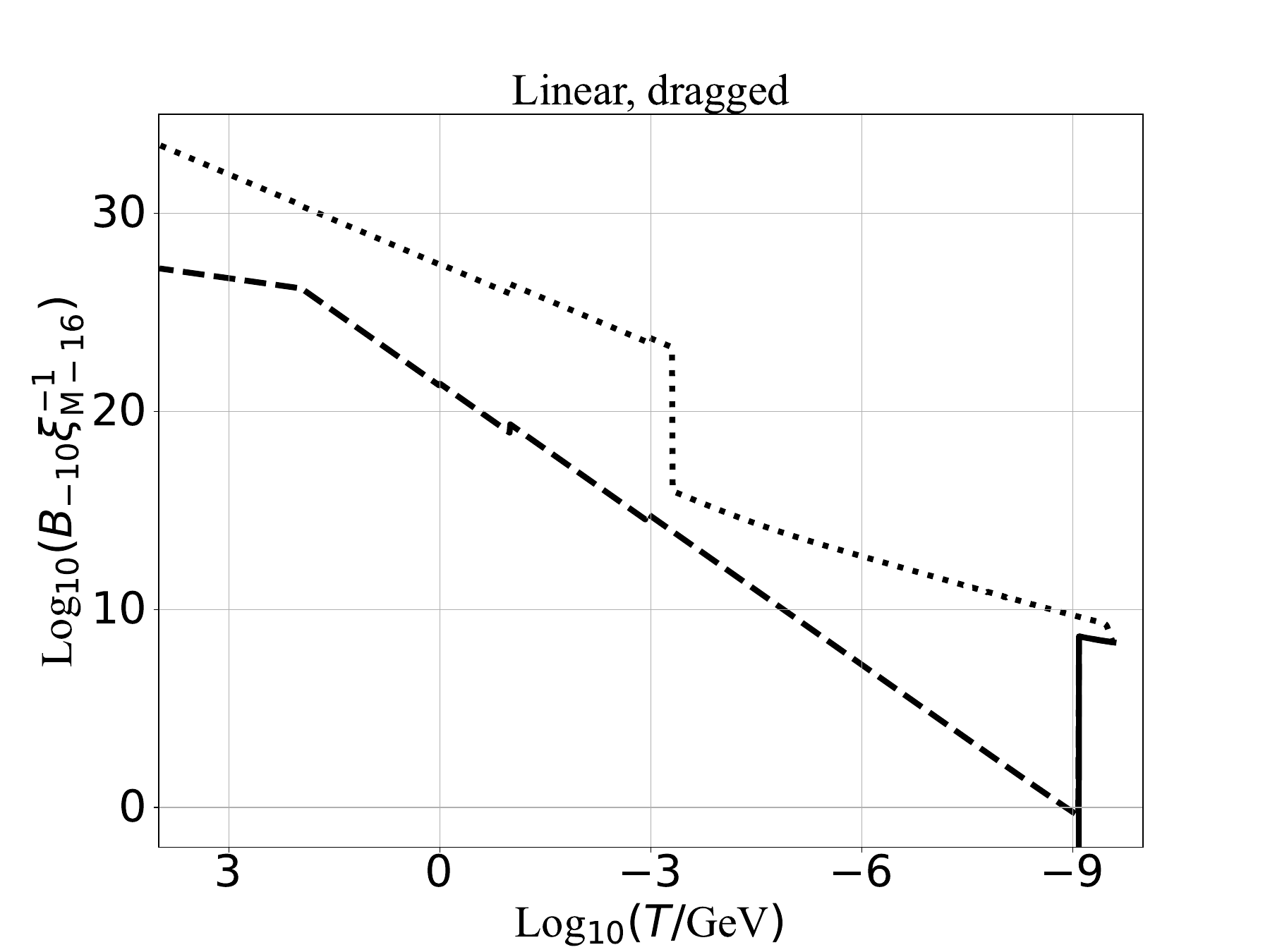}
        \end{minipage}
    \end{tabular}
    \vspace{15mm}
    \caption{\label{fig:TimeScaleLineCoeffs}
    The temperature dependence of products of $B$ and $\xi_{\rm M}$ that is determined by Eq.~\eqref{eq:TimeScaleCondition_general}. $B_{-10}:=B/10^{-10}\,{\rm G}$ and $\xi_{{\rm M}-16}:=\xi_{\rm M}/10^{-16}\,{\rm Mpc}$. $B\xi_{\rm M}^{-3}$ for the viscous Sweet--Parker regime (top left), $B\xi_{\rm M}^{-1}$ for the viscous fast reconnection regime (top right), $B\xi_{\rm M}^{-2}$ for the dragged Sweet--Parker regime (middle left), $B\xi_{\rm M}^{-1}$ for the dragged fast reconnection regime (middle right), $B$ for the linear and viscous regime (bottom left), and $B\xi_{\rm M}^{-1}$ for the linear and dragged regime (bottom right) are plotted. For each plot, the solid line is for the case when both neutrinos and photons decouple from the fluid, the dashed line is for the cases when only neutrinos decouple, and the dotted line is for the cases when both neutrinos and photons are still coupled with the fluid.}
\end{figure}

In this appendix, we list up the expressions of dissipation coefficients employed in Sec.~\ref{sec:application}.
Values of these coefficients are shown in Fig.~\ref{fig:DissipatioCoefficients}.
In particular, in the bottom-right panel of Fig~\ref{fig:DissipatioCoefficients}, we plot the magnetic Prandtl number, ${\rm Pr}_{\rm M}:=\sigma\eta$, as a function of temperature, which is
always larger than unity before recombination.
In addition, we plot in panels of Fig.~\ref{fig:TimeScaleLineCoeffs} the regime-dependent temperature
dependence of products of $B$ and $\xi_{\rm M}$, which is determined by Eq.~\eqref{eq:TimeScaleCondition_general}, or Eqs.~\eqref{eq:TimeScaleCondition_MDNL_MHVSP}, \eqref{eq:MDNL_MHVfast_Decaytime}, \eqref{eq:MDNL_MHDSP_Decaytime}, \eqref{eq:MDNL_MHDfast_DecayTime}, \eqref{eq:Decaytime_MDL_V}, and \eqref{eq:Decaytime_MDL_D}.\\

\noindent
{\it Electric conductivity.} The electric conductivity in the early universe before electrons become massive is approximated as \cite{arnold2000transport}
\begin{align}
    \sigma_{Y}&=\dfrac{6^4\zeta(3)^2\pi^{-3}}{\frac{\pi^2}{8}+\frac{20}{3}+\frac{2}{3}}\dfrac{a(T)T}{g'^2\ln g'^{-1}},\hspace{16mm}
    T> T_{\mathrm{EW}}, \label{eq:Conductivity1}\\[7pt]
    \sigma&=\dfrac{12^4\zeta(3)^2\pi^{-3}N_{\mathrm{leptons}}}{3\pi^2+32N_{\mathrm{species}}}\dfrac{a(T)T}{e^2\ln e^{-1}},\quad
    m_{e}< T< T_{\mathrm{EW}},
    \label{eq:Conductivity2}
\end{align}
where $e=0.30$ and $g'=0.36$ are the coupling constant of the ${\rm U}(1)_{\rm em}$ and ${\rm U}(1)_Y$ gauge fields, respectively.
$m_e= 0.5\,\mathrm{MeV}$ is the electron mass, and $T_{\mathrm{EW}}\sim 0.1\,\mathrm{TeV}$ is the electroweak scale.
$N_{\mathrm{leptons}}$ and $N_{\mathrm{species}}$ count the number of relevant charge careers, which are summarized in Table \ref{tb:Parameters}.
At lower temperatures, nonrelativistic electron-ion scattering is modeled by the Spitzer theory \cite{1962pfig.book.....S}
\begin{equation}
    \sigma
    =\dfrac{\kappa_\sigma e^2 a(T) n_{e,\mathrm{p}}(T)}{\nu_{ei,\mathrm{p}}(T) m_e}
    =\dfrac{\kappa_\sigma a(T)T^{\frac{3}{2}}}{e^2 m_e^{\frac{1}{2}}\ln\Lambda_{ei}}
    ,\quad\quad
    T< m_e, \label{eq:Conductivity3}
\end{equation}
where $n_{e,\mathrm{p}}$ is the physical electron number density, $\nu_{ei}$ is the electron-ion collision frequency, and $\ln\Lambda_{ei}\simeq20$ is the Coulomb logarithm.
The coefficient $\kappa_\sigma$ depends on approximations, and we adopt $\kappa_\sigma\simeq1\times10^2$ here \cite{2024arXiv241103183U}.
$\sigma$ as a function of temperature is plotted in the top-left panel of Fig~\ref{fig:DissipatioCoefficients}.\\

\noindent
{\it Shear viscosity.}
The shear viscosity in the early universe above $T>m_e$ is estimated in Ref.~\cite{arnold2000transport}, which neglected the contribution from neutrinos.
When the mean free path of neutrinos is shorter than the characteristic scales of the system, they contribute to the shear viscosity as well.
In that case, the dimensionful Fermi constant $G_F=1.17\times10^{-5}\,\mathrm{GeV}^{-2}$ comes into the expression, introducing non-trivial $T$ dependence \cite{Weinberg:1971mx, heckler1993neutrino}.
When neutrinos contribute to the shear viscosity, their contribution becomes dominant. 
\begin{align}
    \eta&=\dfrac{1}{5}\dfrac{g_{\nu}}{g_{\mathrm{fluid}}}l_{\nu\mathrm{mfp}},\hspace{52mm}\text{with neutrinos included}\label{eq:ShearViscosity2},\\[5pt]
    \eta
    &=\dfrac{\frac{45}{2\pi^2}\left(\frac{5}{2}\right)^3\zeta(5)^2\left(\frac{12}{\pi}\right)^5\cdot\frac{3}{2}}{9\pi^2+224\left(5+\frac{1}{2}\right)}\dfrac{1}{g'^4\ln g'^{-1}}\dfrac{1}{g_{\mathrm{fluid}}a(T)T},\notag\\
    &\hspace{69mm}T> T_{\mathrm{EW}}\quad\text{without neutrinos}, \label{eq:ShearViscosity1}\\[3pt]
    \eta
    &=\dfrac{\frac{45}{2\pi^2}\left(\frac{5}{2}\right)^3\zeta(5)^2\left(\frac{12}{\pi}\right)^5N_{\mathrm{leptons}}}{9\pi^2+224N_{\mathrm{species}}}\dfrac{1}{e^4\ln e^{-1}}\dfrac{1}{g_{\mathrm{fluid}}a(T)T},\notag\\
    &\hspace{58mm}
    m_{e}< T< T_{\text{EW}}\quad\text{without neutrinos}
    \label{eq:ShearViscosity3},
\end{align}
where $g_\nu$ and $g_\mathrm{fluid}$ are the number of relativistic freedom of neutrinos and fluid constituent particles, respectively, and the neutrino mean free path is \cite{heckler1993neutrino}
\begin{align}
    l_{\nu\mathrm{mfp}}&=\dfrac{1}{a(T)G_{\mathrm{F}}^2T^5}.
\end{align}
After electrons become massive, non-relativistic ion-ion scattering \cite{Hosking+22} and scattering between neutral hydrogens \cite{Banerjee+04} are relevant before and after recombination, respectively.
Similarly to neutrinos, photons contribute to the shear viscosity when their mean free path is shorter than the characteristic scales of the system.
\begin{align}
    \eta 
    &=\dfrac{1}{5}\dfrac{g_{\gamma}}{g_{\mathrm{fluid}}}l_{\gamma\mathrm{mfp}},
    \hspace{52.5mm}\text{with photons included},\label{eq:ShearViscosity4}\\[5pt]
    \eta
    &=\dfrac{\kappa_\eta T^{\frac{5}{2}}}{a(T)e^4n_{i,\mathrm{p}}(T)m_p^{\frac{1}{2}}\ln\Lambda_{ii}},
    \hspace{12.5mm}T_{\mathrm{rec}}< T< m_e \quad\text{without photons},\label{eq:ShearViscosity5}\\
    \eta
    &=\dfrac{T^{\frac{1}{2}}}{\pi a_{\mathrm{B}}^2a(T)n_{H,\mathrm{p}}(T) m_p^{\frac{1}{2}}},\hspace{28mm}
    T< T_{\mathrm{rec}}\quad\text{without photons},\label{eq:ShearViscosity6}
\end{align}
where $g_\gamma=2$ is the number of relativistic degrees of freedom of photon, and the mean free path  is given by 
\begin{align}
    l_{\gamma\mathrm{mfp}}
    &=\dfrac{a(T)^2}{\sigma_{\mathrm{T}}n_e(T)}.
\end{align}
In the above expressions, $m_p\cong 1\,\mathrm{GeV}$ is the proton mass, $\ln\Lambda_{ii}\simeq20$ is the Coulomb logarithm, $a_{\mathrm{B}}:=4\pi/(e^2m_e)$ is the Bohr radius,
and $\sigma_{\mathrm{T}}:=(8\pi/3)(e^2/4\pi m_e)^2$ is the Thomson cross section.
The coefficient $\kappa_\eta$ depends on approximations, and we adopt $\kappa_\eta\simeq30$ here \cite{2024arXiv241103183U}.
$\eta$ as a function of temperature is plotted in the top-right panel of Fig~\ref{fig:DissipatioCoefficients}.\\

\noindent
{\it Drag force.}
When the mean free paths of particles, {\it i.e.,} photons and neutrinos, are larger than the characteristic length scales of the system,
the particles act as a homogeneous background on which the fluid is dissipated via drag (friction) term, $-\alpha \boldsymbol{v}$. 
The drag coefficients due to neutrinos and photons are given by \cite{Banerjee+04}
\begin{align}
    \alpha_{\nu}&=\dfrac{g_{\nu}}{g_{\mathrm{fluid}}-g_{\nu}}l_{\nu\mathrm{mfp}}^{-1},\\[10pt]
    \alpha_{\gamma}&=\dfrac{4}{3}\dfrac{\rho_{\gamma}}{\rho_{\mathrm{b}}}l_{\gamma\mathrm{mfp}}^{-1}, 
\end{align}
where we have subtracted the relativistic degree of freedom of neutrinos, $g_{\nu}= 5.25$, from that for fluid constituent particles. $\rho_\gamma$ and $\rho_\mathrm{b}$ are the photon and baryon densities, respectively.
Also, when the hydrogen atoms decouple from the plasma because of their neutralness, they account for the ambipolar drag \cite{Banerjee+04,1961ApJ...134..270O},
\begin{equation}
    \alpha_{\mathrm{amb}}=2\pi \left(\dfrac{e^2a_{\text{B}}^3}{m_{p}}\right)^{\frac{1}{2}}a(T)n_{H,p}.
\end{equation}
In the matter-dominated era, the Hubble friction should also be taken into account,
\begin{equation}
    \alpha_{\mathrm{Hubble}}:=aH.
\end{equation}
$\alpha$ as a function of temperature is plotted in the bottom-left panel of Fig~\ref{fig:DissipatioCoefficients}.

\begin{table*}
    \begin{center}
    Table of degrees of freedom
    \end{center}
    \hspace{-4mm}
    \begin{tabularx}{1.03\textwidth}{c||c|c|c|c|c|c}\hline
    \hspace{-3mm}$T$ & $T_0-$ & $T_{\mathrm{rec}},T_{\gamma\mathrm{dec}}-$ & $m_e, T_{\nu\mathrm{dec}}-$ & $\begin{matrix}m_{\mu},m_u, m_d,\\ m_s, T_{\mathrm{QCD}}-\end{matrix}$ & $\begin{matrix}m_{\tau}, m_c, m_b\\\hspace{14mm}-\end{matrix}$ & $\begin{matrix}m_t, m_W,\\\hspace{5mm} T_{\mathrm{EW}}-\end{matrix}$ \\ 
    \hspace{-3mm}& $0.1\,\mathrm{meV}-$ & $1\,\mathrm{eV}-$ & $1\,\mathrm{MeV}-$ & $0.1\,\mathrm{GeV}-$ & $1\,\mathrm{GeV}-$ & $0.1\,\mathrm{TeV}-$ \\\hline
    $\begin{matrix}\text{Relativistic}\\ \text{contents of}\\\text{the plasma}\end{matrix}$ & $-$ & $\gamma$ & $\begin{matrix}\cdots,\\e,\nu\end{matrix}$ & $\begin{matrix}\cdots,\\ \mu, u, d, s, g\end{matrix}$ & $\begin{matrix}\cdots,\\ \tau, c, b\end{matrix}$ & $\begin{matrix}\cdots,\\ t, H, W, Z\end{matrix}$ \\ \hline\hline
    \hspace{-3mm}$\begin{matrix}g\\g_s\\g_{\mathrm{fluid}}\end{matrix}$ & $\begin{matrix}3.36\\3.91\\0\end{matrix}$& $\begin{matrix}3.36\\3.91\\2\end{matrix}$ & $10.75$ & $61.75$ &$86.25$&$106.75$\\\hline
    \hspace{-3mm}$\begin{matrix}N_{\mathrm{leptons}}\\N_{\mathrm{species}}\end{matrix}$ &
    $\begin{matrix}0\\0\end{matrix}$ &
    $\begin{matrix}0\\0\end{matrix}$ &
    $\begin{matrix}1\\1\end{matrix}$ &
    $\begin{matrix}2\\4\end{matrix}$ &
    $\begin{matrix}3\\20/3\end{matrix}$ &
    $\begin{matrix}\left(3/2\right)\\\left(5+1/2\right)\end{matrix}$\\\hline
  \end{tabularx}
  \begin{center}
  \caption{The values of $N_{\mathrm{leptons}}, N_{\mathrm{species}}$ from Ref.~\cite{arnold2000transport} as well as the values of $g, g_s, g_{\mathrm{fluid}}$, as functions of the radiation temperature $T$. Both $N_{\mathrm{leptons}}$ and $N_{\mathrm{species}}$ count the number of fermions but are weighted by their lepton charges and electromagnetic charges. The symmetric phase belongs to a different regime but the substitution of the values in the table into Eq.~\eqref{eq:Conductivity2} with replacing the coupling coefficient $e\to g'$ reproduces Eq.~\eqref{eq:Conductivity1} \cite{arnold2000transport}.}
  \label{tb:Parameters}
  \end{center}
\end{table*}

\section{Hosking integral conservation}
\label{sec:Hosking_integral}
In this appendix, we give an explanation on the Hosking integral, whose conservation is essential to determine the evolution of the non-helical magnetic fields.
\subsection{Magnetic helicity}
\label{sec:MagneticHelicity}
To avoid confusion, let us start by defining several quantities that are almost the same except essential differences in their properties.
They are not always defined separately in the literature, but we have to introduce them for clarity and later purpose.
Magnetic helicity density $h$ is a local quantity defined as
\begin{align}
    h({\bm x}):=\bm{A}({\bm x})\cdot\bm{B}({\bm x}),
    \label{eq:LocalMagneticHelicityDensity_Def}
\end{align}
where $\bm{A}$ is the vector potential.
However, this quantity is gauge-dependent, and its local value at a given spatial point has no physical meaning by itself.
Instead, one can define ${\rm U}(1)$-gauge-invariant magnetic helicity $H_V$ for a volume $V$ such that the normal component of $\bm{B}$ is zero on its boundary as
\begin{align}
    H_V:=\int_V d^3 r h({\bm r}),
    \label{eq:Def_MagneticHelicity}
\end{align}
and furthermore one may define gauge-invariant magnetic helicity density $h_V$,
\begin{align}
    h_V:=\dfrac{H_V}{V}.
    \label{eq:SemiLocalMagneticHelicityDensity_Def}
\end{align}
If we take small $V$ compared to the whole system size, $h_V$ is a semi-local quantity and quantifies the magnetic helicity density within the volume $V$.
On the contrary, if we take the large-volume limit,
\begin{align}
    \langle h\rangle:=\lim_{V\to\infty} h_V
    \label{eq:GlobalMagneticHelicityDensity_Def}
\end{align}
quantifies a global parity violation of the magnetic field configuration.
When the system size, {\it e.g.,} the Hubble horizon of the present universe, is much larger than the coherence length of the magnetic field, the infinite-volume average can practically be the volume average over the system size. 
One may identify the infinite-volume average with the ensemble average, as is sometimes referred to as the ergodic hypothesis in this context \cite{Durrer+13}, although the original hypothesis is on the relation between temporal (rather than spatial) averages and ensemble averages.
The identification is justified by the homogeneity of the probability distribution over the space and the suppressed correlation beyond the coherence length.

Let us check the gauge-invariance of $H_V$.
Under a gauge transformation by an arbitrary function $\Gamma$, which transforms the vector potential as ${\bm A}\to{\bm A}+{\bm\nabla}\Gamma$, magnetic helicity in a closed volume $V$ transforms as
\begin{align}
    H_V
    &\to \int_V d^3r \left({\bm A}+{\bm\nabla}\Gamma\right)\cdot{\bm B}\notag\\
    &=H_V+ \int_{\partial V} d^2{\bm s}\cdot (\Gamma{\bm B})\notag\\
    &=H_V,
    \label{eq:GaugeInvariance_Helicity}
\end{align}
where we have used ${\boldsymbol{\nabla}}\cdot{\boldsymbol{B}}=0$ in the second line and a condition that $V$ is closed, {\it i.e.,} ${\bm B}\cdot d^2{\bm s}=0$ on its boundary $\partial V$ in the last line. 
Equation \eqref{eq:GaugeInvariance_Helicity} implies that $H_V$ is gauge-invariant, and hence $h_V$ and $\langle h\rangle$ are gauge-invariant as well.

Magnetic helicity is an approximate conserved quantity in the large electric conductivity limit \cite{1958PNAS...44..489W}.
To prove that, we begin with taking the time derivative of magnetic helicity density $h$.
\begin{align}
    \dot{h}
    &=\dot{\boldsymbol{A}}\cdot{\boldsymbol{B}}+{\boldsymbol{A}}\cdot\dot{\boldsymbol{B}}\notag\\
    &=2\dot{\boldsymbol{A}}\cdot{\boldsymbol{B}}+{\boldsymbol{\nabla}}\cdot(\dot{\boldsymbol{A}}\times{\boldsymbol{A}})\notag\\
    &=2(-{\boldsymbol{E}}-{\boldsymbol{\nabla}}A_0)\cdot{\boldsymbol{B}}+{\boldsymbol{\nabla}}\cdot(\dot{\boldsymbol{A}}\times{\boldsymbol{A}})\notag\\
    &=-2{\boldsymbol{E}}\cdot{\boldsymbol{B}}+{\boldsymbol{\nabla}}\cdot(\dot{\boldsymbol{A}}\times{\boldsymbol{A}}-2A_0{\boldsymbol{B}})\notag\\
    &=-2{\boldsymbol{E}}\cdot{\boldsymbol{B}}-{\boldsymbol{\nabla}}\cdot({\boldsymbol{E}}\times{\boldsymbol{A}}+A_0{\boldsymbol{B}}),
    \label{eq:small_h_dot}
\end{align}  
where $A_0$ denotes the electric potential and we have used $\boldsymbol{B}=\boldsymbol{\nabla}\times\boldsymbol{A}$ in the second line, $\boldsymbol{E}=-{\boldsymbol{\nabla}}A_0-\dot{\boldsymbol{A}}$ in the third and the last lines, and ${\boldsymbol{\nabla}}\cdot{\boldsymbol{B}}=0$ in the fourth line.
Then, for a time-dependent closed volume $V$ that moves with velocity ${\boldsymbol v}$ according to the motion of the fluid, 
\begin{align}
    \dot{H}_V
    &=\int_V d^3r \dot{h}
    +\int_{\partial V} d^2{\boldsymbol s}\cdot {\boldsymbol v}h
    \notag\\
    &=-2\int_V d^3r {\boldsymbol{E}}\cdot{\boldsymbol{B}}
    +\int_{\partial V} d^2{\boldsymbol s}\cdot
    \left[  {\boldsymbol v}({\boldsymbol{A}}\cdot{\boldsymbol{B}}) - ({\boldsymbol{E}}\times{\boldsymbol{A}}+A_0{\boldsymbol{B}})\right]
    \notag\\
    &=-2\sigma^{-1}\int_V d^3r {\boldsymbol{j}}\cdot{\boldsymbol{B}}
    -\sigma^{-1}\int_{\partial V} d^2{\boldsymbol s}\cdot({\boldsymbol{j}}\times{\boldsymbol{A}})
    +\int_{\partial V} d^2{\boldsymbol s}\cdot\boldsymbol{B}(\boldsymbol{v}\cdot\boldsymbol{A}-A_0)\notag\\
    &=-\sigma^{-1}\left[2\int_V d^3r {\boldsymbol{j}}\cdot{\boldsymbol{B}}
    +\int_{\partial V} d^2{\boldsymbol s}\cdot({\boldsymbol{j}}\times{\boldsymbol{A}})\right],
    \label{eq:MagneticHelicityDot}
\end{align}
where we have included the advection term in the first line, substituted Eqs.~\eqref{eq:Def_MagneticHelicity} and \eqref{eq:small_h_dot} in the second line, used Ohm's law in the third line, and used the condition that $V$ is closed in the last line.
From the proportionality to the electric resistivity $\sigma^{-1}$, we see that $H_V$ is approximately conserved in the large conductivity limit, $\sigma\to\infty$.
If the fluid is incompressible, $\dot{V}=0$, and then $h_V=H_V/V$ is conserved in the large conductivity limit as well.
When taking the infinite-volume limit, the random motion of fluid on the boundary of the volume $\partial V$ would cancel so that $\dot{V}=\int_{\partial V} d^2{\boldsymbol s}\cdot {\boldsymbol{v}}\to 0$ even without the assumption of local incompressibility, ${\boldsymbol\nabla}\cdot{\boldsymbol v}=0$.
Therefore, $\langle h\rangle$ is also conserved in the large conductivity limit.

To understand why $H_V$ is conserved and also for the later purpose, it would be good to explain Alfv\'{e}n's theorem \cite{1942Natur.150..405A,1943ArMAF..29B...1A}, which proves that magnetic field lines are frozen in the fluid in the large conductivity limit.
For any surface $S$ that moves with velocity ${\boldsymbol v}$ according to the motion of the fluid, 
\begin{align}
    \dfrac{d}{d\tau}\int_{S}d^2{\boldsymbol s}\cdot{\boldsymbol{B}}
    &=\int_{S}d^2{\boldsymbol s}\cdot{\dot{\boldsymbol{B}}}
    -\int_{\partial S}(d{\boldsymbol{l}}\times{\boldsymbol{v}})\cdot{\boldsymbol B}\notag\\
    &=\int_{S}d^2{\boldsymbol s}\cdot{\dot{\boldsymbol{B}}}
    -\int_{\partial S}d{\boldsymbol{l}}\cdot({\boldsymbol{v}}\times{\boldsymbol B})\notag\\
    &=\int_{S}d^2{\boldsymbol s}\cdot\left[{\dot{\boldsymbol{B}}}-{\boldsymbol\nabla}\times({\boldsymbol{v}}\times{\boldsymbol B})\right]\notag\\
    &=\sigma^{-1}\int_{S}d^2{\boldsymbol s}\cdot\nabla^2{\boldsymbol B},
\end{align}
where we have included the advection term in the first line and used the induction equation \eqref{eq:Faraday's induction equation} in the last line.
The final expression becomes zero in the large conductivity limit, $\sigma\to\infty$, implying that magnetic flux across the surface $S$ is conserved.
Since $S$ is chosen arbitrarily, we understand that magnetic field lines are advected, keeping their flux, by the fluid motion.

\begin{figure}[t]\begin{center}
    \begin{tabular}{cc}
        \begin{minipage}[h]{0.45\hsize}
            \includegraphics[keepaspectratio, width=0.8\textwidth]{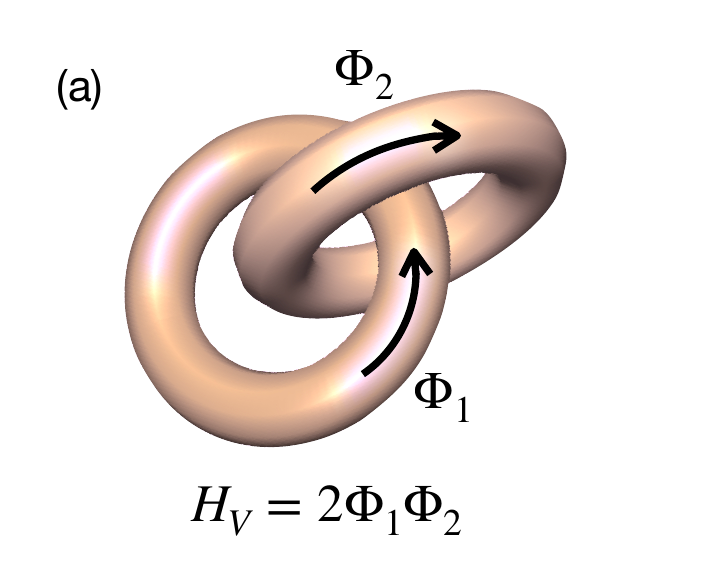}
        \end{minipage}&
        \begin{minipage}[h]{0.45\hsize}
            \includegraphics[keepaspectratio, width=0.8\textwidth]{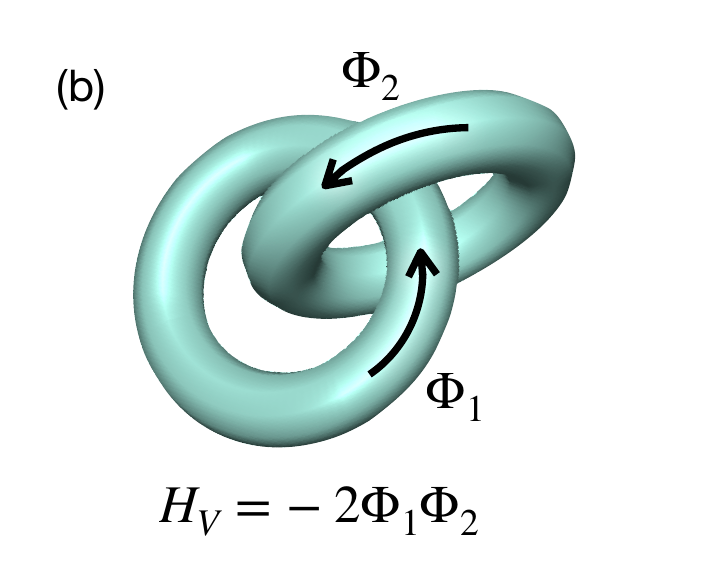}
        \end{minipage}\\
        \begin{minipage}[h]{0.45\hsize}
            \includegraphics[keepaspectratio, width=0.8\textwidth]{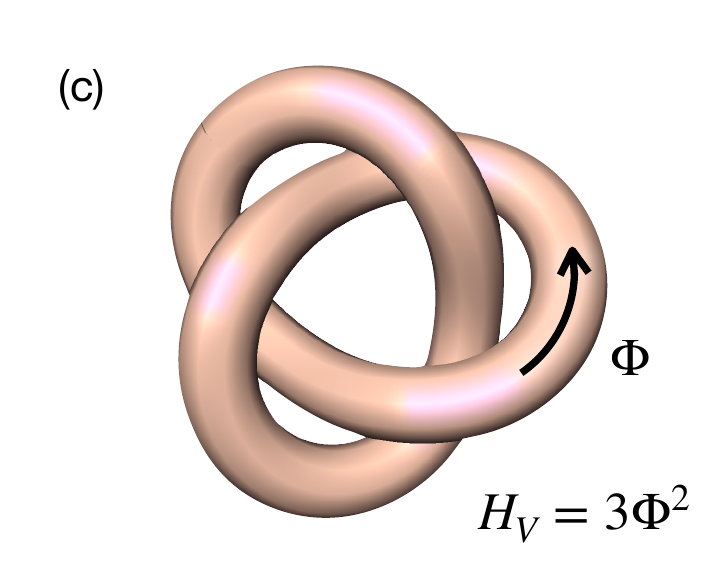}
        \end{minipage}&
        \begin{minipage}[h]{0.45\hsize}
            \includegraphics[keepaspectratio, width=0.8\textwidth]{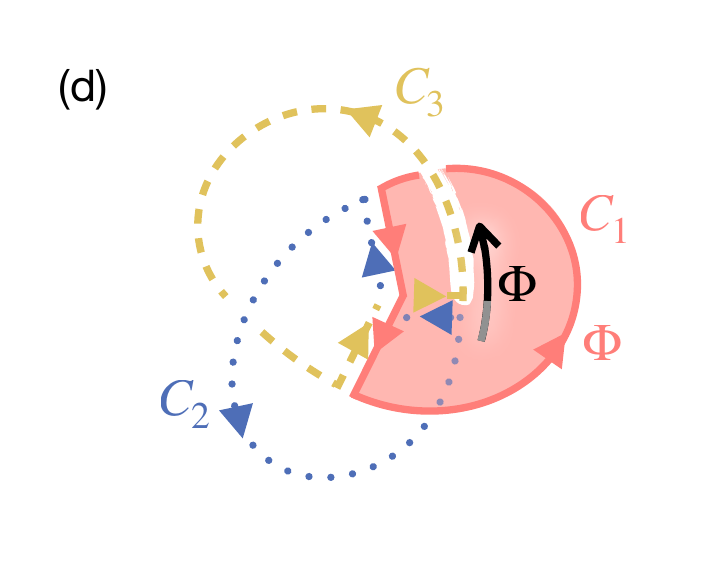}
        \end{minipage}
    \end{tabular}
    \caption{\label{fig:KnotandLink} Simplest magnetic field configurations that have finite magnetic helicity.
    (a) A link of magnetic fluxes $\Phi_1$ and $\Phi_2$, whose magnetic helicity is $2\Phi_1\Phi_2$.
    (b) An anti-link of magnetic fluxes $\Phi_1$ and $\Phi_2$, whose magnetic helicity is $-2\Phi_1\Phi_2$.
    (c) A trefoil knot of magnetic flux $\Phi$, whose magnetic helicity is $3\Phi^2$. 
    (d) In the calculation of magnetic helicity of the trefoil configuration, we separate the integration contour into three loops $C_1$, $C_2$, and $C_3$. Each loop has a contribution $\Phi^2$.}\end{center}
\end{figure}

Then, the topology of magnetic field lines is unchanged because any continuous transformation by the fluid motion cannot untie links and knots of the field lines.
Actually, magnetic helicity $H_V$ quantifies nothing but the topology of magnetic field lines \cite{1969JFM....35..117M,moreau1961constantes}.
To understand that, probably the most common example \cite{moreau1990magnetohydrodynamics,1993noma.book.....B,goedbloed2004principles,galtier2016introduction} to demonstrate is a link between two flux loops (panel (a) and (b) in Fig.~\ref{fig:KnotandLink}).
However, we are going to provide another example with a trefoil knot (panel (c) in Fig.~\ref{fig:KnotandLink}), which seems not too common to demonstrate the explicit calculation here.
Let us consider a trefoil knot of a magnetic flux $\Phi$. 
The volume integral is reduced to a contour integral along the magnetic flux tube as
\begin{align}
    H_V&=\int_{\text{trefoil}} d^3 r {\boldsymbol A}\cdot{\boldsymbol B}\notag\\
    &=\Phi\int_{C_1+C_2+C_3} d{\boldsymbol l} \cdot{\boldsymbol A}.
    \label{eq:TrefoilHelicity}
\end{align}
We have used the definition of magnetic flux, $\Phi=\int_{S}d^2s B$, where $S$ is the cross-section normal to ${\boldsymbol B}$ of the tube.
Also, we have decomposed the integration contour into three loops $C_1$, $C_2$, and $C_3$, which are illustrated in panel (d) in Fig.~\ref{fig:KnotandLink}.
Let us focus on the contribution from $C_1$.
\begin{align}
    \Phi\int_{C_1} d{\boldsymbol l} \cdot{\boldsymbol A}
    &=\Phi\int_{S_1} d^2{\boldsymbol s}\cdot({\boldsymbol \nabla}\times{\boldsymbol A})\notag\\
    &=\Phi\int_{S_1} d^2{\boldsymbol s}\cdot{\boldsymbol B}\notag\\
    &=\Phi^2,
\end{align}
where $S_1$ is a surface surrounded by the loop $C_1$, so that $\partial S_1=C_1$.
Parallel calculations apply to contributions from $C_2$ and $C_3$.
Then, we reproduce the same result as the one in Ref.~\cite{1984JFM...147..133B},
\begin{align}
    \text{Eq.~\eqref{eq:TrefoilHelicity}}=3\Phi^2.
\end{align}
We have learned one of the simplest examples to understand that magnetic helicity quantifies the topology of the magnetic field configuration.
Then, it is quite understandable that magnetic helicity is conserved because of the conservation of the topology in the large conductivity limit.

A unique property of $\langle h\rangle$ is the realizability condition \cite{1978mfge.book.....M}.
We may parametrize to what extent parity is violated by introducing the magnetic helicity fraction
\begin{align}
    \epsilon:=\dfrac{2\pi\langle h\rangle}{B^2\xi_{\rm M}},
    \label{eq:HelicityFraction_Def}
\end{align}
and then, by considering $\langle h\rangle=\langle{\boldsymbol{A}}\cdot{\boldsymbol{B}}\rangle$ as the inner product of two fields ${\boldsymbol{A}}({\boldsymbol{x}})$ and ${\boldsymbol{B}}({\boldsymbol{x}})$, Schwarz inequality implies
\begin{align}
    \vert\epsilon\vert\leq1.
    \label{eq:Realizability_Integrated}
\end{align}
Later in Sec.~\ref{sec:HoskingIntegral}, we will see that this is nothing but an integrated version of Eq.~\eqref{eq:Realizability}.
Since $\langle h\rangle$ is an approximate conserved quantity, the realizability condition \eqref{eq:Realizability_Integrated} imposes
\begin{align}
    B(\tau)^2\xi_{\rm M}(\tau)
    \geq 2\pi \vert\langle h\rangle\vert
    =\vert\epsilon_{\rm ini}\vert B_{\rm ini}^2\xi_{\rm M, ini}
    \label{eq:Constraint_MagneticHelicity}
\end{align} 
on the evolution of the system.
In particular, when the helicity fraction is $\vert\epsilon(\tau)\vert=1$, we say that the magnetic field is maximally helical, and the evolution of the system is constrained by the condition \cite{1975JFM....68..769F}
\begin{align}
    B(\tau)^2\xi_{\rm M}(\tau)
    =B_{\rm ini}^2\xi_{\rm M, ini}
    \quad\text{for maximally helical magnetic field}.
    \label{eq:Constraint_MaximallyHelical}
\end{align} 

One may ask whether the approximate conservation of magnetic helicity holds or not in the early universe.
From Eq.~\eqref{eq:MagneticHelicityDot} and by dimensional analysis, one can estimate the decay time scale of $H_V$ as
\begin{align}
    \dfrac{H_V}{\dot{H}_V}
    \sim \dfrac{B^2\xi_M}{\sigma^{-1} jB}
    \sim \dfrac{B^2\xi^2_M}{\sigma^{-1} B^2}
    \sim \sigma\xi^2_M,
\end{align}
where we have omitted the helicity fraction $\epsilon$ because the degree of parity violation would appear both in the numerator or denominator and estimated $j\sim B/\xi_{\rm M}$.
From this estimate, we obtain a condition
\begin{align}
    \xi_{\rm M}\gg \sigma^{-\frac{1}{2}}\tau^\frac{1}{2},
    \label{eq:ConditionToBelieveHelicityConservation}
\end{align}
with which we can use $\langle h\rangle$ as a conserved quantity at a given time $\tau$ and use Eqs.~\eqref{eq:Constraint_MagneticHelicity} and \eqref{eq:Constraint_MaximallyHelical} to analyze the evolution of the system in the early universe.

\subsection{Hosking integral}
\label{sec:HoskingIntegral}
Recently, another approximate conserved quantity of magneto-hydrodynamics has been discovered in Ref.~\cite{Hosking+21}.
Since theoretical understanding of its gauge-invariance and conservation has not been established rigorously, we give an ambiguous definition as is often seen in the literature.
The Hosking integral is defined as \cite{Hosking+21}
\begin{align}
    I_{{\rm H}}
    :=\int d^3r \langle h({\bm x})h({\bm x}+{\bm r})\rangle,
    \label{eq:HI_Definition1}
\end{align}
where the angle bracket denotes an ensemble average.
In the dedicated numerical study \cite{zhou2022scaling}, the authors proposed several practical ways, which are not equivalent to each other, to incorporate the finite system size and the difficulty of taking the ensemble average into the definition \eqref{eq:HI_Definition1}.

In this manuscript, to highlight the physical meaning of the Hosking integral, we approximate the ensemble average by a volume average over a large volume $V$ and define
\begin{align}
    I_{{\rm H}V,W}
    :=\dfrac{1}{V}\int_V d^3x \int_W d^3r h({\bm x})h({\bm x}+{\bm r}),
    \label{eq:HI_Definition2}
\end{align}
as is introduced as the correlation-integral method in Ref.~\cite{zhou2022scaling}.
We rewrite it as
\begin{align}
    I_{{\rm H}V,W}
    &=\dfrac{1}{V}\int_V d^3x h({\bm x}) \int_W d^3r h({\bm x}+{\bm r})\notag\\
    &=\dfrac{1}{V}\int_V d^3x h({\bm x}) \int_{W(\boldsymbol{x})} d^3y h({\bm y}),
    \label{eq:IH_Decomposed}
\end{align}
where ${W(\boldsymbol{x})}$ is the volume $W$ up to a shift ${\boldsymbol{x}}$ from the origin.
The point in the final expression is that one could expect $I_{{\rm H}V,V}\simeq H_V^2/V$ by taking the volumes of $V$ and $W$ the same.
Theoretical support of finiteness, gauge-invariance, and conservation of Hosking integral \cite{Hosking+21,Hosking+22} in the literature is based on this expectation, parallelly to the long-standing argument on the Saffman integral \cite{1967JFM....27..581S}.
However, to be rigorous, we should not overlook the ${\boldsymbol x}$ dependence of the second factor.
To overcome this issue, we refine the argument in the following paragraphs.\footnote{This is an extension of the explanation about the conservation of the Saffman integral in Ref.~\cite{1967JFM....27..581S}, but has never been completely provided in the literature as far as we know.}

As a lemma, we remind the readers of the property of a random walk.
When we have $N$ random variables that are independent and identically distributed, the deviation of their sum from the expectation is proportional to $\sqrt{N}$.

Let us assume $\langle h\rangle=0$.
Otherwise, we roughly evaluate Eq.~\eqref{eq:HI_Definition1} to be $W \langle h\rangle^2$, which diverges in the large $W$ limit.
In addition, suppose that $V$ and $W$ are spheres of radius $R\gg\xi_{\rm M}$ for simplicity.
Then we expect
\begin{enumerate}
    \item The second factor in Eq.~\eqref{eq:IH_Decomposed}, which we denote as $H_{W({\boldsymbol x})}$ hereafter, is almost gauge-invariant and conserved, and scales as $H_{W({\boldsymbol x})}\propto R^{3/2}$.
    \item Eq.~\eqref{eq:IH_Decomposed} is almost gauge-invariant and conserved, and scales as $\propto R^{0}$.
    \item In the large $R$ limit, Eq.~\eqref{eq:IH_Decomposed} is finite, gauge-invariant, and conserved.
\end{enumerate}
These statements are justified under the assumption that magnetic fields behave randomly beyond the coherence length $\xi_M$.
Let us explain one by one from the first statement.
If we regard $H_{W({\boldsymbol x})}$ as a summation of $N_1\propto W \propto R^3$ random variables, which are magnetic helicities within coherent sub-volumes filling $W$, then the lemma implies $H_{W({\boldsymbol x})}\propto \sqrt{N_1}\propto R^{3/2}$.
Its gauge-invariance and conservation hold, if we can neglect the surface terms in Eqs.~\eqref{eq:GaugeInvariance_Helicity} and \eqref{eq:MagneticHelicityDot}.
This is actually the case because we have $N_2\propto {\rm Area}(\partial W({\boldsymbol x}))\propto R^2$ random contributions on the boundary, which cancel up to fluctuations $\propto \sqrt{N_2}\propto R$.
This contribution is negligible compared to $H_{W({\boldsymbol x})}\propto R^{3/2}$ in the large $R$ limit.
Let us move on to the second statement.
As to the integral $\int_V h({\boldsymbol x})H_{W({\boldsymbol x})}$, if we approximate that $h(\boldsymbol{x})$ and $H_{W({\boldsymbol x})}$ are independent random variables, the latter typicaly scales as $R^{3/2}$ and the former contributes as $\propto \sqrt{V}\propto R^{3/2}$.
Then, we expect the scaling Eq.~\eqref{eq:IH_Decomposed} $\propto R^0$.
Its gauge-invariance and conservation are more subtle than that of $H_{W({\boldsymbol x})}$.
By parallel computations to Eqs.~\eqref{eq:GaugeInvariance_Helicity} and \eqref{eq:MagneticHelicityDot}, we have to neglect not only surface terms but also contributions in the form $\int_V ({\text{terms}})\cdot{\boldsymbol\nabla} H_{W({\boldsymbol x})}$, where $(\text{terms})$ denotes $\Gamma {\boldsymbol B}$ for Eq.~\eqref{eq:GaugeInvariance_Helicity} and ${\boldsymbol{E}}\times{\boldsymbol{A}}+A_0{\boldsymbol{B}}$ for Eq.~\eqref{eq:MagneticHelicityDot}.
These contributions are negligible because
\begin{align}
    {\boldsymbol\nabla} H_{W({\boldsymbol x})}= \int_{\partial W({\boldsymbol x})} d^2{\boldsymbol s} h({\boldsymbol y}),
\end{align}
which scales as $\propto \sqrt{N_2}\propto R$, and is negligible compared to $H_{W({\boldsymbol x})}\propto R^{3/2}$ in the large $R$ limit.
Straightforwardly from the second statement, we expect the third one.

\begin{figure}[ht]
    \begin{tabular}{cc}
        \begin{minipage}[h]{0.55\hsize}
            \includegraphics[keepaspectratio, width=0.85\textwidth]{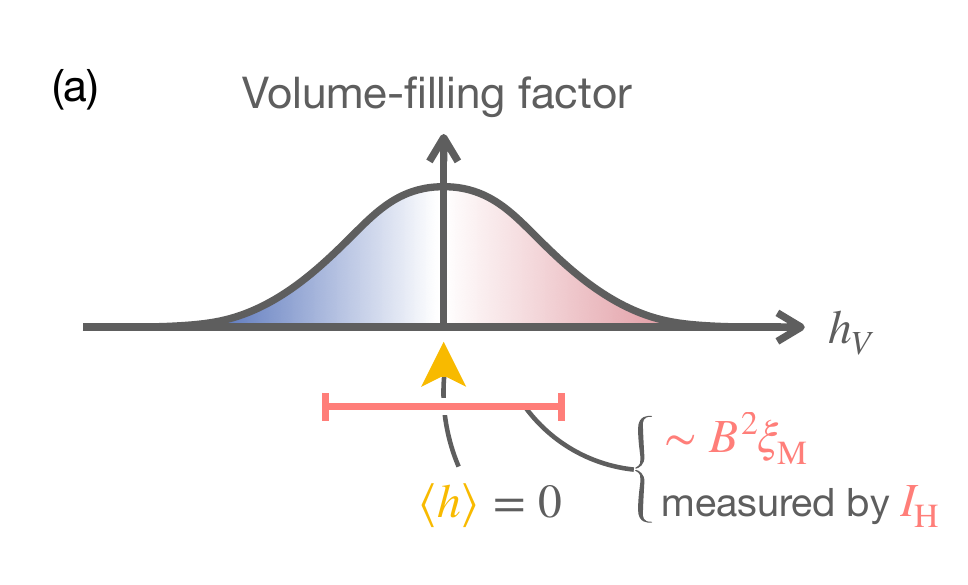}
        \end{minipage}&\hspace{-10mm}
        \begin{minipage}[h]{0.55\hsize}
            \includegraphics[keepaspectratio, width=0.85\textwidth]{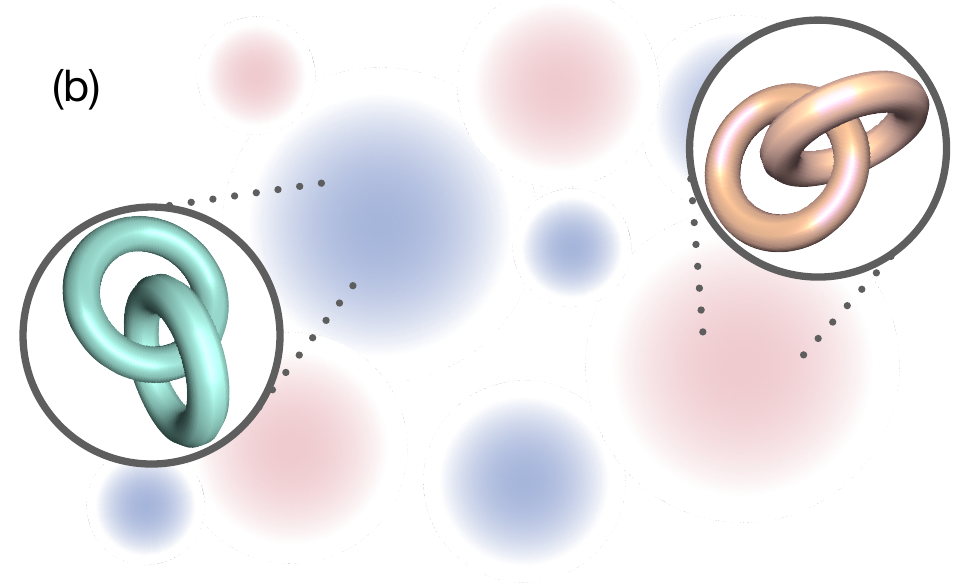}
        \end{minipage}
    \end{tabular}
    \caption{\label{fig:HelicityDistribution} (a) The probability distribution or the volume-filling factor of magnetic helicity density. It fluctuates around the expectation value $\langle h\rangle$.
    The deviation from $\langle h\rangle$ is typically of the order of $B^2\xi_{\rm M}$, as shown in Eq.~\eqref{eq:HelicityFluctuation}.
    When $\langle h\rangle=0$, Hosking integral $I_{\rm H}$ has a physical meaning that it measures the fluctuation.
    (b) Spatial distribution of magnetic helicity density. Local structures such as links or anti-links contribute to positive or negative magnetic helicity density in the local regions.}
\end{figure}

Now we can identify
\begin{align}
    I_{\rm H}=\lim_{V\to \infty} I_{{\rm H}V,V}
\end{align}
and use it as finite, gauge-invariant, and conserved quantity for the case $\langle h\rangle=0$.
Its gauge-invariance and conservation are observed numerically in Ref.~\cite{zhou2022scaling}, where they identify $I_{{\rm H}V,W}$ as Hosking integral, by taking $V$ to be the simulation box size and $W$ to be a sufficiently large volume.
As is clear from the above discussion for the justification, Hosking integral $I_{\rm H}$ quantifies the local fluctuations of magnetic helicity from zero, as seen in Fig.~\ref{fig:HelicityDistribution}.

One can estimate the value of $I_{\rm H}$ in terms of $B$ and $\xi_{\rm M}$ \cite{Hosking+21}.
Since the two-point function of magnetic helicity density in the integrand of the expression \eqref{eq:HI_Definition1} is essentially the four-point function of the magnetic field, we can decompose it into products of two-point functions of the magnetic field that are characterized by $B$ and $\xi_{\rm M}$ if we assume quasi-Gaussianity in the Fourier space \cite{Kamada:2020bmb}.
By inserting volume averaging that does not affect ensemble averaging, which is position-independent, we find an expression in the Fourier space,
\begin{align}
    I_{{\rm H}}
    &=\int d^3r \langle A_i({\bm x})B_i({\bm x})A_j({\bm x}+{\bm r})B_j({\bm x}+{\bm r})\rangle\notag\\
    &=\dfrac{1}{V}\int d^3x \int d^3r \langle A_i({\bm x})B_i({\bm x})A_j({\bm x}+{\bm r})B_j({\bm x}+{\bm r})\rangle\notag\\
    &=\dfrac{1}{V}\int d^3x \int d^3r\int \dfrac{d^3k}{(2\pi)^3}\int \dfrac{d^3l}{(2\pi)^3}\int \dfrac{d^3p}{(2\pi)^3}\int \dfrac{d^3q}{(2\pi)^3}\notag\\
    &\hspace{13.3mm}\times e^{i{\bm k}\cdot{\bm x}+i{\bm l}\cdot{\bm x}+i{\bm p}\cdot({\bm x}+{\bm r})+i{\bm q}\cdot({\bm x}+{\bm r})}\langle A_i({\bm k})B_i({\bm l})A_j({\bm p})B_j({\bm q}) \rangle\notag\\
    &=\dfrac{1}{V}\int \dfrac{d^3k}{(2\pi)^3}\int \dfrac{d^3p}{(2\pi)^3} \langle A_i({\bm k})B_i(-{\bm k})A_j({\bm p})B_j(-{\bm p}) \rangle,
    \label{eq:IH_FourierSpace}
\end{align}
where $V=(2\pi)^2\delta^3({\bm k}={\bm 0})$ formally denotes the volume of the whole space.
By assuming quasi-Gaussianity and using isotropy, the integrand of Eq.~\eqref{eq:IH_FourierSpace} is decomposed as
\begin{align}
    \langle A_i({\bm k})B_i(-{\bm k})A_j({\bm p})B_j(-{\bm p}) \rangle
    &=\langle A_i({\bm k})B_i(-{\bm k})\rangle\langle A_j({\bm p})B_j(-{\bm p}) \rangle\notag\\
    &\;\quad+ \langle A_i({\bm k})A_j({\bm p})\rangle\langle B_i(-{\bm k})B_j(-{\bm p}) \rangle\notag\\
    &\;\quad+ \langle A_i({\bm k})B_j(-{\bm p})\rangle\langle B_i(-{\bm k})A_j({\bm p}) \rangle\notag\\
    &= V^2 k^{-2} A_B(k)A_B(p)\notag\\
    &\;\quad+ (2\pi)^3\delta^3({\bm k}+{\bm p})Vk^{-2}\dfrac{P^2_B(k)+A^2_B(k)}{2}\notag\\
    &\;\quad+ (2\pi)^3\delta^3({\bm k}-{\bm p})Vk^{-2}\dfrac{P^2_B(k)+A^2_B(k)}{2}.
    \label{eq:IH_Decomposition}
\end{align}
As for the contribution from the first term in the final expression of Eq.~\eqref{eq:IH_Decomposition}, a useful relationship between the parity-violating contributions is
\begin{align}
    \langle h\rangle
    &=\dfrac{1}{V}\int d^3x\langle {\bm A}(\bm x)\cdot{\bm B}(\bm x)\rangle\notag\\
    &=\dfrac{1}{V}\int\dfrac{d^3k}{(2\pi)^3}\langle {\bm A}({\bm k})\cdot{\bm B}(-{\bm k})\rangle\notag\\
    &=\int\dfrac{d^3k}{(2\pi)^3} k^{-1} A_B(k),
    \label{eq:HelicityInTermsOfPowerSpectrum}
\end{align}
from which we understand that the realizability condition \eqref{eq:Realizability_Integrated} is saturated when the mode-wise inequality \eqref{eq:Realizability} is saturated.
By substituting Eq.~\eqref{eq:HelicityInTermsOfPowerSpectrum}, we have a divergent contribution to $I_{\rm H}$ for nonzero helicity fraction, namely
\begin{align}
    V\langle h\rangle^2
    =\dfrac{\epsilon^2}{4\pi^2} B^4\xi^2_{\rm M} V,
    \label{eq:DivergentContributionToHelicity}
\end{align}
in the large $V$ limit.
This is consistent with the heuristic argument above in this section \ref{sec:HoskingIntegral} and confirms that $\epsilon=0$ is necessary for the Hosking integral to be well-defined.
We then assume $\epsilon=0$.
As for the contribution from the second and the last terms in the final expression of Eq.~\eqref{eq:IH_Decomposition}, we obtain
\begin{align}
    I_{\rm H}=\int\dfrac{d^3k}{(2\pi)^3} k^{-2}(P^2_B(k)+A^2_B(k)).
    \label{eq:IHInTermsOfPowerSpectrum}
\end{align}
Because of the realizability condition, Eq.~\eqref{eq:Realizability}, the contribution from $A_B$ is always subdominant.
In particular, if we suppose $A_B(k)\propto P_B(k)$, which is not the case in general, $\epsilon=0$ implies $A_B(k)=0$.
The dominant contribution from $P_B(k)$ is, if we assume a peaky spectrum, approximated by the dimensional analysis, implying \cite{Hosking+21}
\begin{align}
    I_{\rm H}\sim B^4\xi_{\rm M}^5.
    \label{eq:IHDimensionalAnalysis}
\end{align}
Of course, this approximation involves a numerical coefficient depending on the shapes of the spectrum.
However, importantly, self-similar decay does not change the coefficient in time.
From the evaluation, Eq.~\eqref{eq:IHDimensionalAnalysis}, of the Hosking integral, we expect that the evolution of non-helical magnetic fields is constrained by the condition \cite{Hosking+21}
\begin{align}
    B(\tau)^4\xi_{\rm M}(\tau)^5=B_{\rm ini}^4\xi_{\rm M,ini}^5\quad
    \text{for non-helical magnetic field},
    \label{eq:Constraint_NonHelical}
\end{align}
when the electric conductivity is sufficiently large.

Let us mention an implication of the evaluation Eq.~\eqref{eq:IHDimensionalAnalysis}.
If we approximate that the two-point correlation of magnetic helicity density is $\langle h_V^2\rangle$ inside each coherent patch and vanishes beyond the coherence length $\xi_{\rm M}$, and then we obtain $I_{\rm H}\sim \xi_{\rm M}^3 \langle h_V^2\rangle$ \cite{zhou2022scaling} from the expression in Eq.~\eqref{eq:HI_Definition1}.
By comparing this evaluation with the one in Eq.~\eqref{eq:IHDimensionalAnalysis}, we obtain
\begin{align}
    \langle h_V^2\rangle\sim B^4\xi_{\rm M}^2,
    \label{eq:HelicityFluctuation}
\end{align}
implying that magnetic helicity density locally fluctuates with the amplitude $\sim B^2\xi_{\rm M}$ even if the magnetic field is globally non-helical \cite{Kamada:2020bmb} (panel (a) in Fig.~\ref{fig:HelicityDistribution}).

\section{Details of the regime-dependent analyses \label{appx:RegimeDependentAnalyses}}
Based on the classification of the regimes, we analyze the evolution of the system in each regime.
While the analyses for the regimes in Secs.~\ref{sec:MDNL_NHVfast} and \ref{sec:KDMH} follow the ones in Refs.~\cite{Hosking+21,Hosking+22}, the analyses in the remaining regimes are studied in our previous paper \cite{Uchida:2022vue} and this work, in which we complete all the possible regimes for the first time.
We explain every regime here to make this manuscript self-contained.
In the derivation of the scaling solutions, we assume that the system is not frozen ever since $\tau_{\rm ini}$.
Then, Eq.~\eqref{eq:TimeScaleCondition_general2} is reduced to Eq.~\eqref{eq:TimeScaleCondition_general}.

\subsection{Magnetically dominated and non-linear regimes}
\label{sec:MDNL}
\begin{figure}[ht]
    \begin{tabular}{ll}
        \begin{minipage}[h]{0.48\hsize}
            \includegraphics[keepaspectratio, width=0.85\textwidth]{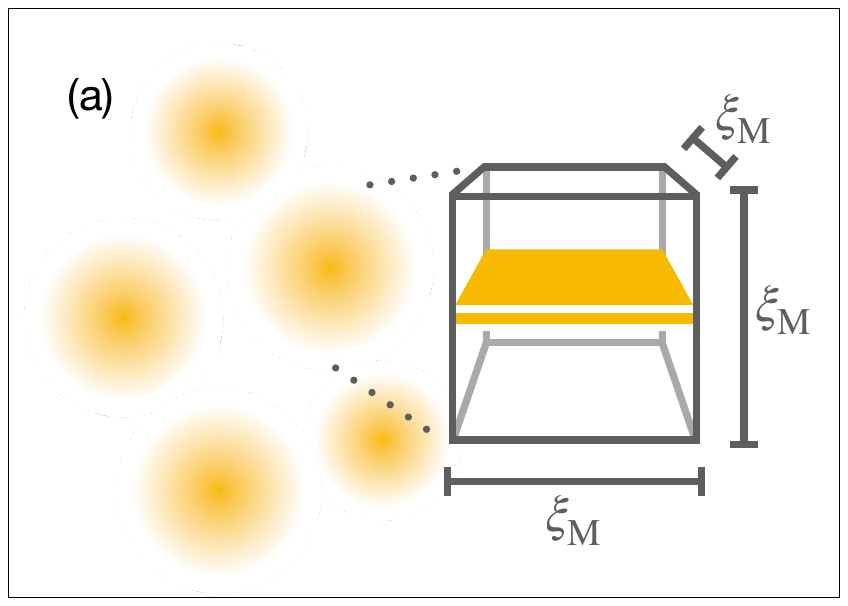}
        \end{minipage}&
        \begin{minipage}[h]{0.48\hsize}
            \includegraphics[keepaspectratio, width=0.85\textwidth]{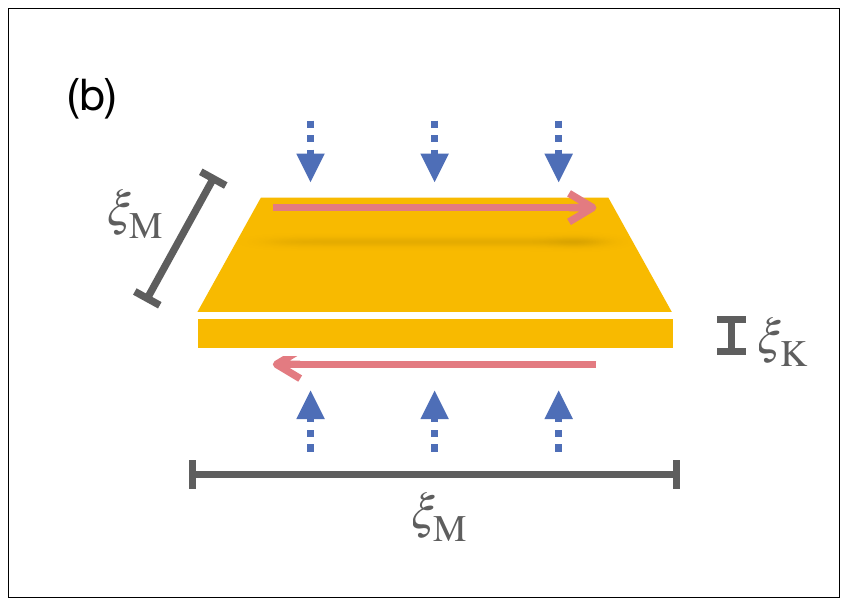}
        \end{minipage}\\
        \begin{minipage}[h]{0.48\hsize}
            \includegraphics[keepaspectratio, width=0.85\textwidth]{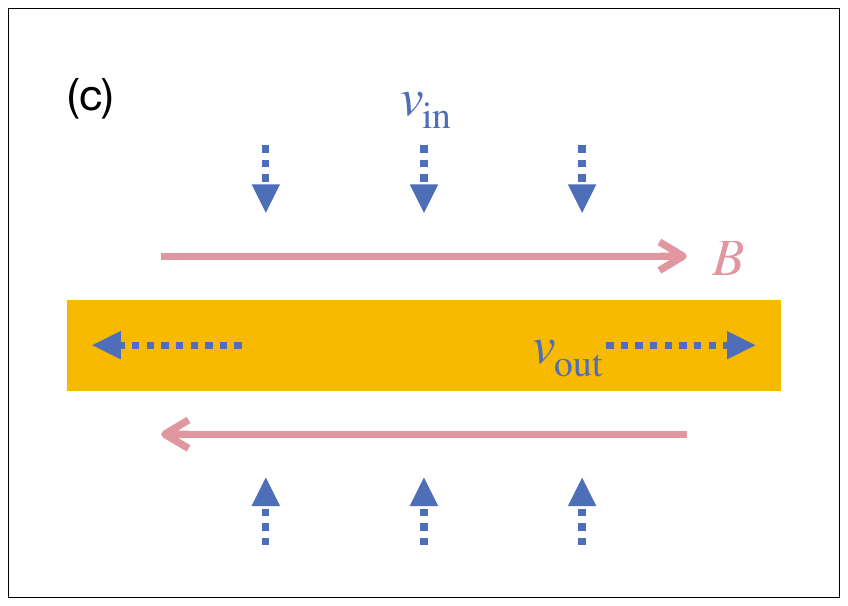}
        \end{minipage}&
        \begin{minipage}[h]{0.48\hsize}
            \includegraphics[keepaspectratio, width=0.85\textwidth]{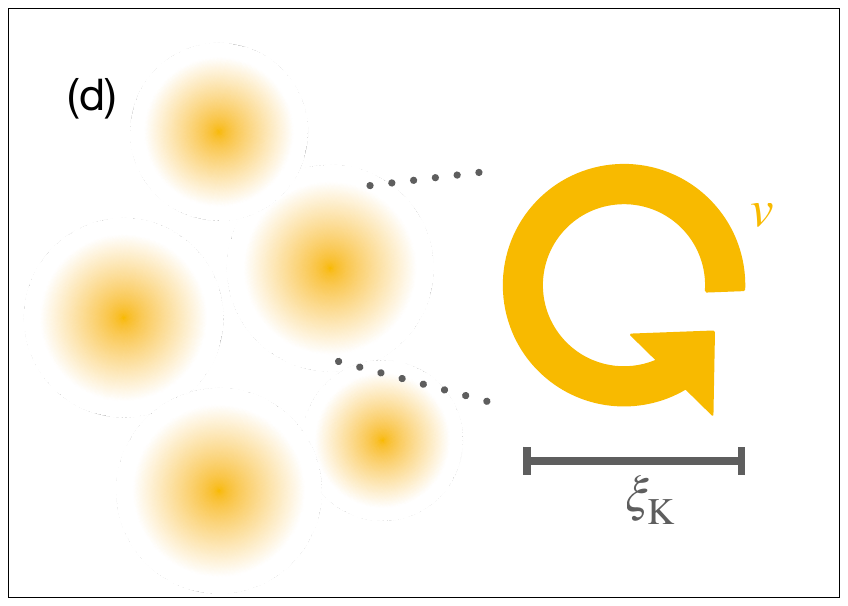}
        \end{minipage}
    \end{tabular}
    \caption{\label{fig:NonlinearPicture} Mechanisms of the energy decay in non-linear regimes. 
    (a--c) Magnetic reconnection is operating on a current sheet (orange-shaded) in each coherent patch (gray-outlined box). Anti-parallel magnetic fields on both sides of the sheet (red solid arrows) are inflowing into the sheet (vertical blue-dotted arrows), and a portion of the inflowing energy is converted into outflow from the sheet (horizontal blue-dotted arrows). 
    (d) The time scale of the energy cascade of the velocity field is controlled by the largest eddies (thick orange arrow) in the cascading process.}
\end{figure}

In this branch of regimes, magnetic energy dominates over kinetic energy, and magnetic reconnection continuously dissipates magnetic energy into both heat and outflow velocity.
Conditions
\begin{align}
    \Gamma\ll1,\quad
    {\rm Re}_{\rm M}\gg1
    \label{eq:MDNL_condition}
\end{align}
are satisfied.
As depicted in panels (a), (b), and (c) of Fig.~\ref{fig:NonlinearPicture}, the system is characterized by six variables, $B$, $\xi_{\rm M}$, $v$, $\xi_{\rm K}$, $v_{\rm in}$, and $v_{\rm out}$.
We have several regimes within this branch according to the helicity fraction $\epsilon$, the ratio of dissipation terms $r_{\rm diss}$, and the Lundquist number $S$.
We extend the analysis in the literature \cite{Parker57,1958IAUS....6..123S,park+84,Hosking+21,Hosking+22} to find six equations for each subregime.
By solving them, we find formulae that describe the evolution of the system in each regime.

\subsubsection{Maximally helical and viscous Sweet--Parker regime (\ctext{1})}
\label{sec:MDNL_MHVSP}
In this regime, the magnetic field is maximally helical, shear viscosity dominates over drag force, and the Lundquist number is relatively small.
In addition to Eq.~\eqref{eq:MDNL_condition},
\begin{align}
    \vert\epsilon\vert=1,\quad
    r_{\rm diss}(\xi_{\rm K})\ll1,\quad
    S\ll S_{\rm c}
    \label{eq:MHVSP_condition}
\end{align}
are satisfied.
The mechanism that governs the evolution of the system is the Sweet--Parker reconnection with shear viscosity.

\begin{figure}[ht]\begin{center}
    \begin{tabular}{rl}
        \begin{minipage}[h]{0.60\hsize}
            \includegraphics[keepaspectratio, width=0.85\textwidth]{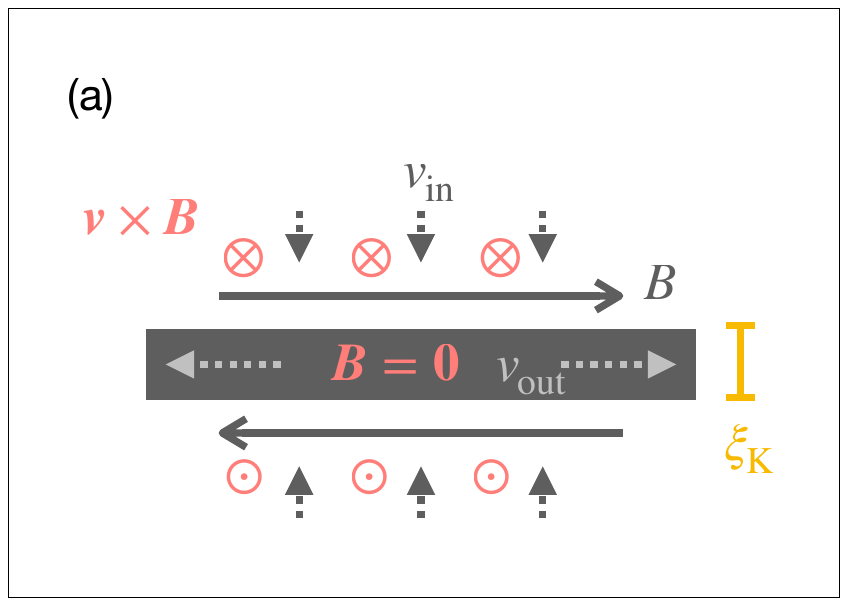}
        \end{minipage}&
        \begin{minipage}[h]{0.60\hsize}
            \hspace{-15mm}
            \includegraphics[keepaspectratio, width=0.85\textwidth]{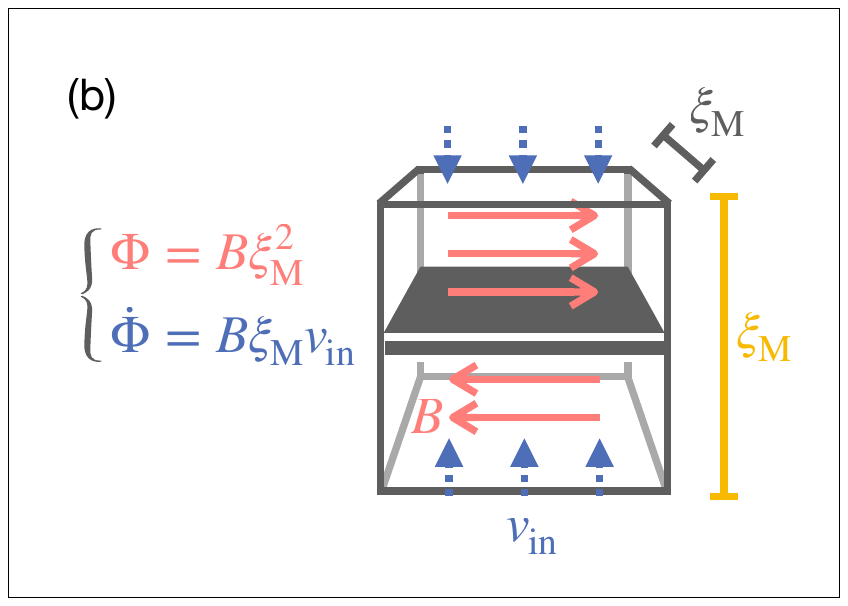}
        \end{minipage}
    \end{tabular}
    \caption{\label{fig:SPReconnection} Illustrations of Sweet--Parker reconnection. (a) Magnetic field (gray solid arrows) is anti-parallel on both sides of the current sheet (gray-shaded region) and vanishes inside the sheet. Since the inflow velocity (gray-dotted arrows) is there, the Lorentz force ${\bm v}\times {\bm B}$ can be estimated to be $\sim v_{\rm in}B$ outside the sheet. (b) The time scale of the reconnection process can be identified as the rate of processing magnetic flux $\Phi$. We have magnetic fluxes $\sim B\xi_{\rm M}^2$ (red solid arrows) in each half of the box separated by the current sheet (gray-shaded region). Because of the inflow velocity (blue-dotted arrows), each flux is processed as $\dot{\Phi}\sim B\xi_{\rm M}v_{\rm in}$, from which we can estimate the time scale to be $\Phi/\dot{\Phi}\sim \xi_{\rm M}/v_{\rm in}=:\tau_{\rm rec}$.}\end{center}
\end{figure}

Let us follow Refs.~\cite{Parker57,1958IAUS....6..123S,park+84} to find relations between the six variables.
Look at panel (c) of Fig.~\ref{fig:NonlinearPicture}.
The mass conservation on each current sheet implies \cite{Parker57}
\begin{align}
    v_{\rm in}\xi_{\rm M}=v_{\rm out}\xi_{\rm K},
    \label{eq:MassConservation}
\end{align}
where the left and right-hand sides represent the rate of incoming and outgoing mass, respectively.
Another condition is the stationarity of the current sheet.
If we focus only on the structure across the current sheet, by assuming a large aspect ratio $\xi_{\rm M}\gg\xi_{\rm K}$, we can replace the spatial derivative ${\boldsymbol \nabla}$ in the induction equation \eqref{eq:Faraday's induction equation} by $1/\xi_{\rm K}$ (panel (a) of Fig.~\ref{fig:SPReconnection}).
As for the term ${\bm\nabla}\times({\bm v}\times{\bm B})$, we have $|{\bm v}\times{\bm B}|\sim v_{\rm in}B$ outside the sheet, while $|{\bm v}\times{\bm B}|\sim 0$ inside the sheet.
Since the outside and the inside are separated by the width of the sheet $\xi_{\rm K}$, we estimate $|{\bm\nabla}\times({\bm v}\times{\bm B})|\sim v_{\rm in}B/\xi_{\rm K}$.
As for the term $\sigma^{-1}\nabla^2{\bm B}$, we have opposite directions of magnetic field $|{\bm B}|\sim B$ on both sides of the sheet, while $|{\bm B}|\sim 0$ inside.
Therefore, we estimate $|\sigma^{-1}\nabla^2{\bm B}|\sim B/\sigma\xi_{\rm K}^2$.
Then, the balance between terms in the induction equation \eqref{eq:Faraday's induction equation} implies \cite{Parker57}
\begin{align}
    \dfrac{v_{\rm in}B}{\xi_{\rm K}}=\dfrac{B}{\sigma\xi_{\rm K}^2}.
    \label{eq:Stationarity}
\end{align}
For convenience, we use Eqs.~\eqref{eq:MassConservation} and \eqref{eq:Stationarity} to obtain an expression of the outflow velocity in terms of the coherence lengths,
\begin{align}
    v_{\rm out}=\dfrac{\xi_{\rm M}}{\sigma\xi_{\rm K}^2}.
    \label{eq:OutflowVelocity}
\end{align}
Energy dissipation of the outflow along the current sheet implies
\begin{align}
    \dfrac{B^2}{2}
    =\dfrac{\rho+p}{2}v_{\rm out}^2+
    \int_0^{\xi_{\rm M}/2}\dfrac{d\xi}{v_{\rm out}}\dfrac{(\rho+p)\eta v_{\rm out}^2}{\xi_{\rm K}^2},
    \label{eq:Dissipation_1}
\end{align}
which is reduced to \cite{park+84}
\begin{align}
    B^2=(\rho+p) (1+{\rm Pr}_{\rm M})v_{\rm out}^2,
    \label{eq:Dissipation_2}
\end{align}
by approximating the integrand in Eq.~\eqref{eq:Dissipation_1} to be constant and using Eq.~\eqref{eq:OutflowVelocity}.
The first term in the right-hand side of Eq.~\eqref{eq:Dissipation_1} or the term $1$ in Eq.~\eqref{eq:Dissipation_2} represents the energy of the outflow, while the second term in the right-hand side of Eq.~\eqref{eq:Dissipation_1} or the term ${\rm Pr}_{\rm M}$ in Eq.~\eqref{eq:Dissipation_2} represents the energy dissipation by the shear viscosity on the current sheet.
Although the original idea of the Sweet--Parker model \cite{1958IAUS....6..123S,Parker57} neglects the latter contribution, large Prandtl number ${\rm Pr}_{\rm M}\gg1$ in the early universe implies that the former contribution is negligible but rather the latter one is the dominant.
Then we approximate Eq.~\eqref{eq:Dissipation_2} as
\begin{align}
    B^2=(\rho+p) {\rm Pr}_{\rm M}v_{\rm out}^2.
    \label{eq:Dissipation_3}
\end{align}
The condition about decay time scales, Eq.~\eqref{eq:TimeScaleCondition_general}, implies \cite{Hosking+21}
\begin{align}
    \tau_{\rm rec}=\tau,
    \label{eq:Decaytime_Rec}
\end{align}
where the reconnection time scale is given by $\tau_{\rm rec}:=\xi_{\rm M}/v_{\rm in}$ (panel (b) of Fig.~\ref{fig:SPReconnection}).
Finally, the kinetic energy in each coherent patch is dominated by the outflow, and the root-mean-square velocity $v$ can be approximated by the outflow energy as \cite{Hosking+21}
\begin{align}
    \dfrac{v^2}{2}=\dfrac{\xi_{\rm K}}{\xi_{\rm M}}\dfrac{v_{\rm out}^2}{2},
    \label{eq:Dilution}
\end{align}
by taking into account dilution by the volume ratio of a current sheet to a unit patch.
Neglecting the contribution of the inflow velocity is justified by confirming $v_{\rm in}\ll v$ in the final expression.

Let us summarize what we have to solve.
We need to solve Eqs.~\eqref{eq:MassConservation}, \eqref{eq:Stationarity}, \eqref{eq:Dissipation_3}, \eqref{eq:Decaytime_Rec}, and \eqref{eq:Dilution} together with the helicity conservation, Eq.~\eqref{eq:Constraint_MaximallyHelical}.
In addition, we have two assumptions for the solution, which guarantee that the set of equations is reasonable.
One is the large aspect ratio of the current sheet,
\begin{align}
    \dfrac{\xi_{\rm M}}{\xi_{\rm K}}\gg1,
    \label{eq:LargeAspectRatio}
\end{align}
and the other is the small inflow velocity,
\begin{align}
    \dfrac{v_{\rm in}}{v}\ll1.
    \label{eq:SmallInflowVelocity}
\end{align}
They will be justified after solving the set of equations.

To solve the relevant equations, let us go stepwise and see which equations lead to which results.
First, we express quantities about the velocity field in terms of $B$ and $\xi_{\rm M}$.
By using the condition about the decay time scale, Eq.~\eqref{eq:Decaytime_Rec}, we obtain
\begin{align}
    v_{\rm in}=\tau^{-1}\xi_{\rm M}.
    \label{eq:InflowVelocityInTermsOfBXiM}
\end{align}
We substitute this expression into the stationarity condition, Eq.~\eqref{eq:Stationarity}, to obtain an expression of the kinetic coherence length $\xi_{\rm K}$, substitute the resultant expression into Eq.~\eqref{eq:OutflowVelocity} to obtain an expression of the outflow velocity $v_{\rm out}$, and finally obtain an expression of the typical velocity $v$ by considering the energy dilution, Eq.~\eqref{eq:Dilution}.
Then we obtain
\begin{align}
    \xi_{\rm K}=\sigma^{-1}\tau\xi_{\rm M}^{-1},\quad
    v_{\rm out}=\sigma\tau^{-2}\xi_{\rm M}^3,\quad
    v=\sigma^{\frac{1}{2}}\tau^{-\frac{3}{2}}\xi_{\rm M}^2.
    \label{eq:MDNL_MHVSP_Vel_InTermsOfBXiM}
\end{align}
Before going to the next step, we express the order parameters in terms of $B$ and $\xi_{\rm M}$ as well, by substituting Eqs.~\eqref{eq:InflowVelocityInTermsOfBXiM} and \eqref{eq:MDNL_MHVSP_Vel_InTermsOfBXiM} into the definitions.
The energy ratio and the magnetic Reynolds number are
\begin{align}
    \Gamma=(\rho+p)\sigma\tau^{-3}B^{-2}\xi_{\rm M}^4,\quad
    {\rm Re}_{\rm M}
    =\left(\dfrac{\xi_{\rm M}}{\sigma^{-\frac{1}{2}}\tau^{\frac{1}{2}}}\right)^5,
    \label{eq:MDNL_MHVSP_GammaAndReM}
\end{align}
in terms of which the ratio of dissipation terms evaluated at the kinetic coherence length and the Lundquist number are
\begin{align}
    r_{\rm diss}(\xi_{\rm K})
    ={\rm Pr}_{\rm M}^{-1}{\rm Re}_{\rm M}^{-\frac{2}{5}}\alpha\tau,
    \quad
    S
    ={\rm Re}_{\rm M}^{\frac{4}{5}},
    \label{eq:MDNL_MHVSP_rdissAndS}
\end{align}
where we define the Lundquist number as
\begin{align}
    S:=\dfrac{\tau_\sigma(\xi_{\rm M})}{\xi_{\rm M}/v_{\rm out}}=\sigma v_{\rm out} \xi_{\rm M}.
    \label{eq:Def_LundquistNumber}
\end{align}
Also, the largeness of the aspect ratio, Eq.~\eqref{eq:LargeAspectRatio}, and the smallness of the inflow velocity, Eq.~\eqref{eq:SmallInflowVelocity}, are guaranteed by the large magnetic Reynolds number because
\begin{align}
    \dfrac{\xi_{\rm M}}{\xi_{\rm K}}={\rm Re}_{\rm M}^{\frac{2}{5}},\quad
    \dfrac{v_{\rm in}}{v}={\rm Re}_{\rm M}^{-\frac{1}{5}}
    \label{eq:MDNL_MHVSP_Ratios}
\end{align}
are derived from Eqs.~\eqref{eq:InflowVelocityInTermsOfBXiM} and \eqref{eq:MDNL_MHVSP_Vel_InTermsOfBXiM}.

The next step is to use  Eq.~\eqref{eq:Dissipation_3}.
By substituting the expression of the outflow velocity $v_{\rm out}$, we obtain 
\begin{align}
    (\rho+p)^{\frac{1}{4}}\sigma^{\frac{3}{4}}\eta^{\frac{1}{4}}
    B^{-\frac{1}{2}}\xi_{\rm M}^{\frac{3}{2}}=\tau,
    \label{eq:TimeScaleCondition_MDNL_MHVSP}
\end{align}
which corresponds to Eq.~\eqref{eq:TimeScaleCondition_general}.
We can relate $\Gamma$ and ${\rm Re}_{\rm M}$ by employing this condition and the expression in Eq.~\eqref{eq:MDNL_MHVSP_GammaAndReM} as
\begin{align}
    \Gamma={\rm Pr}_{\rm M}^{-1}{\rm Re}_{\rm M}^{-\frac{2}{5}}.
    \label{eq:MDNL_MHVSP_Gamma}
\end{align}

Finally, by solving the conservation law, Eq.~\eqref{eq:Constraint_MaximallyHelical}, and the time-dependent condition, Eq.~\eqref{eq:TimeScaleCondition_MDNL_MHVSP}, we obtain 
\begin{align}
    B&=B_{\rm ini}^{\frac{6}{7}}\xi_{\rm M,ini}^{\frac{3}{7}}(\rho+p)^{\frac{1}{14}}\sigma^{\frac{3}{14}}\eta^{\frac{1}{14}}\tau^{-\frac{2}{7}},
    \label{eq:B_MDNL_MHVSP}\\
    \xi_{\rm M}&=B_{\rm ini}^{\frac{2}{7}}\xi_{\rm M,ini}^{\frac{1}{7}}(\rho+p)^{-\frac{1}{7}}\sigma^{-\frac{3}{7}}\eta^{-\frac{1}{7}}\tau^{\frac{4}{7}}.
    \label{eq:XiM_MDNL_MHVSP}
\end{align}

As Ref.~\cite{Hosking+22} pointed out for a drag force dominant regime, we should be careful about the effect of dissipation on the inflow.
We have implicitly assumed in the above discussion that the Sweet--Parker reconnection is the rate-determining process, although the inflow velocity can be limited rather by the dissipation of the inflow.
To justify the assumption, let us consider the region away from the current sheet and suppose that the inflow velocity $v'_{\rm in}$ is determined by the dissipation effect.
Then we will compare the two inflow speeds, $v'_{\rm in}$ and $v_{\rm in}$.

In the region where the flow is laminar and stationary but the dissipation is significant, the energy injection from the magnetic field should balance the energy dissipation \cite{Banerjee+04}.
By approximating the relevant terms in the Navier--Stokes equation \eqref{eq:Navier--Stokes equation} as
\begin{align}
    \left\vert\dfrac{({\bm\nabla}\times{\bm B})\times{\bm B}}{\rho+p}\right\vert\sim \dfrac{B^2}{(\rho+p)\xi_{\rm M}},\quad
    \vert\eta[\nabla^2{\bm v}+\cdots]\vert\sim \dfrac{\eta v'_{\rm in}}{\xi_{\rm M}^2}
    \label{eq:MDNL_MHVSP_EstimateInNSE}
\end{align}
and balancing these terms, we obtain an estimate of the inflow velocity
\begin{align}
    v'_{\rm in}=\dfrac{B^2\xi_{\rm M}}{(\rho+p)\eta}
    ={\rm Re}_{\rm M}^{\frac{6}{5}}v_{\rm in}
    \gg v_{\rm in},
    \label{eq:Smallness_DissipatedInflowVelocity}
\end{align}
where we have used Eqs.~\eqref{eq:EnergyRatio_Def}, \eqref{eq:SmallInflowVelocity}, \eqref{eq:InflowVelocityInTermsOfBXiM}, \eqref{eq:MDNL_MHVSP_GammaAndReM}, and \eqref{eq:MDNL_MHVSP_Gamma} in the second equality.
The hierarchy between $v_{\rm in}$ and $v'_{\rm in}$ supports our assumption that the rate-determining process is magnetic reconnection.
The inflow velocity $v_{\rm in}$ is so small that the estimated magnitudes of the two terms in Eq.~\eqref{eq:MDNL_MHVSP_EstimateInNSE} do not balance, implying that the magnetic field actually takes a force-free configuration,
\begin{align}
    ({\bm\nabla}\times{\bm B})\times{\bm B}\ll \dfrac{B^2}{\xi_{\rm M}},
    \label{eq:ForceFree}
\end{align}
of which observation in numerical studies can support our analysis.

\subsubsection{Non-helical and viscous Sweet--Parker regime (\ctext{2})}
\label{sec:MDNL_NHVSP}
In this regime, the magnetic field is non-helical, shear viscosity dominates over drag force, and the Lundquist number is relatively small.
In addition to Eq.~\eqref{eq:MDNL_condition},
\begin{align}
    \epsilon=0,\quad
    r_{\rm diss}(\xi_{\rm K})\ll1,\quad
    S\ll S_{\rm c}
    \label{eq:NHVSP_condition}
\end{align}
are satisfied.
The mechanism that governs the evolution of the system is the same as the one in Sec.~\ref{sec:MDNL_MHVSP}, except for that the dynamics is constrained by the Hosking integral conservation \eqref{eq:Constraint_NonHelical}, instead of the helicity conservation Eq.~\eqref{eq:Constraint_MaximallyHelical}.

By solving Eqs.~\eqref{eq:MassConservation}, \eqref{eq:Stationarity}, \eqref{eq:Dissipation_3}, \eqref{eq:Decaytime_Rec}, and \eqref{eq:Dilution}, together with Eq.~\eqref{eq:Constraint_MaximallyHelical}, we obtain
\begin{align}
    B&=B_{\rm ini}^{\frac{12}{17}}\xi_{\rm M,ini}^{\frac{15}{17}}(\rho+p)^{\frac{5}{34}}\sigma^{\frac{15}{34}}\eta^{\frac{5}{34}}\tau^{-\frac{10}{17}},
    \label{eq:B_MDNL_NHVSP}\\
    \xi_{\rm M}&=B_{\rm ini}^{\frac{4}{17}}\xi_{\rm M,ini}^{\frac{5}{17}}(\rho+p)^{-\frac{2}{17}}\sigma^{-\frac{6}{17}}\eta^{-\frac{2}{17}}\tau^{\frac{8}{17}}.
    \label{eq:XiM_MDNL_NHVSP}
\end{align}

Expressions in Eqs.~\eqref{eq:InflowVelocityInTermsOfBXiM}, \eqref{eq:MDNL_MHVSP_Vel_InTermsOfBXiM}, \eqref{eq:MDNL_MHVSP_GammaAndReM}, \eqref{eq:MDNL_MHVSP_rdissAndS}, \eqref{eq:MDNL_MHVSP_Ratios}, \eqref{eq:TimeScaleCondition_MDNL_MHVSP}, \eqref{eq:MDNL_MHVSP_Gamma}, and \eqref{eq:Smallness_DissipatedInflowVelocity} hold independently of the initial conditions, $B_{\rm ini}$ and $\xi_{\rm M,ini}$.

\subsubsection{Maximally helical and dragged Sweet--Parker regime (\ctext{3})}
\label{sec:MDNL_MHDSP}
In this regime, the magnetic field is maximally helical, drag force dominates over shear viscosity, and the Lundquist number is relatively small.
In addition to Eq.~\eqref{eq:MDNL_condition},
\begin{align}
    \vert\epsilon\vert=1,\quad
    r_{\rm diss}(\xi_{\rm K})\gg1,\quad
    S\ll S_{\rm c}
    \label{eq:MHDSP_condition}
\end{align}
are satisfied.
The mechanism that governs the evolution of the system is basically the Sweet--Parker reconnection, but the dissipation is due to the drag force rather than the shear viscosity \cite{Uchida:2022vue}.
Note that we are considering a different regime from the so-called collisionless regime \cite{ji2011phase,Hosking+22}.
We assume that the fluid approximation of the plasma still holds.

We do the same discussion as in Sec.~\ref{sec:MDNL_MHVSP}, except for the estimate of energy dissipation in Eq.~\eqref{eq:Dissipation_1}.
When the drag force dominates over shear viscosity, energy dissipation of the outflow along the current sheet implies
\begin{align}
    \dfrac{B^2}{2}
    =\dfrac{\rho+p}{2}v_{\rm out}^2+
    \int_0^{\xi_{\rm M}/2}\dfrac{d\xi}{v_{\rm out}}(\rho+p)\alpha v_{\rm out}^2,
    \label{eq:Dissipation_4}
\end{align}
which is reduced to
\begin{align}
    B^2=(\rho+p) (v_{\rm out}^2+\alpha\xi_{\rm M}v_{\rm out}),
    \label{eq:Dissipation_5}
\end{align}
by approximating the integrand in Eq.~\eqref{eq:Dissipation_4} to be constant.
The first term in the right-hand side of Eq.~\eqref{eq:Dissipation_4} or the term $v_{\rm out}^2$ in Eq.~\eqref{eq:Dissipation_2} represents the energy of the outflow, while the second term in the right-hand side of Eq.~\eqref{eq:Dissipation_4} or the term $\alpha\xi_{\rm M}v_{\rm out}$ in Eq.~\eqref{eq:Dissipation_5} represents the energy dissipation by the shear viscosity on the current sheet.
As is the case in Sec.~\ref{sec:MDNL_MHVSP}, the former contribution can be assumed to be negligible.
This assumption is justified later by confirming
\begin{align}
    \alpha\xi_{\rm M}\gg v_{\rm out}.
    \label{eq:Validity_Dissipation6}
\end{align}
With this assumption, we approximate Eq.~\eqref{eq:Dissipation_6} as
\begin{align}
    B^2=(\rho+p) \alpha\xi_{\rm M}v_{\rm out}.
    \label{eq:Dissipation_6}
\end{align}

We replace Eq.~\eqref{eq:Dissipation_3} in Sec.~\ref{sec:MDNL_MHVSP} with Eq.~\eqref{eq:Dissipation_6}.
Then, we need to solve Eqs.~\eqref{eq:MassConservation}, \eqref{eq:Stationarity}, \eqref{eq:Dissipation_6}, \eqref{eq:Decaytime_Rec}, \eqref{eq:Dilution}, and \eqref{eq:Constraint_MaximallyHelical}.
If we solve the set of equations step by step, the expressions we find in Sec.~\ref{sec:MDNL_MHVSP}, namely Eqs.~\eqref{eq:InflowVelocityInTermsOfBXiM}, \eqref{eq:MDNL_MHVSP_Vel_InTermsOfBXiM}, \eqref{eq:MDNL_MHVSP_GammaAndReM}, \eqref{eq:MDNL_MHVSP_rdissAndS}, and \eqref{eq:MDNL_MHVSP_Ratios}, hold.
We can justify the assumption, Eq.~\eqref{eq:Validity_Dissipation6}, by confirming
\begin{align}
    \dfrac{\alpha\xi_{\rm M}}{v_{\rm out}}={\rm Pr}_{\rm M}r_{\rm diss}\gg1.
    \label{eq:Justify_Dissipation6}
\end{align}
By substituting the expression of the outflow into Eq.~\eqref{eq:Dissipation_6}, we obtain
\begin{align}
    (\rho+p)^{\frac{1}{2}}\sigma^{\frac{1}{2}}\alpha^{\frac{1}{2}}B^{-1}\xi_{\rm M}^2=\tau,
    \label{eq:MDNL_MHDSP_Decaytime}
\end{align} 
which leads to expressions of the energy ratio
\begin{align}
    \Gamma={\rm Pr}_{\rm M}^{-1}{\rm Re}_{\rm M}^{-\frac{2}{5}}r_{\rm diss}(\xi_{\rm K})^{-1}
    \label{eq:MDNL_MHDSP_Gamma}.
\end{align}
We solve Eq.~\eqref{eq:MDNL_MHDSP_Decaytime} with the helicity conservation constraint, Eq.~\eqref{eq:Constraint_MaximallyHelical}, to obtain
\begin{align}
    B&=B_{\rm ini}^{\frac{4}{5}}\xi_{\rm M,ini}^{\frac{2}{5}}(\rho+p)^{\frac{1}{10}}\sigma^{\frac{1}{10}}\alpha^{\frac{1}{10}}\tau^{-\frac{1}{5}},
    \label{eq:B_MDNL_MHDSP}\\
    \xi_{\rm M}&=B_{\rm ini}^{\frac{2}{5}}\xi_{\rm M,ini}^{\frac{1}{5}}(\rho+p)^{-\frac{1}{5}}\sigma^{-\frac{1}{5}}\alpha^{-\frac{1}{5}}\tau^{\frac{2}{5}}.
    \label{eq:XiM_MDNL_MHDSP}
\end{align}

Again, we should be careful about the effect of the dissipation on the inflow.
Let us consider the region away from the current sheet and suppose that the inflow velocity $v'_{\rm in}$ is determined by the drag force.
In that region, the energy injection from the magnetic field should balance with the energy dissipation by the drag force \cite{Banerjee+04}.
By approximating the relevant terms in Navier--Stokes equation \eqref{eq:Navier--Stokes equation} as
\begin{align}
    \left\vert\dfrac{({\bm\nabla}\times{\bm B})\times{\bm B}}{\rho+p}\right\vert\sim \dfrac{B^2}{(\rho+p)\xi_{\rm M}},\quad
    \vert-\alpha{\bm v}\vert\sim \alpha v'_{\rm in}
    \label{eq:MDNL_MHDSP_EstimateInNSE}
\end{align}
and balancing these terms, we obtain an estimate of the inflow velocity
\begin{align}
    v'_{\rm in}=\dfrac{B^2}{(\rho+p)\alpha\xi_{\rm M}}
    ={\rm Re}_{\rm M}^{\frac{2}{5}}v_{\rm in}
    \gg v_{\rm in},
    \label{eq:Smallness_DissipatedInflowVelocity_Drag}
\end{align}
which supports our assumption that the rate-determining process is magnetic reconnection on the current sheets.

\subsubsection{Non-helical and dragged Sweet--Parker regime (\ctext{4})}
\label{sec:MDNL_NHDSP}
In this regime, the magnetic field is non-helical, drag force dominates over shear viscosity, and the Lundquist number is relatively small.
In addition to Eq.~\eqref{eq:MDNL_condition},
\begin{align}
    \epsilon=0,\quad
    r_{\rm diss}(\xi_{\rm K})\gg1,\quad
    S\ll S_{\rm c}
    \label{eq:NHDSP_condition}
\end{align}
are satisfied.
The only difference from Sec.~\ref{sec:MDNL_MHDSP} is that the constraint from the Hosking integral conservation \eqref{eq:Constraint_NonHelical}, instead of the helicity conservation Eq.~\eqref{eq:Constraint_MaximallyHelical}.

As in Sec.~\ref{sec:MDNL_MHDSP}, expressions in Eqs.~\eqref{eq:InflowVelocityInTermsOfBXiM}, \eqref{eq:MDNL_MHVSP_Vel_InTermsOfBXiM}, \eqref{eq:MDNL_MHVSP_GammaAndReM}, \eqref{eq:MDNL_MHVSP_rdissAndS}, \eqref{eq:MDNL_MHVSP_Ratios}, \eqref{eq:Justify_Dissipation6}, \eqref{eq:MDNL_MHDSP_Decaytime}, \eqref{eq:MDNL_MHDSP_Gamma}, and \eqref{eq:Smallness_DissipatedInflowVelocity_Drag} are derived independently of the initial conditions, $B_{\rm ini}$ and $\xi_{\rm M,ini}$.
In combination with the Hosking integral conservation, we obtain
\begin{align}
    B&=B_{\rm ini}^{\frac{8}{13}}\xi_{\rm M,ini}^{\frac{10}{13}}(\rho+p)^{\frac{5}{26}}\sigma^{\frac{5}{26}}\alpha^{\frac{5}{26}}\tau^{-\frac{5}{13}},
    \label{eq:B_MDNL_NHDSP}\\
    \xi_{\rm M}&=B_{\rm ini}^{\frac{4}{13}}\xi_{\rm M,ini}^{\frac{5}{13}}(\rho+p)^{-\frac{2}{13}}\sigma^{-\frac{2}{13}}\alpha^{-\frac{2}{13}}\tau^{\frac{4}{13}}.
    \label{eq:XiM_MDNL_NHDSP}
\end{align}

\subsubsection{Maximally helical and viscous fast reconnection regime (\ctext{5})}
\label{sec:MDNL_MHVfast}
In this regime, the magnetic field is maximally helical, shear viscosity dominates over drag force, and the Lundquist number is large.
In addition to Eq.~\eqref{eq:MDNL_condition},
\begin{align}
    \vert\epsilon\vert=1,\quad
    r_{\rm diss}(\xi_{\rm K})\ll1,\quad
    S\gg S_{\rm c}
    \label{eq:MHVfast_condition}
\end{align}
are satisfied.
The mechanism that governs the evolution of the system is no longer the Sweet--Parker reconnection because the current sheet becomes unstable to form plasmoids that are connected by smaller current sheets \cite{1986PhFl...29.1520B,2005PhRvL..95w5003L,2007PhPl...14j0703L,ji2011phase}.
The ratio of dissipation terms may be evaluated at the width of the smallest current sheet involved in the mechanism, $\delta_{\rm c}$, which is explained in this section.
Such a reconnection regime is referred to as the fast magnetic reconnection \cite{2022JPlPh..88e1501S,Hosking+22}.

\begin{figure}[thbp]\begin{center}
    \begin{tabular}{cc}
        \begin{minipage}[h]{0.60\hsize}
            \includegraphics[keepaspectratio, width=0.83\textwidth]{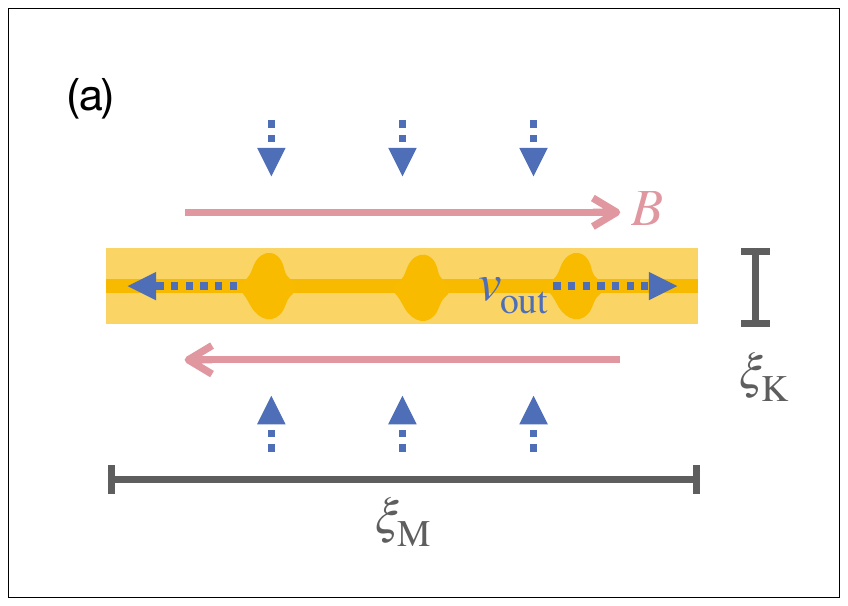}
        \end{minipage}&
        \begin{minipage}[h]{0.60\hsize}
            \hspace{-18mm}
            \includegraphics[keepaspectratio, width=0.83\textwidth]{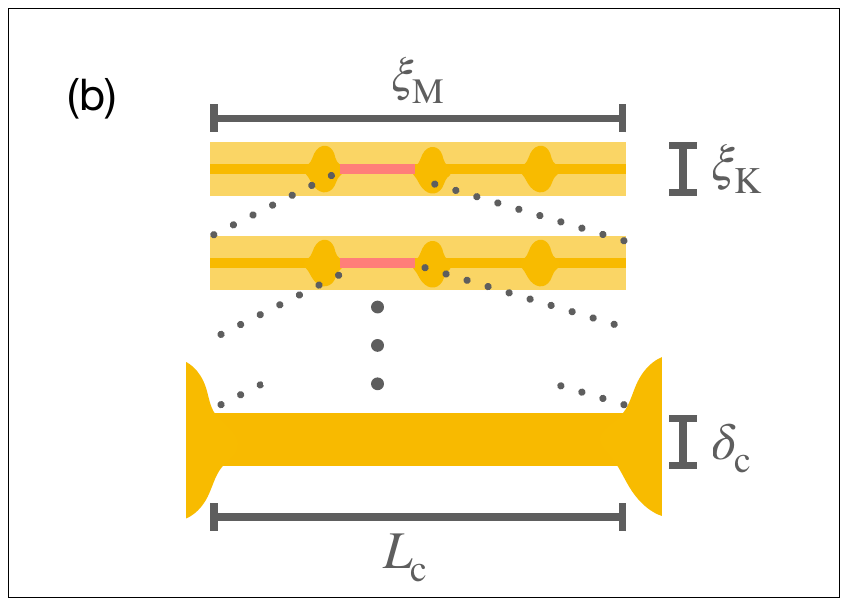}
        \end{minipage}\\
        \begin{minipage}[h]{0.57\hsize}\vspace{0mm}
            \hspace{40mm}
            \includegraphics[keepaspectratio, width=0.83\textwidth]{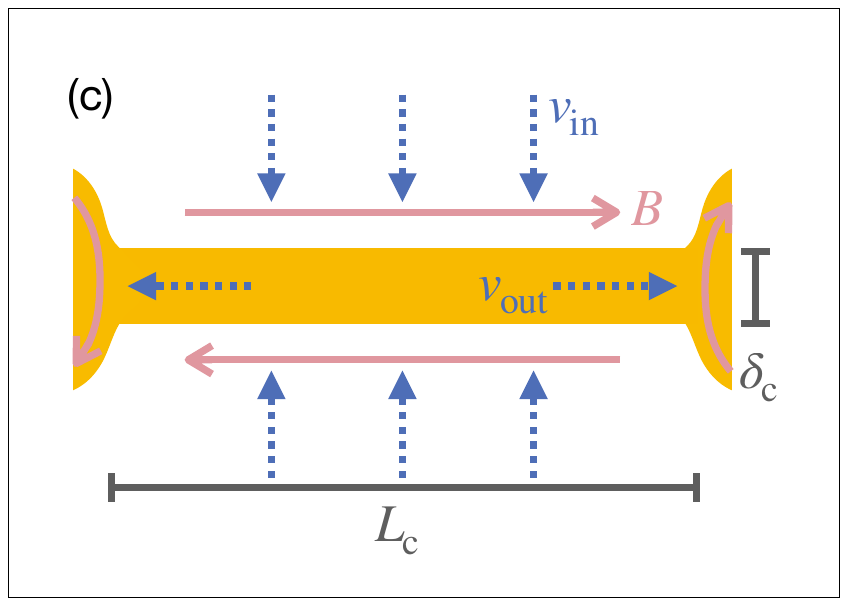}
        \end{minipage}&
    \end{tabular}
    \caption{\label{fig:fastReconnection} The structure of current sheets of the fast reconnection. 
    (a) Sheets in the lowest hierarchy (orange-shaded region) determine the magnetic and kinetic coherence lengths. (b) The sheets have a hierarchical structure, in which each sheet has plasmoids (dark blobs) and smaller sheets in the next hierarchy until the size reaches down to the critical length, $L_{\rm c}$. (c) The critical sheets are similar to the Sweet--Parker current sheet because they do not suffer from the plasmoid instability.}\end{center}
\end{figure}

Nevertheless, a similar analysis to the one in Sec.~\ref{sec:MDNL_MHVSP} is still possible by applying the same procedure to the so-called critical current sheets \cite{ji2011phase,2022JPlPh..88e1501S,Hosking+22}.
We model the system as depicted in Fig.~\ref{fig:fastReconnection}, following Refs.~\cite{2001EP&S...53..473S,ji2011phase}.
Macroscopically, the system looks similar to the Sweet--Parker one.
We have a sheet-like structure of thickness $\xi_{\rm K}$ and area $\xi^2_{\rm M}$ (orange-shaded region in panel (a)) in each unit patch of volume $\xi^3_{\rm M}$.
Magnetic field (red solid arrows) is anti-parallel on both sides of the structure, the fluid is inflowing onto the faces of the current sheet (vertical blue-dotted arrows), and it is outflowing from the narrow sides of the sheet (horizontal blue-dotted arrows).
However, the sheet has a hierarchical structure inside it (panel (b)).
The largest sheet in the hierarchy is of the size $\xi^2_{\rm M}\times \xi_{\rm K}$ (the top orange-shaded region).
It has plasmoids (dark blobs) inside it and is separated into smaller sheets of the second hierarchy (the middle orange-shaded region).
The largest sheet is unstable and plasmoids inside it are continuously formed and ejected by the outflow. 
Similarly, each sheet in a hierarchy has smaller sheets in the next hierarchy because of the instability.
If we assign Lundquist numbers for sheets in the $i$-th hierarchy as
\begin{align}
    S_i:=\sigma v_{\rm out} L_i,
    \label{eq:ithLundquistNumber_Def}
\end{align}
where $L_i$ is the typical length of the sheets, the hierarchical structure continues to the $(i+1)$-th hierarchy if $S_i\gg S_{\rm c}$.
However, for a given outflow velocity $v_{\rm out}$, the hierarchy terminates at a length scale $L_{\rm c}$ because the condition
\begin{align}
    S_{\rm c}=\sigma v_{\rm out} L_{\rm c},
    \label{eq:CriticalLundquistNumber}
\end{align}
determines the size of the smallest sheets, which we refer to as the critical current sheet (the bottom orange-shaded region). 
If we parametrize the thickness of the critical current sheets as $\delta_{\rm c}$ and the inflow velocity onto them as $v_{\rm in}$, a parallel discussion to the Sweet--Parker one holds (panel (c)).

We assume that the reconnecting magnetic field $B$ and the outflow velocity $v_{\rm out}$ are common in the hierarchy of the current sheets.
From Eqs.~\eqref{eq:Def_LundquistNumber} and \eqref{eq:CriticalLundquistNumber}, we obtain the critical size of the current sheets
\begin{align}
    L_{\rm c}=\dfrac{S_{\rm c}}{S}\xi_{\rm M}.
    \label{eq:CriticalSheetLength}
\end{align}
The consideration about mass conservation and (quasi-)stationarity in Sec.~\ref{sec:MDNL_MHVSP} for the Sweet--Parker current sheet applies to the critical sheets, and Eqs.~\eqref{eq:MassConservation}, \eqref{eq:Stationarity}, and \eqref{eq:OutflowVelocity} analogously hold by replacing $\xi_{\rm M}$ with $L_{\rm c}$ and $\xi_{\rm K}$ with $\delta_{\rm c}$.
Conditions about mass conservation and (quasi-)stationarity of the critical sheets imply
\begin{align}
    v_{\rm in}L_{\rm c}=v_{\rm out}\delta_{\rm c},\quad
    \dfrac{v_{\rm in}B}{\delta_{\rm c}}=\dfrac{B}{\sigma\delta^2_{\rm c}},
    \label{eq:CriticalSheet_Conditions}
\end{align}
from which we obtain an expression of the outflow velocity 
\begin{align}
    v_{\rm out}=\dfrac{L_{\rm c}}{\sigma \delta_{\rm c}^2},
    \label{eq:OutflowVelocity_CriticalSheet}
\end{align}
where we have substituted Eq.~\eqref{eq:CriticalSheetLength} in the second equality.
We relate this with the magnetic field strength by considering macroscopic energy budget, Eq.~\eqref{eq:Dissipation_3}.
In addition, we have conditions about the decay time scale, Eq.~\eqref{eq:Decaytime_Rec}, and the dilution of kinetic energy, Eq.~\eqref{eq:Dilution}.

Before solving them together with the helicity conservation, Eq.~\eqref{eq:Constraint_MaximallyHelical}, let us estimate the kinetic coherence length $\xi_{\rm K}$.
We assume that the aspect ratio of the unstable sheets is determined as if they are stable Sweet--Parker ones.
In particular, for the lowest hierarchy, we employ analogues of Eqs.~\eqref{eq:MassConservation} and \eqref{eq:Stationarity} with an unknown inflow velocity that can be different from the inflow velocity onto the critical sheets, and hence we obtain Eq.~\eqref{eq:OutflowVelocity}.
Then the kinetic coherence length can be expressed as \cite{ji2011phase}
\begin{align}
    \xi_{\rm K}=S^{-\frac{1}{2}}\xi_{\rm M}.
    \label{eq:AspectRatio_Fast}
\end{align}
If we express $\delta_{\rm c}$ in terms of $\xi_{\rm K}$,
\begin{align}
    \delta_{\rm c}
    =S_{\rm c}^{-\frac{1}{2}}L_{\rm c}
    =\sqrt{\dfrac{S_{\rm c}}{S}}\xi_{\rm K}.
    \label{eq:CriticalSheetWidth}
\end{align}
From Eqs.~\eqref{eq:CriticalSheetLength} and \eqref{eq:CriticalSheetWidth}, it is clear that, when $S=S_{\rm c}$, $L_{\rm c}=\xi_{\rm M}$ and $\delta_{\rm c}=\xi_{\rm K}$ hold.
Then, the solution reduces to the one in Sec.~\ref{sec:MDNL_MHVSP} because the difference of the analyses is just the replacement of $\xi_{\rm M}$ with $L_{\rm c}$ and $\xi_{\rm K}$ with $\delta_{\rm c}$.

Now let us solve the set of equations, \eqref{eq:CriticalSheet_Conditions}, \eqref{eq:Dissipation_3}, \eqref{eq:Decaytime_Rec}, \eqref{eq:Dilution}, and \eqref{eq:Constraint_MaximallyHelical}.
Since we are interested in finding an expression of $\xi_{\rm K}$ rather than $\delta_{\rm c}$, we replace Eq.~\eqref{eq:CriticalSheet_Conditions} with Eq.~\eqref{eq:AspectRatio_Fast} and
\begin{align}
    v_{\rm in}
    &=S^{-\frac{1}{2}}_{\rm c}v_{\rm out},
    \label{eq:CriticalInflowVelocity}
\end{align}
which is derived from Eqs.~\eqref{eq:CriticalSheet_Conditions} and \eqref{eq:CriticalLundquistNumber}.

We solve the set of equations step by step as we did in Sec.~\ref{sec:MDNL_MHVSP}.
First, we use the condition about the decay time scale, Eq.~\eqref{eq:Decaytime_Rec}, to obtain Eq.~\eqref{eq:InflowVelocityInTermsOfBXiM}.
We substitute this expression into Eq.~\eqref{eq:CriticalInflowVelocity} to obtain
\begin{align}
    v_{\rm out}=S_{\rm c}^{\frac{1}{2}}\tau^{-1}\xi_{\rm M}.
    \label{eq:OutflowVelocity_Fast}
\end{align}
Then the Lundquist number becomes $S=S_{\rm c}^{\frac{1}{2}}\sigma\tau^{-1}\xi^2_{\rm M}$, and we obtain expressions of the kinetic coherence length $\xi_{\rm K}$, by using Eq.~\eqref{eq:AspectRatio_Fast}, and of the typical velocity $v$, by considering the energy dilution, Eq.~\eqref{eq:Dilution}.
We obtain
\begin{align}
    \xi_{\rm K}=S_{\rm c}^{-\frac{1}{4}}\sigma^{-\frac{1}{2}}\tau^{\frac{1}{2}},\quad
    v=S_{\rm c}^{\frac{3}{8}}\sigma^{-\frac{1}{4}}\tau^{-\frac{3}{4}}\xi_{\rm M}^{\frac{1}{2}}.
    \label{eq:MDNL_MHVfast_Vel_IntermsOfBXiM}
\end{align}
We express the order parameters in terms of $B$ and $\xi_{\rm M}$ as well, by substituting Eqs.~\eqref{eq:InflowVelocityInTermsOfBXiM}, \eqref{eq:OutflowVelocity_Fast}, and \eqref{eq:MDNL_MHVfast_Vel_IntermsOfBXiM} into the definitions.
The energy ratio, the magnetic Reynolds number, and the ratio of dissipation terms become
\begin{align}\hspace{-3mm}
    \Gamma&=(\rho+p)S_{\rm c}^{\frac{3}{4}}\sigma^{-\frac{1}{2}}\tau^{-\frac{3}{2}}B^{-2}\xi_{\rm M},\quad
    {\rm Re}_{\rm M}
    =\left(\dfrac{\xi_{\rm M}}{S_{\rm c}^{-\frac{1}{4}}\sigma^{-\frac{1}{2}}\tau^{\frac{1}{2}}}\right)^{\frac{5}{2}},
    \notag\\
    &r_{\rm diss}(\delta_{\rm c})
    =\left(\dfrac{\xi_{\rm M}}{\sigma^{-1}\eta^{-\frac{1}{2}}\alpha^{\frac{1}{2}}\tau}\right)^{-2},\quad
    r_{\rm diss}(\xi_{\rm K})
    =S_{\rm c}^{-\frac{1}{2}}{\rm Pr}_{\rm M}^{-1}\alpha\tau,
    \label{eq:MDNL_MHVfast_GammaAndReM}
\end{align}
in terms of which the Lundquist number has the same expression as in Sec.~\ref{sec:MDNL_MHVSP},
\begin{align}
    S
    ={\rm Re}_{\rm M}^{\frac{4}{5}}.
    \label{eq:MDNL_MHVfast_rdissAndS}
\end{align}
The largeness of the aspect ratio of the sheets and the smallness of the macroscopic inflow velocity are guaranteed because
\begin{align}
    \dfrac{\xi_{\rm M}}{\xi_{\rm K}}\left(=S^{\frac{1}{2}}\right)\gg \dfrac{L_{\rm c}}{\delta_{\rm c}}=S_{\rm c}^{\frac{1}{2}}\gg 1,\quad
    \dfrac{(S_{\rm c}/S)^{\frac{1}{2}}v_{\rm in}}{v}=S^{-\frac{1}{4}}\ll1
    \label{eq:MDNL_MHVfast_Ratios}
\end{align}
in this regime.
Next, we substitute the expression of the outflow velocity, Eq.~\eqref{eq:OutflowVelocity_Fast}, into Eq.~\eqref{eq:Dissipation_3}, to obtain \cite{2022JPlPh..88e1501S,Hosking+22}
\begin{align}
    S_{\rm c}^{\frac{1}{2}}(\rho+p)^{\frac{1}{2}}{\rm Pr}_{\rm M}^{\frac{1}{2}}B^{-1}\xi_{\rm M}=\tau.
    \label{eq:MDNL_MHVfast_Decaytime}
\end{align}
By using this, we can relate $\Gamma$ with the magnetic Reynolds number in the same way as Eq.~\eqref{eq:MDNL_MHVSP_Gamma} in Sec.~\ref{sec:MDNL_MHVSP}.
Finally, we solve Eqs.~\eqref{eq:MDNL_MHVfast_Decaytime} and \eqref{eq:Constraint_MaximallyHelical} to obtain
\begin{align}
    B&=B_{\rm ini}^{\frac{2}{3}}\xi_{\rm ini}^{\frac{1}{3}}S_{\rm c}^{\frac{1}{6}}(\rho+p)^{\frac{1}{6}}\sigma^{\frac{1}{6}}\eta^{\frac{1}{6}}\tau^{-\frac{1}{3}},\label{eq:MDNL_MHVfast_B}\\
    \xi_{\rm M}&=B_{\rm ini}^{\frac{2}{3}}\xi_{\rm ini}^{\frac{1}{3}}S_{\rm c}^{-\frac{1}{3}}(\rho+p)^{-\frac{1}{3}}\sigma^{-\frac{1}{3}}\eta^{-\frac{1}{3}}\tau^{\frac{2}{3}}.\label{eq:MDNL_MHVfast_XiM}
\end{align}

We should be careful about the dissipation of the inflow \cite{Hosking+22}. The discussion here is parallel to the one in Sec.~\ref{sec:MDNL_MHVSP}. 
Let us consider the region away from the current sheet and suppose that the inflow velocity $v'_{\rm in}$ is determined by the energy dissipation.
In such a region, the energy injection from the magnetic field should balance with the energy dissipation \cite{Banerjee+04}.
By approximating the relevant terms in Navier--Stokes equation \eqref{eq:Navier--Stokes equation} as we do in Eq.~\eqref{eq:MDNL_MHVSP_EstimateInNSE} and balancing these terms, we obtain the same estimate of the inflow velocity as Eq.~\eqref{eq:Smallness_DissipatedInflowVelocity}, which implies $v'_{\rm in}\gg v_{\rm in}$ and supports our assumption that the rate-determining process is magnetic reconnection.

\subsubsection{Non-helical and viscous fast reconnection regime (\ctext{6})}
\label{sec:MDNL_NHVfast}
In this regime, the magnetic field is non-helical, shear viscosity dominates over drag force, and the Lundquist number is large.
In addition to Eq.~\eqref{eq:MDNL_condition},
\begin{align}
    \epsilon=0,\quad
    r_{\rm diss}(\xi_{\rm K})\ll1,\quad
    S\gg S_{\rm c}
    \label{eq:NHVfast_condition}
\end{align}
are satisfied.
The only difference from Sec.~\ref{sec:MDNL_MHVfast} is that the constraint from the Hosking integral conservation \eqref{eq:Constraint_NonHelical}, instead of the helicity conservation, Eq.~\eqref{eq:Constraint_MaximallyHelical}.

By using the same equations as the ones in Sec.~\ref{sec:MDNL_MHVfast} except the helicity conservation, expressions in Eqs.~\eqref{eq:InflowVelocityInTermsOfBXiM}, \eqref{eq:OutflowVelocity_Fast}, \eqref{eq:MDNL_MHVfast_Vel_IntermsOfBXiM}, \eqref{eq:MDNL_MHVfast_GammaAndReM}, \eqref{eq:MDNL_MHVfast_rdissAndS}, \eqref{eq:MDNL_MHVfast_Ratios}, \eqref{eq:MDNL_MHVfast_Decaytime}, \eqref{eq:MDNL_MHVSP_Gamma}, and \eqref{eq:Smallness_DissipatedInflowVelocity} are derived independently of the initial conditions, $B_{\rm ini}$ and $\xi_{\rm M,ini}$.
In combination with the Hosking integral conservation, we obtain
\begin{align}
    B&=B_{\rm ini}^{\frac{4}{9}}\xi_{\rm ini}^{\frac{5}{9}}S_{\rm c}^{\frac{5}{18}}(\rho+p)^{\frac{5}{18}}\sigma^{\frac{5}{18}}\eta^{\frac{5}{18}}\tau^{-\frac{5}{9}},
    \label{eq:B_MDNL_NHVfast}\\
    \xi_{\rm M}&=B_{\rm ini}^{\frac{4}{9}}\xi_{\rm ini}^{\frac{5}{9}}S_{\rm c}^{-\frac{2}{9}}(\rho+p)^{-\frac{2}{9}}\sigma^{-\frac{2}{9}}\eta^{-\frac{2}{9}}\tau^{\frac{4}{9}}.
    \label{eq:XiM_MDNL_NHVfast}
\end{align}

\subsubsection{Maximally helical and dragged fast reconnection regime (\ctext{7})}
\label{sec:MDNL_MHDfast}
In this regime, the magnetic field is maximally helical, drag force dominates over shear viscosity, and the Lundquist number is large.
In addition to Eq.~\eqref{eq:MDNL_condition},
\begin{align}
    \vert\epsilon\vert=1,\quad
    r_{\rm diss}(\delta_{\rm c})\gg1,\quad
    S\gg S_{\rm c}
    \label{eq:MHDfast_condition}
\end{align}
are satisfied.
The mechanism is basically the fast reconnection described in Sec.~\ref{sec:MDNL_MHVfast}, but with drag force rather than shear viscosity.

The discussion in Sec.~\ref{sec:MDNL_MHVfast} is almost valid except for the estimate of the energy budget.
We have, macroscopically, Eq.~\eqref{eq:Dissipation_6}, assuming Eq.~\eqref{eq:Justify_Dissipation6}.
Then, we need to solve the set of equations, \eqref{eq:AspectRatio_Fast}, \eqref{eq:CriticalInflowVelocity}, \eqref{eq:Dissipation_6}, \eqref{eq:Decaytime_Rec}, \eqref{eq:Dilution}, and \eqref{eq:Constraint_MaximallyHelical}.
As in Sec.~\ref{sec:MDNL_MHVfast}, the expressions in Eqs.~\eqref{eq:InflowVelocityInTermsOfBXiM}, \eqref{eq:OutflowVelocity_Fast}, \eqref{eq:MDNL_MHVfast_Vel_IntermsOfBXiM}, \eqref{eq:MDNL_MHVfast_GammaAndReM}, \eqref{eq:MDNL_MHVfast_rdissAndS}, and \eqref{eq:MDNL_MHVfast_Ratios} hold.
The only difference comes from Eq.~\eqref{eq:Dissipation_6}, which becomes
\begin{align}
    S_{\rm c}^{\frac{1}{2}}(\rho+p)\alpha B^{-2}\xi_{\rm M}^2=\tau,
    \label{eq:MDNL_MHDfast_DecayTime}
\end{align}
by substituting the expression of the outflow, Eq.~\eqref{eq:OutflowVelocity_Fast}.
The energy ratio $\Gamma$ can be expressed in the same way as Eq.~\eqref{eq:MDNL_MHDSP_Gamma}.
We solve Eq.~\eqref{eq:MDNL_MHDfast_DecayTime} with the helicity conservation, Eq.~\eqref{eq:Constraint_MaximallyHelical}, and the solutions are
\begin{align}
    B&=B_{\rm ini}^{\frac{2}{3}}\xi_{\rm ini}^{\frac{1}{3}}S_{\rm c}^{\frac{1}{12}}(\rho+p)^{\frac{1}{6}}\alpha^{\frac{1}{6}}\tau^{-\frac{1}{6}},
    \label{eq:B_MDNL_MHDfast}\\
    \xi_{\rm M}&=B_{\rm ini}^{\frac{2}{3}}\xi_{\rm ini}^{\frac{1}{3}}S_{\rm c}^{-\frac{1}{6}}(\rho+p)^{-\frac{1}{3}}\alpha^{-\frac{1}{3}}\tau^{\frac{1}{3}}
    \label{eq:XiM_MDNL_MHDfast}.
\end{align}

As the final remark in this section, we can discuss how significant the effect of the drag force on the inflow is in exactly the same manner as in Sec.~\ref{sec:MDNL_MHDSP}.
The hierarchy between $v_{\rm in}$ and $v'_{\rm in}$, Eq.~\eqref{eq:Smallness_DissipatedInflowVelocity_Drag} implies that the rate-determining process is the reconnection process.

\subsubsection{Non-helical and dragged fast reconnection regime (\ctext{8})}
\label{sec:MDNL_NHDfast}
In this regime, the magnetic field is non-helical, drag force dominates over shear viscosity, and the Lundquist number is large.
In addition to Eq.~\eqref{eq:MDNL_condition},
\begin{align}
    \epsilon=0,\quad
    r_{\rm diss}(\delta_{\rm c})\gg1,\quad
    S\gg S_{\rm c}
    \label{eq:NHDfast_condition}
\end{align}
are satisfied.
The only difference from Sec.~\ref{sec:MDNL_MHDfast} is that the constraint from the Hosking integral conservation \eqref{eq:Constraint_NonHelical}, instead of the helicity conservation, Eq.~\eqref{eq:Constraint_MaximallyHelical}.

As in Sec.~\ref{sec:MDNL_MHDfast}, expressions in Eqs.~\eqref{eq:InflowVelocityInTermsOfBXiM}, \eqref{eq:OutflowVelocity_Fast}, \eqref{eq:MDNL_MHVfast_Vel_IntermsOfBXiM}, \eqref{eq:MDNL_MHVfast_GammaAndReM}, \eqref{eq:MDNL_MHVfast_rdissAndS}, \eqref{eq:MDNL_MHVfast_Ratios}, \eqref{eq:Justify_Dissipation6}, \eqref{eq:MDNL_MHDfast_DecayTime}, \eqref{eq:MDNL_MHDSP_Gamma}, and \eqref{eq:Smallness_DissipatedInflowVelocity_Drag} are derived independently of the initial conditions, $B_{\rm ini}$ and $\xi_{\rm M,ini}$.
In combination with the Hosking integral conservation, we obtain
\begin{align}
    B&=B_{\rm ini}^{\frac{4}{9}}\xi_{\rm ini}^{\frac{5}{9}}S_{\rm c}^{\frac{5}{36}}(\rho+p)^{\frac{5}{18}}\alpha^{\frac{5}{18}}\tau^{-\frac{5}{18}},
    \label{eq:B_MDNL_NHDfast}\\
    \xi_{\rm M}&=B_{\rm ini}^{\frac{4}{9}}\xi_{\rm ini}^{\frac{5}{9}}S_{\rm c}^{-\frac{1}{9}}(\rho+p)^{-\frac{2}{9}}\alpha^{-\frac{2}{9}}\tau^{\frac{2}{9}}.
    \label{eq:XiM_MDNL_NHDfast}
\end{align}

\subsection{Magnetically dominated and linear regimes}
\label{sec:MDL}
In this branch of regimes, magnetic energy dominates over kinetic energy, but the magnetic Reynolds number is small, and magnetic reconnection is not taking place.
Conditions
\begin{align}
    \Gamma\ll1,\quad
    {\rm Re}_{\rm M}\ll1
    \label{eq:MDL_condition}
\end{align}
are satisfied.
In these regimes, the magnetic field is dominant and excites fluid motion.
However, dissipation of the velocity field inhibits a significant excitation that leads to a large kinetic Reynolds number.
Then, the dynamics of the fluid motion is linear and proceeds in a quasi-stationary way \cite{Jedamzik:1996wp,Banerjee+04}.
Although the system is characterized by four parameters, $B$, $\xi_{\rm M}$, $v$, and $\xi_{\rm K}$, since the energy is injected into the velocity field by the magnetic field at the magnetic coherence length, we approximate
\begin{align}
    \xi_{\rm K}=\xi_{\rm M}.
    \label{eq:MDL_XiK}
\end{align}
We extend the analysis in the literature \cite{Banerjee+04}, by incorporating the conservation of the Hosking integral, to find three additional equations for each regime \cite{Uchida:2022vue}.
We have several regimes according to the helicity fraction $\epsilon$ and the ratio of dissipation terms $r_{\rm diss}$, for each of which we find formulae that describe the evolution of the system.

\subsubsection{Maximally helical and viscous regime (\ctext{9})}
\label{sec:MDL_MHV}
In this regime, the magnetic field is maximally helical, and shear viscosity dominates over drag force.
In addition to Eq.~\eqref{eq:MDL_condition},
\begin{align}
    \vert\epsilon\vert=1,\quad
    r_{\rm diss}(\xi_{\rm M})\ll1
    \label{eq:MDL_MHV_condition}
\end{align}
are assumed to hold.

Stationarity implies the balance between terms in the Navier--Stokes equation \eqref{eq:Navier--Stokes equation} \cite{Banerjee+04},
\begin{align}
    \dfrac{B^2}{(\rho+p)\xi_{\rm M}}=\dfrac{\eta v}{\xi_{\rm K}^2},
    \label{eq:MDL_V_Stationarity}
\end{align}
just like the evaluation of $v'_{\rm in}$ in Eq.~\eqref{eq:MDNL_MHDSP_EstimateInNSE}.
As is justified later, we neglected the advection term by assuming a small kinetic Reynolds number
\begin{align}
    {\rm Re}_{\rm K}\ll1.
    \label{eq:SmallKineticReynoldsNumber}
\end{align}
This assumption implies that the characteristic time scale of the process in this regime is the eddy turnover time at the kinetic coherence length, $\tau_{\rm eddy}:=\xi_{\rm K}/v$.
This estimate is supported by a dedicated study \cite{Jedamzik:1996wp}.
Then, the condition on decay time scales, Eq.~\eqref{eq:TimeScaleCondition_general}, implies \cite{Banerjee+04}
\begin{align}
    \tau_{\rm eddy}=\tau.
    \label{eq:Decaytime_Eddy}
\end{align}

To summarize, we are going to solve Eqs.~\eqref{eq:MDL_XiK}, \eqref{eq:MDL_V_Stationarity}, and \eqref{eq:Decaytime_Eddy} with the condition of helicity conservation \eqref{eq:Constraint_MaximallyHelical}.
By using Eqs.~\eqref{eq:MDL_XiK} and \eqref{eq:MDL_V_Stationarity}, we obtain
\begin{align}
    \Gamma={\rm Re}_{\rm K}={\rm Pr}_{\rm M}^{-1}{\rm Re}_{\rm M},
    \label{eq:Relation_MDL_MHV}
\end{align}
which guarantees the assumption of a small kinetic Reynolds number, Eq.~\eqref{eq:SmallKineticReynoldsNumber}.
Equation \eqref{eq:Decaytime_Eddy} can be rewritten as
\begin{align}
    (\rho+p)\eta B^{-2}=\tau.
    \label{eq:Decaytime_MDL_V}
\end{align}
Together with the constraint by the helicity conservation, Eq.~\eqref{eq:Constraint_MagneticHelicity}, we obtain 
\begin{align}
    B&=(\rho+p)^{\frac{1}{2}}\eta^{\frac{1}{2}}\tau^{-\frac{1}{2}},
    \label{eq:B_MDL_MHV}\\
    \xi_{\rm M}&=B_{\rm ini}^2\xi_{\rm M,ini}(\rho+p)^{-1}\eta^{-1}\tau
    \label{eq:XiM_MDL_MHV}
\end{align}
for the magnetic field and $v=\tau^{-1}\xi_{\rm M},\,\xi_{\rm K}=\xi_{\rm M}$ for the velocity field.
If we compare the terms in the left-hand side of the induction equation, Eq.~\eqref{eq:Faraday's induction equation}, they balance with the same magnitude $\sim (\rho+p)^{1/2}\eta^{1/2}\tau^{-3/2}$, implying that the magnetic field is not dissipated by the resistivity, although the magnetic Reynolds number is small \cite{Banerjee+04}.
It suggests that, even when Eq.~\eqref{eq:ConditionToBelieveHelicityConservation} is violated, magnetic helicity is approximately conserved in this regime, where the magnetic field takes a configuration such that
\begin{align}
    \vert\nabla^2{\bm B}\vert\ll \dfrac{B}{\xi^2_{\rm M}}.
    \label{eq:harmonic}
\end{align}

\subsubsection{Non-helical and viscous regime (\ctext{10})}
\label{sec:MDL_NHV}
In this regime, the magnetic field is non-helical, and shear viscosity dominates over drag force.
In addition to Eq.~\eqref{eq:MDL_condition},
\begin{align}
    \epsilon=0,\quad
    r_{\rm diss}(\xi_{\rm M})\ll1
    \label{eq:MDL_NHV_condition}
\end{align}
are satisfied.
The only difference from Sec.~\ref{sec:MDL_MHV} is that the constraint from the Hosking integral conservation \eqref{eq:Constraint_NonHelical}, instead of the helicity conservation Eq.~\eqref{eq:Constraint_MaximallyHelical}.

Since the equations in Sec.~\ref{sec:MDL_MHV} except the helicity conservation are the same as the ones in Sec.~\ref{sec:MDL_MHV}, we obtain Eqs.~\eqref{eq:Relation_MDL_MHV}, \eqref{eq:SmallKineticReynoldsNumber}, and \eqref{eq:Decaytime_MDL_V}.
We solve Eq.~\eqref{eq:Decaytime_MDL_V} and the Hosking integral conservation to obtain
\begin{align}
    B&=(\rho+p)^{\frac{1}{2}}\eta^{\frac{1}{2}}\tau^{-\frac{1}{2}},
    \label{eq:B_MDL_NHV}\\
    \xi_{\rm M}&=B_{\rm ini}^{\frac{4}{5}}\xi_{\rm M,ini}(\rho+p)^{-\frac{2}{5}}\eta^{-\frac{2}{5}}\tau^{\frac{2}{5}}
    \label{eq:XiM_MDL_NHV}
\end{align}
for the magnetic field, and $v=\tau^{-1}\xi_{\rm M},\,\xi_{\rm K}=\xi_{\rm M}$ for the velocity field.
Note that the magnetic field is not dissipated by the resistivity, although the magnetic Reynolds number is small \cite{Banerjee+04}.

\subsubsection{Maximally helical and dragged regime (\ctext{11})}
\label{sec:MDL_MHD}
In this regime, the magnetic field is maximally helical, and drag force dominates over shear viscosity.
In addition to Eq.~\eqref{eq:MDL_condition},
\begin{align}
    \vert\epsilon\vert=1,\quad
    r_{\rm diss}(\xi_{\rm M})\gg1
    \label{eq:MDL_MHD_condition}
\end{align}
are satisfied.

The discussion in Sec.~\ref{sec:MDL_MHV} holds except the estimate of energy dissipation in Eq.~\eqref{eq:MDL_V_Stationarity}.
Stationarity implies the balance between terms in the Navier--Stokes equation \eqref{eq:Navier--Stokes equation} \cite{Banerjee+04},
\begin{align}
    \dfrac{B^2}{(\rho+p)\xi_{\rm M}}=\alpha v,
    \label{eq:MDL_D_Stationarity}
\end{align}
as we did in the evaluation of $v'_{\rm in}$ in Eq.~\eqref{eq:MDNL_MHDSP_EstimateInNSE}.
We assume a small kinetic Reynolds number, and then the condition on decay time scales becomes Eq.~\eqref{eq:Decaytime_Eddy}.
We are going to solve Eqs.~\eqref{eq:MDL_XiK}, \eqref{eq:MDL_D_Stationarity}, and \eqref{eq:Decaytime_Eddy} with the condition of the helicity conservation \eqref{eq:Constraint_MaximallyHelical}.
By using Eqs.~\eqref{eq:MDL_XiK}, \eqref{eq:MDL_D_Stationarity}, we obtain
\begin{align}
    \Gamma={\rm Re}_{\rm K}={\rm Pr}_{\rm M}^{-1}r_{\rm diss}(\xi_{\rm M})^{-1}{\rm Re}_{\rm M}
    \label{eq:Relation_MDL_MHD}
\end{align}
which guarantees the assumption of a small kinetic Reynolds number, Eq.~\eqref{eq:SmallKineticReynoldsNumber}.
Equation \eqref{eq:Decaytime_Eddy} can be rewritten as
\begin{align}
    (\rho+p)\alpha B^{-2}\xi_{\rm M}^2=\tau
    \label{eq:Decaytime_MDL_D}
\end{align}
Together with the constraint by the helicity conservation, Eq.~\eqref{eq:Constraint_MagneticHelicity}, we obtain 
\begin{align}
    B&=B_{\rm ini}^{\frac{2}{3}}\xi_{\rm M,ini}^{\frac{1}{3}}(\rho+p)^{\frac{1}{6}}\alpha^{\frac{1}{6}}\tau^{-\frac{1}{6}},
    \label{eq:B_MDL_MHD}\\
    \xi_{\rm M}&=B_{\rm ini}^{\frac{2}{3}}\xi_{\rm M,ini}^{\frac{1}{3}}(\rho+p)^{-\frac{1}{3}}\alpha^{-\frac{1}{3}}\tau^{\frac{1}{3}}
    \label{eq:XiM_MDL_MHD}
\end{align}
for the magnetic field and $v=\tau^{-1}\xi_{\rm M},\,\xi_{\rm K}=\xi_{\rm M}$ for the velocity field.
Because of Eq.~\eqref{eq:Decaytime_Eddy}, again, the terms in the left-hand side of the induction equation, Eq.~\eqref{eq:Faraday's induction equation}, balance with the same magnitude.
Therefore, the magnetic field is not dissipated by the resistivity, although the magnetic Reynolds number is small \cite{Banerjee+04}.

\subsubsection{Non-helical and dragged regime (\ctext{12})}
\label{sec:MDL_NHD}
In this regime, the magnetic field is non-helical, and drag force dominates over shear viscosity.
In addition to Eq.~\eqref{eq:MDL_condition},
\begin{align}
    \epsilon=0,\quad
    r_{\rm diss}(\xi_{\rm M})\gg1
    \label{eq:MDL_NHD_condition}
\end{align}
are satisfied.
The only difference from Sec.~\ref{sec:MDL_MHD} is that the constraint from the Hosking integral conservation \eqref{eq:Constraint_NonHelical}, instead of the helicity conservation Eq.~\eqref{eq:Constraint_MaximallyHelical}.

Since the equations are the same as those in Sec.~\ref{sec:MDL_MHD} except the helicity conservation, we obtain Eqs.~\eqref{eq:Relation_MDL_MHD}, \eqref{eq:SmallKineticReynoldsNumber}, and \eqref{eq:Decaytime_MDL_D}.
We solve Eq.~\eqref{eq:Decaytime_MDL_D} and the Hosking integral conservation to obtain
\begin{align}
    B&=B_{\rm ini}^{\frac{4}{9}}\xi_{\rm M,ini}^{\frac{5}{9}}(\rho+p)^{\frac{5}{18}}\alpha^{\frac{5}{18}}\tau^{-\frac{5}{18}},
    \label{eq:B_MDL_NHD}\\
    \xi_{\rm M}&=B_{\rm ini}^{\frac{4}{9}}\xi_{\rm M,ini}^{\frac{5}{9}}(\rho+p)^{-\frac{2}{9}}\alpha^{-\frac{2}{9}}\tau^{\frac{2}{9}}
    \label{eq:XiM_MDL_NHD}
\end{align}
for the magnetic field, and $v=\tau^{-1}\xi_{\rm M},\,\xi_{\rm K}=\xi_{\rm M}$ for the velocity field.
Note that the magnetic field is not dissipated by the resistivity, although the magnetic Reynolds number is small \cite{Banerjee+04}.

\subsection{Kinetically dominated and maximally helical regime (\ctext{13})}
\label{sec:KDMH}
In this regime, kinetic energy initially dominates over magnetic energy, and the magnetic field is maximally helical.
Conditions
\begin{align}
    \Gamma_{\rm ini}\gtrsim1,\quad
    \vert\epsilon\vert=1
    \label{eq:KDMH_condition}
\end{align}
are satisfied.
In kinetically dominated regimes, numerical studies \cite{Brandenburg+17, 2021MNRAS.501.3074B} have found that dynamo at small scales excites magnetic fields, and almost equipartition
\begin{align}
    \Gamma\sim1,\quad
    \xi_{\rm M}\sim\xi_{\rm K}
    \label{eq:Equipartition}
\end{align}
is achieved quickly before entering into the scaling regime, even if $\Gamma_{\rm ini}\gg1$.
Hereafter, we suppose an approximate equipartition as the initial condition and classify the regimes according to the magnetic helicity fraction at this stage.

In the subsequent evolution, we have no reason to expect that Eq.~\eqref{eq:Equipartition} is maintained.
Rather, we expect that the system enters into magnetically dominant regimes \cite{Hosking+21}.
On one hand, we expect that the characteristic time scale is $\tau_{\rm eddy}$, provided that the velocity field governs the decay dynamics and that magnetic field energy is just fed by the small-scale dynamo.
On the other hand, the magnetic field can play a crucial role because it involves conserved quantities.
For the maximally helical magnetic field, magnetic helicity conservation \eqref{eq:Constraint_MaximallyHelical} prohibits too fast decay of the magnetic field.
Contrary to the magnetic field, the velocity field should decay faster, if the Saffman cross helicity integral,
\begin{align}
    I_{\rm C}:=\int d^3r \langle h_{\rm C}({\bm x})h_{\rm C}({\bm x}+{\bm r})\rangle,
    \label{eq:SaffmanCrossHelicityIntegral}
\end{align}
constrains the decay.
By dimensional analysis, the constraint becomes
\begin{align}
    B^2v^2\xi_{\rm M}^3={\rm const.}
    \label{eq:SaffmanCrossHelicityIntegral_Conservation}
\end{align}
This constraint, the magnetic helicity conservation Eq.~\eqref{eq:Constraint_MaximallyHelical}, and the decay time scale imply \cite{Hosking+22}
\begin{align}
    \Gamma\propto \tau^{-\frac{1}{2}},
\end{align}
from which we understand that the system unavoidably enters the magnetically dominant regimes (solid arrows in Fig.~\ref{fig:InitiallyKD}).
Therefore, for $\Gamma_{\rm ini}\sim 1$ with a slight dominance of kinetic energy, the system behaves as if it is initially magnetically dominant.

Let us comment on the use of the Saffman cross helicity integral.
First, because of the absence of a magnetogenesis scenario in which generation of cross helicity is the distinctive property, we suppose that the cross helicity is zero on average.
Then, the Saffman cross helicity integral is well-defined.
However, the conservation of local cross helicity, which guarantees the conservation of the Saffman cross helicity integral, is only marginal.
Nevertheless, it is slightly better conserved compared with the total energy because, in the Fourier space, we have
\begin{align}
    \dfrac{d h_{\rm C}(k)}{d\tau}= -(\eta k^2+\alpha +\sigma k^2)h_{\rm C}(k),\\
    \dfrac{d \rho_{\rm tot}(k)}{d\tau}= -2(\eta k^2+\alpha +\sigma k^2)\rho_{\rm tot}(k),
\end{align}
where $h_{\rm C}(k)$ is the Fourier component of $\langle h_{\rm C}({\bm x})\rangle$ and $\rho_{\rm tot}(k):=d\rho_{\rm M}/d\log k+d\rho_{\rm K}/d\log k$.
Finite dissipation terms imply exponential decay of both of the quantities, but the decay of the total energy is slightly faster.
Based on this consideration, we expect the analysis based on the conservation of the Saffman cross helicity integral can capture a qualitative behaviour of the system.

\subsection{Kinetically dominated and non-helical regimes}
\label{sec:KDNH}
In this branch of regimes, kinetic energy initially dominates over magnetic energy, and the magnetic field is non-helical.
Conditions
\begin{align}
    \Gamma_{\rm ini}\gtrsim1,\quad
    \epsilon=0
    \label{eq:KDNH_condition}
\end{align}
are satisfied.
As is explained in Sec.~\ref{sec:KDMH}, quasi-equipartition \eqref{eq:Equipartition} is achieved quickly before entering into the scaling regime.

\begin{figure}[t]\begin{center}
    \begin{minipage}[h]{0.70\hsize}
    \includegraphics[keepaspectratio, width=0.9\textwidth]{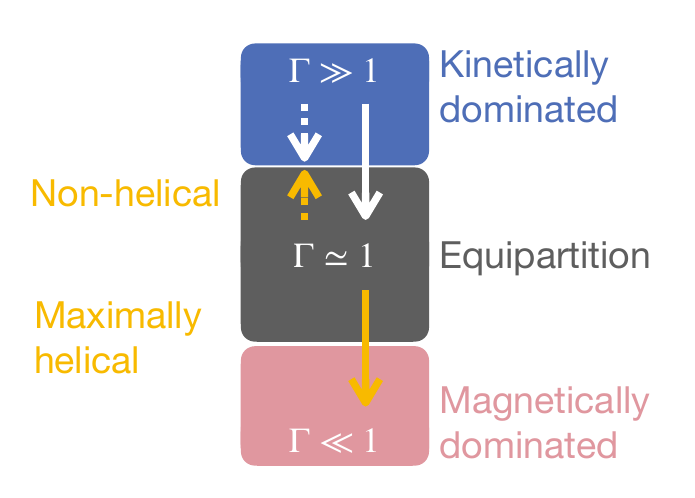}
    \end{minipage}
    \caption{\label{fig:InitiallyKD} The fate of kinetically dominated initial conditions. The small-scale dynamo supports magnetic fields slightly subdominant to the kinetic field (white arrows). If the magnetic field is maximally helical at equipartition, the system enters a magnetically dominant regime (orange solid arrow). If the magnetic field is non-helical at equipartition, the equipartition condition is maintained in the subsequent evolution (orange-dotted arrow).}\end{center}
\end{figure}

At this point, we have no reason to expect that Eq.~\eqref{eq:Equipartition} is stably maintained.
We adopt the eddy turnover time as the decay time scale and use the Hosking integral conservation \eqref{eq:Constraint_NonHelical} and Saffman cross helicity integral conservation \eqref{eq:SaffmanCrossHelicityIntegral_Conservation}.
Then we obtain
\begin{align}
    \Gamma\propto \tau^{\frac{12}{5}},
\end{align}
which implies the entrance into kinetically dominant regimes.
Then, it is reasonable to expect that the competition between small-scale dynamo and the above-explained negative feedback maintains the equipartition condition \eqref{eq:Equipartition} \cite{Hosking+21} (dotted arrows in Fig.~\ref{fig:InitiallyKD}), where the decay dynamics is regarded as a pure hydrodynamics.
In the subsequent scaling regimes, we assume Eqs.~\eqref{eq:Equipartition}, \eqref{eq:MDL_XiK}, and the conservation of kinetic energy at large scales, Eq.~\eqref{eq:Constraint_PureHD}.

In these regimes, Eq.~\eqref{eq:Constraint_NonHelical} is not necessarily satisfied.
However, it does not contradict the magnetic helicity conservation for two reasons.
First, the magnetic field is enhanced at small scales, where helicity dissipation by the term $\sigma^{-1}\nabla^2 {\bm B}$ is significant.
Second, the large velocity field can violate the Gaussianity of magnetic fields through the non-linear term in the induction equation.
Then, the estimate in Eq.~\eqref{eq:IHDimensionalAnalysis} is not as reliable as in the magnetically dominated regimes. 
Based on these considerations, we ignore Eq.~\eqref{eq:Constraint_NonHelical} in the following subsections.

\subsubsection{Turbulent regime (\ctext{14})}
\label{sec:KDNH_NL}
In this regime, the kinetic Reynolds number is large and the decay mechanism is the kinetic energy cascade, which corresponds to the situation discussed in Ref.~\cite{Banerjee+04}.
In addition to Eq.~\eqref{eq:KDNH_condition},
\begin{align}
    {\rm Re}_{\rm K}\gg1
\end{align}
is satisfied.
We assume the quasi-equipartition conditions, Eqs.~\eqref{eq:Equipartition} and \eqref{eq:MDL_XiK}, in the scaling regime.

We solve the conservation of kinetic energy at large scales, 
\begin{align}
    v^2\xi_{\text K}^{n+3}={\rm const.}\quad\text{in pure hydrodynamics,}
    \label{eq:Constraint_PureHD}
\end{align}
and the condition about the decay time, Eq.~\eqref{eq:Decaytime_Eddy}, to obtain \cite{Banerjee+04}
\begin{align}
    v= v_{\rm ini}^{\frac{2}{n+5}}\xi_{\rm K,ini}^{\frac{2(n+3)}{n+5}}\tau^{-\frac{n+3}{n+5}},
    \quad
    \xi_{\rm K}= v_{\rm ini}^{\frac{2}{n+5}}\xi_{\rm K,ini}^{\frac{2(n+3)}{n+5}}\tau^{\frac{2}{n+5}},
    \label{eq:XiK_KDNH_NL}
\end{align}
or equivalently
\begin{align}
    B&=B_{\rm ini}^{\frac{2}{n+5}}\xi_{\rm M, ini}^{\frac{2(n+3)}{n+5}}(\rho+p)^{\frac{n+3}{2n+10}}\tau^{-\frac{n+3}{n+5}},
    \label{eq:B_KDNH_NL}\\
    \xi_{\rm M}&= B_{\rm ini}^{\frac{2}{n+5}}\xi_{\rm M,ini}^{\frac{2(n+3)}{n+5}}(\rho+p)^{-\frac{1}{n+5}}\tau^{\frac{2}{n+5}},
    \label{eq:XiM_KDNH_NL}
\end{align}
in terms of the magnetic field.
These solutions imply the corresponding equation to Eq.~\eqref{eq:TimeScaleCondition_general} in this regime,
\begin{align}
    (\rho+p)^{\frac{1}{2}}B^{-1}\xi_{\rm M}=\tau.
    \label{eq:KDNH_NL_DecayTime}
\end{align}
This relation has been widely employed in the literature \cite{Banerjee+04,Durrer+13,Brandenburg+17,2015JCAP...05..054F}.

\subsubsection{Linear and viscous regime (\ctext{15})}
\label{sec:KDNH_LV}
In this regime, the kinetic Reynolds number is small and the decay is governed by the dissipation by shear viscosity at $\xi_{\rm K}$.
In addition to Eq.~\eqref{eq:KDNH_condition},
\begin{align}
    {\rm Re}_{\rm K}\ll1,\quad
    r_{\rm diss}(\xi_{\rm K})\ll1
\end{align}
are satisfied.
We assume quasi-equipartition conditions, Eqs.~\eqref{eq:Equipartition} and \eqref{eq:MDL_XiK}, in the scaling regime.

We solve the energy conservation at large scales, Eq.~\eqref{eq:Constraint_PureHD}, and the condition
\begin{align}
    \tau_\eta(\xi_{\rm K})=\tau
    \label{eq:Condition_DecayTimeScale_KDL}
\end{align}
for the decay time scale to obtain
\begin{align}
    v= v_{\rm ini}\xi_{\rm K,ini}^{\frac{n+3}{2}}\eta^{-\frac{n+3}{4}}\tau^{-\frac{n+3}{4}},\quad
    \xi_{\rm K}=\tau^{\frac{1}{2}}\eta^{\frac{1}{2}},
\end{align}
or equivalently
\begin{align}
    B&= B_{\rm ini}\xi_{\rm M,ini}^{\frac{n+3}{2}}\eta^{-\frac{n+3}{4}}\tau^{-\frac{n+3}{4}},
    \label{eq:B_KDNH_L_V}\\
    \xi_{\rm M}&=\tau^{\frac{1}{2}}\eta^{\frac{1}{2}},
    \label{eq:XiM_KDNH_L_V}
\end{align}
in terms of the magnetic field.
These solutions imply
\begin{align}
    \eta^{-1}\xi_{\rm M}^2&=\tau,
    \label{eq:KDNH_L_V_DecayTime}
\end{align}
which is the corresponding equation to Eq.~\eqref{eq:TimeScaleCondition_general} in this regime.
Note that, since ${\rm Re}_{\rm K}=\tau_\eta(\xi_{\rm K})/\tau_{\rm eddy}$, Eqs.~\eqref{eq:Decaytime_Eddy} and \eqref{eq:Condition_DecayTimeScale_KDL} coindides to give the same solutions in the linear and nonlinear regimes at the boundary, ${\rm Re}_{\rm K}=1$.

\subsubsection{Linear and dragged regime (\ctext{16})}
\label{sec:KDNH_LD}
In this regime, the kinetic Reynolds number is small and the decay is governed by the dissipation by drag force at $\xi_{\rm K}$.
In addition to Eq.~\eqref{eq:KDNH_condition},
\begin{align}
    {\rm Re}_{\rm K}\ll1,\quad
    r_{\rm diss}(\xi_{\rm K})\gg1
\end{align}
are satisfied.

We would impose $\tau_\alpha=\tau$, if a scaling solution exists in this regime.
However, we cannot employ it to determine the scaling behaviour of the system because $\tau_\alpha:=\alpha^{-1}$ is independent of $v$ and $\xi_{\rm K}$.
Contrary to the other regimes in Sec.~\ref{sec:KDNH}, the decay of the velocity field cannot slow down the decay process.
Since we are assuming that the system is not frozen, we have $\tau_\alpha\ll\tau$.
Then, the velocity field exponentially decays within a Hubble time, implying that the system quickly enters the magnetically dominated regimes.

\subsection{Consistency among the analyses}
\label{sec:ConsistencyoftheAnalyses}
In this section, we will combine all the results that we obtained so far.
Table \ref{tb:ResultsSummary} summarizes the results.
We suppose that the system has never been frozen and employ Eq.~\eqref{eq:TimeScaleCondition_general}, as is already stated in the beginning of this chapter.
At this point, the analysis for each regime is still patchy.
However, by confirming the consistency at the boundaries, we can draw a whole picture of how magnetic fields decay, without worrying about details of intermediate stages between different regimes and possible conflicts between analyses in different regimes.

Let us start by explaining two concerns that will be discussed in what follows.
The concerns originats from that the dichotomous criteria of regime classification, as if it is universal for all the regimes, are not universal as conditions of $B$ and $\xi_{\rm M}$.
For instance, a criterion we employed is whether the magnetic Reynolds number ${\rm Re}_{\rm M}$ is large or not.
However, ${\rm Re}_{\rm M}$ as a function of $B$ and $\xi_{\rm M}$ is not unique.
It is the expression in Eq.~\eqref{eq:MDNL_MHVSP_GammaAndReM} for the Sweet--Parker regimes, while it becomes the expression in Eq.~\eqref{eq:MDNL_MHVfast_GammaAndReM} for the fast reconnection regimes.
Therefore, a concern is that there can exist a configuration characterized by $B$ and $\xi_{\rm M}$ such that it is out of both of the regimes.
The other concern is, on the contrary, a configuration such that the regime it belongs to cannot be uniquely determined.
However, our regime-wise analyses are consistent and do not leave untouched parameter regions, which the two concerns suggest, in the $B$-$\xi_{\rm M}$ plane.

We check, in the below, the consistency at every possible boundary between two regimes, apart from whether such a boundary appears in the early universe.

The magnetically dominated, nonlinear, and viscous Sweet--Parker regimes are discussed in Secs.~\ref{sec:MDNL_MHVSP} and \ref{sec:MDNL_NHVSP}.
Since Eq.~\eqref{eq:TimeScaleCondition_MDNL_MHVSP} is common to these two regimes, we combine it with either of Eq.~\eqref{eq:Constraint_MaximallyHelical} or \eqref{eq:Constraint_NonHelical} to find the solutions.
When the conditions to be in these regimes are violated, the system belongs to other regimes and Eq.~\eqref{eq:TimeScaleCondition_MDNL_MHVSP} no longer holds.
In the following, we will find the boundary of these regimes.

One of the conditions for being in these regimes is the large magnetic Reynolds number ${\rm Re}_{\rm M}\gg1$.
Equations \eqref{eq:MDNL_MHVSP_rdissAndS} and \eqref{eq:MDNL_MHVSP_Gamma} imply that the large magnetic Reynolds number guarantees the smallness of energy ratio $\Gamma$ and the largeness of Lundquist number $S$.
By using the expression in Eq.~\eqref{eq:MDNL_MHVSP_GammaAndReM}, the largeness of the magnetic Reynolds number is rephrased as
\begin{align}
    \xi_{\rm M}\gg \sigma^{-\frac{1}{2}}\tau^{\frac{1}{2}},\quad\text{or equivalently,}\quad
    B\gg (\rho+p)^{\frac{1}{2}}\eta^{\frac{1}{2}}\tau^{-\frac{1}{2}}.
    \label{eq:condition_SP_ToBe_NL}
\end{align}
When these conditions are violated, magnetic reconnection no longer continues, and the system enters the magnetically dominated, linear, and viscous regimes discussed in Secs.~\ref{sec:MDL_MHV} and \ref{sec:MDL_NHV}.

Another condition is that the Lundquist number is not too large.
By using the expression in Eq.~\eqref{eq:MDNL_MHVfast_rdissAndS}, the condition is rephrased as
\begin{align}
    \xi_{\rm M}\ll S_{\rm c}^{\frac{1}{4}}\sigma^{-\frac{1}{2}}\tau^{\frac{1}{2}},\quad\text{or equivalently,}\quad
    B\ll S_{\rm c}^{\frac{3}{4}}(\rho+p)^{\frac{1}{2}}\eta^{\frac{1}{2}}\tau^{-\frac{1}{2}}.
    \label{eq:Boundary_VSPfast1}
\end{align}
When these conditions are violated, the Sweet--Parker current sheets become unstable, and the system enters the magnetically dominated, non-linear, and viscous fast reconnection regimes discussed in Secs.~\ref{sec:MDNL_MHVfast} and \ref{sec:MDNL_NHVfast}.

Also, shear viscosity should be more significant than the drag term at the kinetic coherence length.
By using the expression in Eq.~\eqref{eq:MDNL_MHVfast_rdissAndS}, we rephrase it as
\begin{align}
    \xi_{\rm M}\gg \sigma^{-1}\eta^{-\frac{1}{2}}\alpha^{\frac{1}{2}}\tau,\quad\text{or equivalently,}\quad
    B\gg (\rho+p)^{\frac{1}{2}}\sigma^{-\frac{3}{2}}\eta^{-1}\alpha^{\frac{3}{2}}\tau.
    \label{eq:condition_SP_ToBe_V}
\end{align}
When this condition is violated, the system is in the magnetically dominated, non-linear, and dragged Sweet--Parker regimes discussed in Secs.~\ref{sec:MDNL_MHDSP} and \ref{sec:MDNL_NHDSP}.

We should be careful also that, when mean free paths of particles are longer than the typical length scales of the system, they contribute as rather the drag term than the shear viscosity \cite{Banerjee+04}.
This is because, as opposed to the shear viscosity, which is a macroscopic description of the collision between the constituent particles, the friction between the fluid and the free-streaming particles as the background appears as the drag term.
Then, some portion of the contributions to $\eta$ is switched off and becomes a contribution to $\alpha$ at around $l_{\rm mfp}$, where $l_{\rm mfp}$ is a mean free path of any species of constituent particles of the fluid.
In the early universe, neutrinos and photons decouple from the plasma fluid and then contribute as the drag term. 
Therefore, we have a possibility that
\begin{align}
    \xi_{\rm K}\gg l_{\rm mfp}
\end{align}
determines a boundary with the dragged Sweet--Parker regimes.
Note that, by using Eqs.~\eqref{eq:MDNL_MHVSP_Ratios} and \eqref{eq:MDNL_MHVSP_GammaAndReM}, this condition becomes
\begin{align}
    \xi_{\rm M}\ll \sigma^{-1}\tau l_{\rm mfp}^{-1}
\end{align}
in terms of the magnetic coherence length.
When the magnetic coherence length is large and the magnetic field is strong, the aspect ratio of the current sheet becomes larger and, consequently, shortens the kinetic coherence length. 
When the condition, Eq.~\eqref{eq:condition_SP_ToBe_V}, is considered, this point should be taken into account as well.

Let us focus on the boundary between the viscous Sweet--Parker regimes and the linear and viscous regimes.
The discussion for the former regimes suggests that the boundary is set when the condition, Eq.~\eqref{eq:condition_SP_ToBe_NL}, is violated.
In the latter regimes, on the other hand, we can find the location of the boundary, independently of the analyses in the former regimes.
The expression of the magnetic Reynolds number in Secs.~\ref{sec:MDL_MHV} and \ref{sec:MDL_NHV} is ${\rm Re}_{\rm M}=\sigma v\xi_{\rm M}=\sigma\tau^{-1}\xi_{\rm M}^2$, and the condition for this regime to be linear becomes
\begin{align}
    \xi_{\rm M}\ll \sigma^{-\frac{1}{2}}\tau^{\frac{1}{2}},\quad\text{or equivalently,}\quad
    B\ll (\rho+p)^{\frac{1}{2}}\eta^{\frac{1}{2}}\tau^{-\frac{1}{2}},
    \label{eq:condition_MDLV_ToBe_L}
\end{align}
where we used Eq.~\eqref{eq:Decaytime_MDL_V}.
This is complementary to Eq.~\eqref{eq:condition_SP_ToBe_NL}, and we can set the boundary at
\begin{align}
    \xi_{\rm M}=\sigma^{-\frac{1}{2}}\tau^{\frac{1}{2}},\quad
    B= (\rho+p)^{\frac{1}{2}}\eta^{\frac{1}{2}}\tau^{-\frac{1}{2}}
    \label{eq:Boundary1}
\end{align}
consistently for both of the regimes.

In a parallel manner, we can find all the possible boundaries.
The viscous Sweet--Parker regimes (Secs.~\ref{sec:MDNL_MHVSP} and \ref{sec:MDNL_NHVSP}) and the viscous fast reconnection regimes (Secs.~\ref{sec:MDNL_MHVfast} and \ref{sec:MDNL_NHVfast}) are connected at
\begin{align}
    \xi_{\rm M}= S_{\rm c}^{\frac{1}{4}}\sigma^{-\frac{1}{2}}\tau^{\frac{1}{2}},\quad
    B= S_{\rm c}^{\frac{3}{4}}(\rho+p)^{\frac{1}{2}}\eta^{\frac{1}{2}}\tau^{-\frac{1}{2}},
    \label{eq:Boundary2}
\end{align}
when the Lundquist number becomes the critical value, $S=S_{\rm c}$.
The viscous Sweet--Parker regimes and the dragged Sweet--Parker regimes (Secs.~\ref{sec:MDNL_MHDSP} and \ref{sec:MDNL_NHDSP}) are connected at
\begin{align}
    \xi_{\rm M}=\sigma^{-1}\eta^{-\frac{1}{2}}\alpha^\frac{1}{2}\tau,\quad
    B=(\rho+p)^\frac{1}{2}\sigma^{-\frac{3}{2}}\eta^{-1}\alpha^{\frac{3}{2}}\tau
    \label{eq:Boundry3}
\end{align}
when the ratio of dissipation terms at the kinetic coherence length becomes unity, $r_{\rm diss}(\xi_{\rm K})=1$.
The viscous fast reconnection regimes (Secs.~\ref{sec:MDNL_MHVfast} and \ref{sec:MDNL_NHVfast}) and the dragged fast reconnection regimes (Secs.~\ref{sec:MDNL_MHDfast} and \ref{sec:MDNL_NHDfast}) are not complementary from the construction, because we have a hierarchy of larger to smaller current sheets in these regimes.
When $r_{\rm diss}(\delta_{\rm c})\ll1\ll r_{\rm diss}(\xi_{\rm K})$, the largest structure feels the drag force, but the critical current sheets feel shear viscosity.
Such an intermediate fast reconnection regime is not established, as far as we understand.
However, conversely to the basic motivation we have in this section, we employ the consistency at the regime boundaries as the requirement that supports the reasonableness of analyses and conservatively choose $r_{\rm diss}(\xi_{\rm K})=1$, corresponding to a condition
\begin{align}
    \alpha\tau=S_{\rm c}^{\frac{1}{2}}{\rm Pr}_{\rm M},
    \label{eq:Boundary4}
\end{align}
as the boundary because Eqs.~\eqref{eq:MDNL_MHVfast_Decaytime} and \eqref{eq:MDNL_MHDfast_DecayTime} coincide when this condition holds.
The dragged Sweet--Parker regimes (Secs.~\ref{sec:MDNL_MHDSP} and \ref{sec:MDNL_NHDSP}) and the dragged fast reconnection regimes (Secs.~\ref{sec:MDNL_MHDfast} and \ref{sec:MDNL_NHDfast}) are connected at
\begin{align}
    \xi_{\rm M}=S_{\rm c}^{\frac{1}{4}}\sigma^{-\frac{1}{2}}\tau^\frac{1}{2},\quad
    B=S_{\rm c}^{\frac{1}{2}}(\rho+p)^{\frac{1}{2}}\sigma^{-\frac{1}{2}}\alpha^\frac{1}{2},
    \label{eq:Boundary5}
\end{align}
when the Lundquist number takes the critical value.
The magnetically dominated linear viscous regimes (Secs.~\ref{sec:MDL_MHV} and \ref{sec:MDL_NHV}) and the magnetically dominated linear dragged regimes (Secs.~\ref{sec:MDL_MHD} and \ref{sec:MDL_NHD}) are connected at
\begin{align}
    \xi_{\rm M}=\eta^{\frac{1}{2}}\alpha^{-\frac{1}{2}},\quad
    B=(\rho+p)^{\frac{1}{2}}\eta^{\frac{1}{2}}\tau^{-\frac{1}{2}},
    \label{eq:Boundary6}
\end{align}
when the dissipation terms become comparable at the kinetic coherence length becomes unity, $r_{\rm diss}(\xi_{\rm K})=1$.
The magnetically dominated linear dragged regimes (Secs.~\ref{sec:MDL_MHD} and \ref{sec:MDL_NHD}) and the dragged Sweet--Parker regimes (Secs.~\ref{sec:MDNL_MHDSP} and \ref{sec:MDNL_NHDSP}) are connected at
\begin{align}
    \xi_{\rm M}=\sigma^{-\frac{1}{2}}\tau^{\frac{1}{2}},\quad
    B=(\rho+p)^{\frac{1}{2}}\sigma^{-\frac{1}{2}}\alpha^{\frac{1}{2}},
    \label{eq:Boundary7}
\end{align}
when the magnetic Reynolds number becomes unity.

\begin{figure}[ht]\begin{center}
    \begin{minipage}[h]{0.95\hsize}
    \includegraphics[keepaspectratio, width=0.95\textwidth]{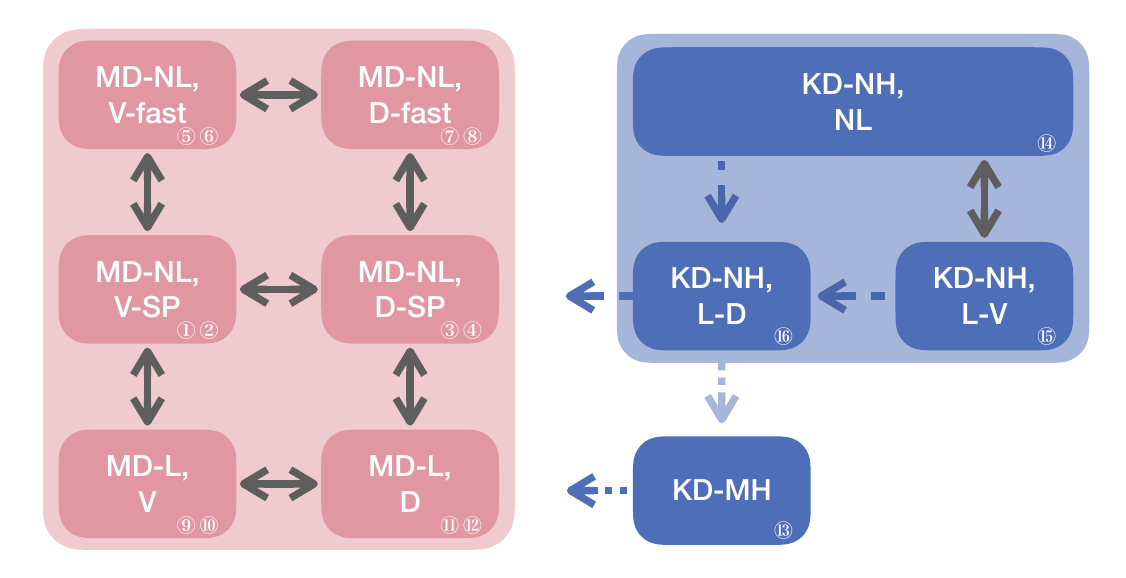}
    \end{minipage}
    \caption{\label{fig:RegimesTransition} Summary of Sec.~\ref{sec:ConsistencyoftheAnalyses}. As for the possible transitions between different regimes, our regime-wise analyses are consistent at the boundaries of the regimes (gray solid arrows). When $\alpha\tau\gg1$, one-way transitions from kinetically dominated regimes to a magnetically dominated regime is expected (blue-dashed arrows). When the magnetic field becomes maximally helical in a kinetically dominated regime, one-way transitions to a magnetically dominated regime are expected (blue-dotted arrows).}\end{center}
    \vspace{5mm}
\end{figure}

See the red-shaded region in Fig.~\ref{fig:RegimesTransition}.
We summarized the possible transitions between magnetically dominant regimes (gray solid arrows).
Importantly, the network of the transitions is closed and no transition from a magnetically dominated regime to a kinetically dominated regime exists because a small energy ratio, $\Gamma\ll1$, is always satisfied in magnetically dominant regimes, even at their boundaries.

Let us discuss the cases where the kinetic energy is initially dominant.
The non-helical and turbulent regime (Sec.~\ref{sec:KDNH_NL}) and the non-helical, linear, and viscous regime (Sec.~\ref{sec:KDNH_LV}) have scaling solutions.
These regimes are consistent at their boundary
\begin{align}
    \xi_{\rm M}=\eta^{\frac{1}{2}}\tau^{\frac{1}{2}},\quad
    B=(\rho+p)^{\frac{1}{2}}\eta^{\frac{1}{2}}\tau^{\frac{3}{2}},
    \label{eq:Boundary8}
\end{align}
when the kinetic Reynolds number becomes unity.
We have another important boundary for the kinetically dominated regimes at
\begin{align}
    \alpha\tau=1.
    \label{eq:Boundary9}
\end{align}
When $\alpha\tau\gg1$, the non-helical, linear, and viscous regime becomes the non-helical, linear, and dragged regime (Sec.~\ref{sec:KDNH_LD}) and quickly enters a magnetically dominant regime because $r_{\rm diss}=\alpha\tau$.
The non-helical and turbulent regime also becomes the non-helical, linear, and dragged regime and is quickly dominated by the magnetic energy because ${\rm Re}_{\rm K}=(\alpha\tau)^{-1}$.
These transitions into the magnetically dominated regimes are represented by the blue-dashed arrows in Fig.~\ref{fig:RegimesTransition}.

Finally, suppose that the magnetic field is partially helical with a tiny helicity fraction, $0<\vert\epsilon\vert\ll1$.
Then there exists a possibility that the non-helical evolution terminates at some point, when the inequality \eqref{eq:Realizability_Integrated} is saturated.
This is likely to happen, assuming a constant enthalpy density $\rho+p$ and the reasonable slope of the velocity IR spectrum $n>-2$, because $B^2\xi_{\rm M}$ decreases as the magnetic field decays.
After that, the magnetic field becomes maximally helical and enters the maximally helical regime (Sec.~\ref{sec:KDMH}).
Then, the system quickly becomes magnetically dominant (blue-dotted arrows in Fig.~\ref{fig:RegimesTransition}). 

\subsection{Applicability}
\label{sec:applicability}
Let us figure out the range of applicability.
Regimes out of the applicability range are beyond the scope of this manuscript.

First, we assumed a large electric conductivity.
We have two conditions on this point.
One is that a large magnetic Prandtl number is necessary to justify the approximations we have made.
This condition is always satisfied in the early universe before the recombination.
Because of the large magnetic Prandtl number in the early universe, we have safely approximated in the reconnection regimes that a large fraction of the injected magnetic energy is almost dissipated in the current sheet.
The other is that, to treat magnetic helicity as a conserved quantity and employ Eqs.~\eqref{eq:Constraint_MaximallyHelical} and \eqref{eq:Constraint_NonHelical}, the condition in Eq.~\eqref{eq:ConditionToBelieveHelicityConservation} is necessary unless Eq.~\eqref{eq:harmonic} holds.
If this condition is violated, we do not have any trustable conserved quantity in magneto-hydrodynamics.
To be conservative, a possible option may be employing the magnetic energy at large scales as a conserved quantity \cite{Banerjee+04}, since we do not have any reasons to expect the inverse transfer without helicity conservation.

Second, the fluid approximation we made is violated if the length scales involved in the analysis are too small.
The smallest length scales that appeared in our analysis are $\delta_{\rm c}$ for the fast reconnection regimes and $\xi_{\rm K}$ for the other regimes.
Before the recombination epoch, these length scales are to be compared with the mean free paths of electrons and positrons.
If the mean free paths are longer, the fluid approximation completely breaks down, and the system is no longer within the framework of magneto-hydrodynamics.
After the recombination epoch, these length scales are to be compared with the ion inertial length.
If protons decouple from the fluid of electrons and violate the single-fluid approximation, then the system becomes the so-called kinetic or collisionless regime \cite{galtier2016introduction}.
The reconnection time scale in the collisionless regime is empirically known to be $\tau_{\rm collisionless}\sim 10 \rho^\frac{1}{2} B^{-1}\xi_{\rm M}$ \cite{galtier2016introduction,Hosking+22}.
In Ref.~\cite{Hosking+22}, they considered such a regime and concluded that this reconnection is faster than the photon drag on the inflow discussed around Eq.~\eqref{eq:Smallness_DissipatedInflowVelocity_Drag}, implying that the photon drag determines the time scale of the decay in this regime.

Finally, our analysis could be updated, with the strategy kept the same.
While we are conservative in this manuscript, a trial of discussing the effect of strong magnetic fields on the transport coefficients \cite{1965RvPP....1..205B} and taking the collisionless regime into account can be found in the literature \cite{Hosking+22}.
Also, apart from the Sweet--Parker model, theoretical understandings of magnetic reconnection are not well-established \cite{ji2011phase}.
Numerical or experimental studies will provide us with both support for the existing understanding and hints for deeper theoretical understandings, and magnetic reconnection models that are not established may turn out to play a crucial role in the evolution of primordial magnetic fields.
If some regimes that are not considered in this paper turn out to be important, they could be incorporated into the analysis.

\bibliographystyle{unsrt}
\bibliography{Revise}
\end{document}